\documentclass[10pt,journal]{IEEEtran}

\usepackage{multirow}
\usepackage[dvips]{graphicx}
\usepackage{amsmath}
\usepackage[psamsfonts]{amssymb}
\usepackage{amsxtra}
\usepackage{threeparttable}
\usepackage{relsize}

\usepackage{stmaryrd}
\usepackage{amsfonts,amsmath,amssymb}

\usepackage{color}
\usepackage[table]{xcolor}

\usepackage{rotating}

\usepackage{cite}

\usepackage{algorithm}
\usepackage{algorithmic}
\usepackage{booktabs}
\usepackage{url}

\usepackage{colortbl}

\definecolor{shadingcolor}{rgb} {.87,    0.92,    .98}
\newcommand{\shadingbox}[1]{   %% shading for table
    \fboxsep 0pt
    \colorbox{shadingcolor}{
        {\hskip -2pt #1}\hskip -2.5pt
    }
}

\newcommand{\minitab}[2][l]{\begin{tabular}{@{}#1}#2\end{tabular}}

\input{def1.set}

\usepackage{mathpazo}

\usepackage[mathscr]{eucal}

\newcommand{\matn}[2][n]{\ensuremath{\mathbf{#2}^{(#1)}}}
\usepackage[font=small,labelfont=bf,tableposition=top]{caption}
\usepackage[font=footnotesize]{subcaption}

\begin{document}
\title{Era of Big Data Processing: A New Approach  via Tensor Networks and Tensor Decompositions}

\author{Andrzej CICHOCKI\\
RIKEN Brain Science Institute, Japan \\
and  Systems Research Institute of the  Polish Academy of Science, Poland\\
a.cichocki@riken.jp\\
Part of this work was presented on the International Workshop on Smart Info-Media Systems in Asia,
 (invited talk - SISA-2013) Sept.30--Oct.2, 2013, Nagoya, Japan}

\maketitle

%\title{Era of Big Data Processing: A New Approach  via Tensor Networks and Tensor Decompositions}
%
%
%
%\author{%
%\authorblockN{%
%Andrzej CICHOCKI \\
% %International Workshop on Smart Info-Media Systems in Asia \\
%% (SISA-2013) Sept.30--Oct.2,2013, Nagoya, Japan (invited talk)\\ %\authorrefmark{1}
%}
%%
%\authorblockA{%
%%\authorrefmark{1}
%RIKEN Brain Science Institute, JAPAN \\
%and  Systems Research Institute of the  Polish Academy of Science, POLAND\\
%a.cichocki@riken.jp\\
%Part of this work was presented on the International Workshop on Smart Info-Media Systems in Asia,
% (SISA-2013) Sept.30--Oct.2,2013, Nagoya, Japan (invited talk)\\ %\authorrefmark{1
%%+81-48-467-9668
%}
%}

%\maketitle

\begin{abstract}
Many problems  in  computational neuroscience, neuroinformatics, pattern/image recognition, signal processing and machine learning generate massive amounts of multidimensional data with multiple aspects and high dimensionality. Tensors (i.e., multi-way arrays)  provide often a natural and compact representation for such massive multidimensional data via suitable low-rank approximations.
Big data analytics  require novel technologies to efficiently process huge datasets within tolerable elapsed times. Such a new emerging technology for multidimensional big data is a multiway analysis via tensor networks (TNs) and tensor decompositions (TDs) which represent tensors by sets of factor (component) matrices and lower-order (core) tensors.   Dynamic tensor analysis allows us to discover meaningful hidden structures of complex data and to perform generalizations by capturing multi-linear and multi-aspect relationships.   We will discuss some fundamental TN models, their mathematical and graphical descriptions and associated learning algorithms for large-scale TDs and TNs, with many potential applications including: Anomaly detection, feature extraction, classification, cluster analysis, data fusion and integration, pattern recognition,  predictive modeling, regression, time series analysis and  multiway component analysis.\\

\end{abstract}

\section{\bf Introduction and Motivations}

\begin{figure}[ht]
\centering
\includegraphics[width=8.99cm,]{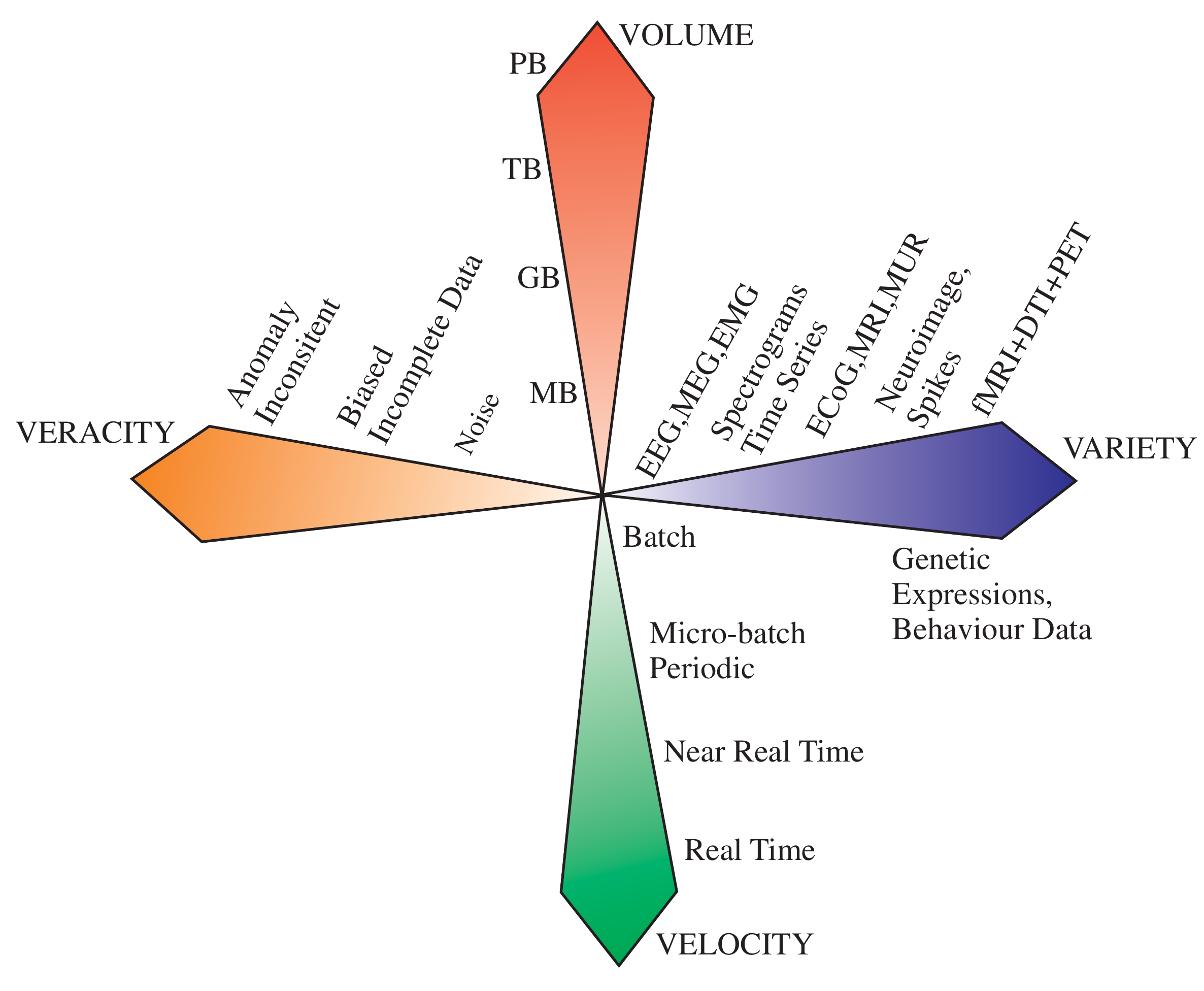}
\caption{Big Data analysis for neuroscience recordings.  Brain data can be recorded by electroencephalography (EEG), electrocorticography (ECoG), magnetoencephalography (MEG), fMRI, DTI, PET, Multi Unit Recording (MUR). This involves analysis of multiple modalities/multiple subjects neuroimages, spectrograms, time series, genetic and behavior data. One of the challenges in computational and system neuroscience is to make fusion (assimilation) of such data and to understand the multiple  relationships  among them in such tasks as  perception, cognition and social interactions. The our ``V''s  of big data: Volume - scale of data, Variety - different forms of data, Veracity - uncertainty of data, and Velocity - analysis of streaming data, comprise the challenges ahead of us.}
\label{Fig:Big-data}
\end{figure}

Big Data consists of multidimensional, multi-modal data-sets that are  so huge and complex that they  cannot be easily stored or processed by using standard computers. Big data are characterized not only by big Volume but also another specific ``V'' features (see Fig. \ref{Fig:Big-data}). High Volume implies the need for algorithms that are scalable; High Velocity address the challenges  related to process data in near real-time, or virtually real-time; High Veracity  demands robust and predictive algorithms for noisy, incomplete or inconsistent data, and finally, high Variety may require integration across  different kind of data, e.g., neuroimages, time series, spiking trains, genetic and behavior data.
 Many challenging problems for big data  are related to capture, manage, search, visualize, cluster, classify, assimilate, merge, and process
 the data within a tolerable elapsed time, hence demanding new innovative solutions and technologies. Such  emerging technology  is  Tensor Decompositions (TDs) and Tensor Networks (TNs) via low-rank matrix/tensor approximations.
The challenge is how to analyze large-scale, multiway data sets. Data explosion creates deep research challenges that require new scalable, TD and TN algorithms.

\begin{figure}%[h]
\centering
\includegraphics[width=4.6cm]{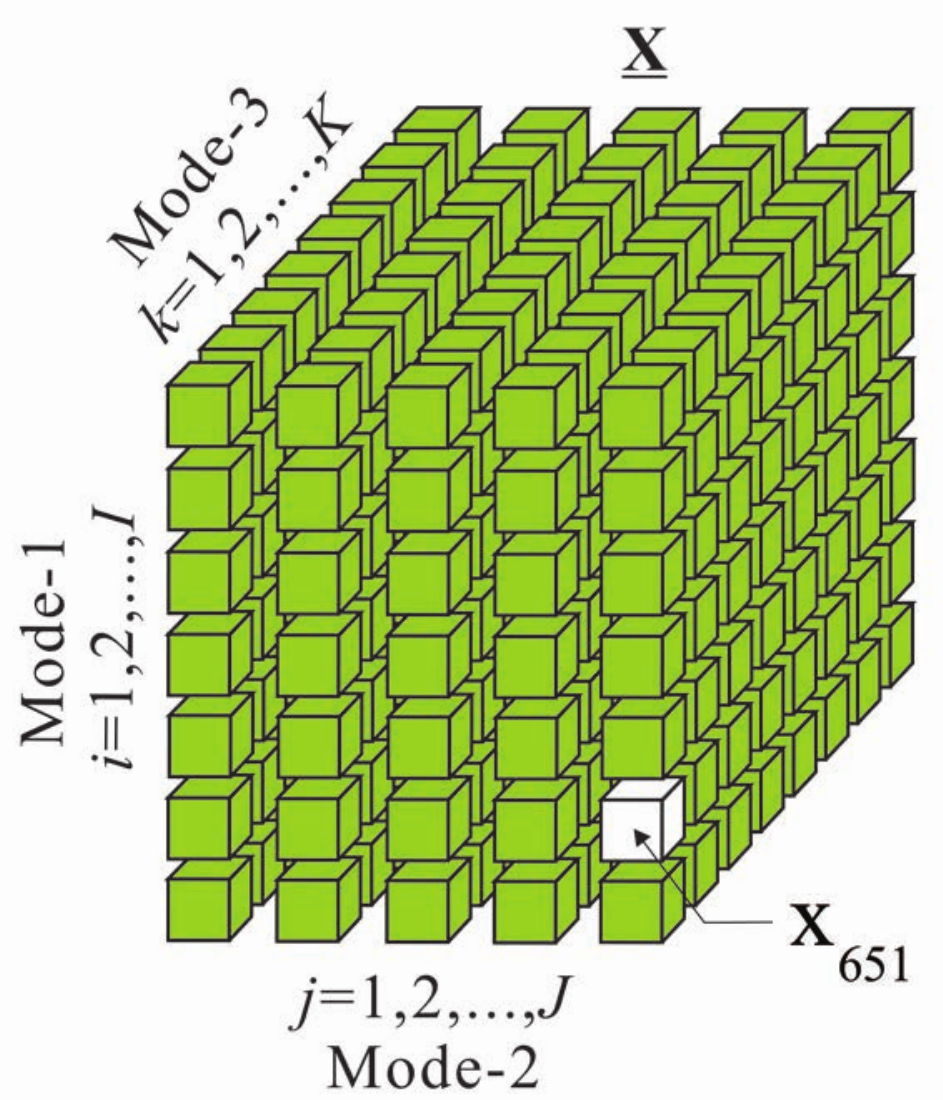}\\
\vspace{0.4cm}
\includegraphics[width=8.6cm]{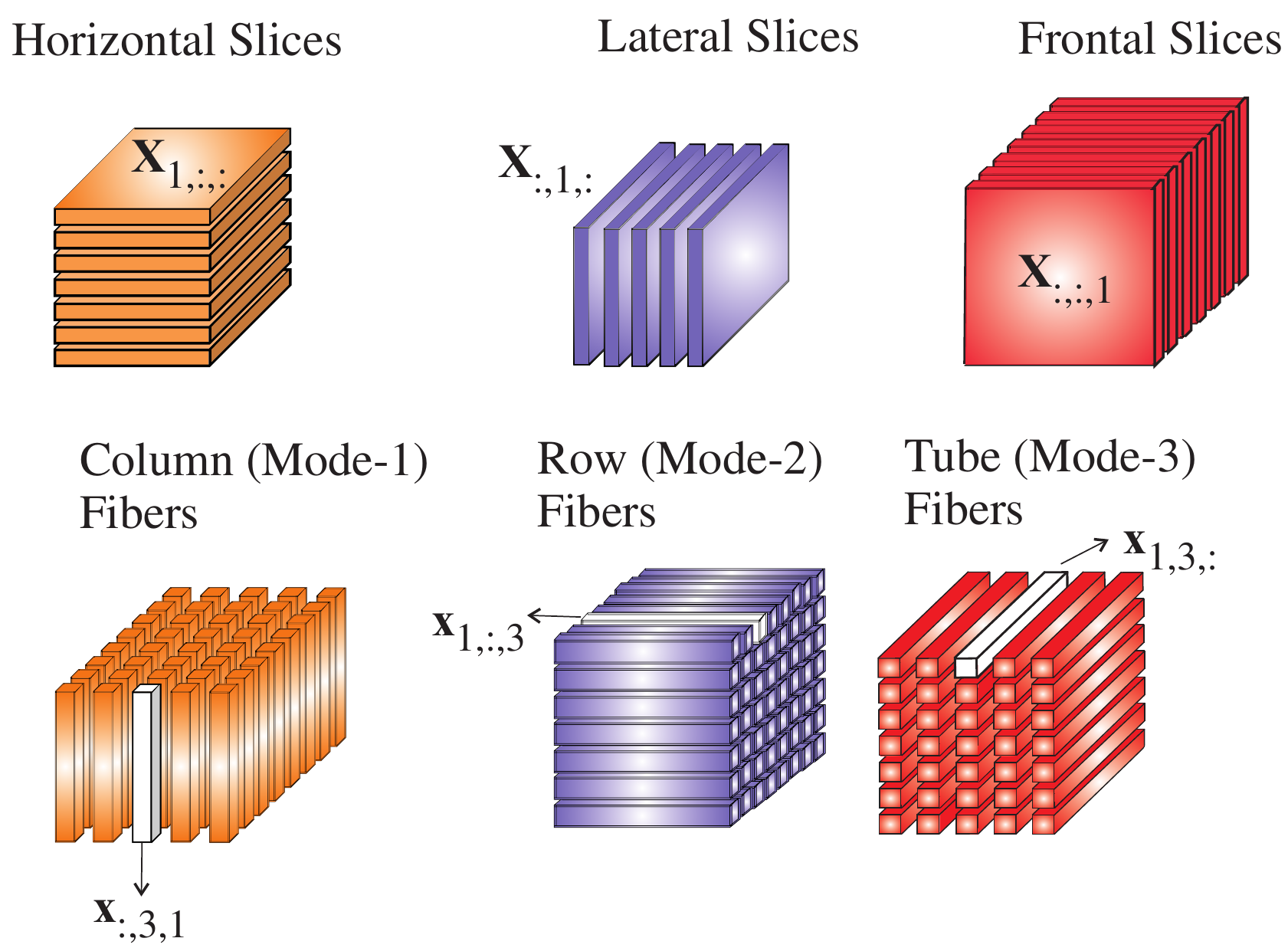}
%\\
%\includegraphics[width=7.1cm,height=4.1cm]{matrix.eps}
\caption{A 3rd-order tensor  $\underline \bX \in \Real^{I \times J \times K}$, with
entries $x_{i, j, k}=\underline \bX(i,j,k)$ and its sub-tensors: Slices  and  fibers.
All  fibers are treated as column vectors.}
\label{Fig:fibers}
\end{figure}

\begin{figure}[ht]
(a)\\
\begin{center}
\hspace{-.8cm}
\includegraphics[width=4.6cm]{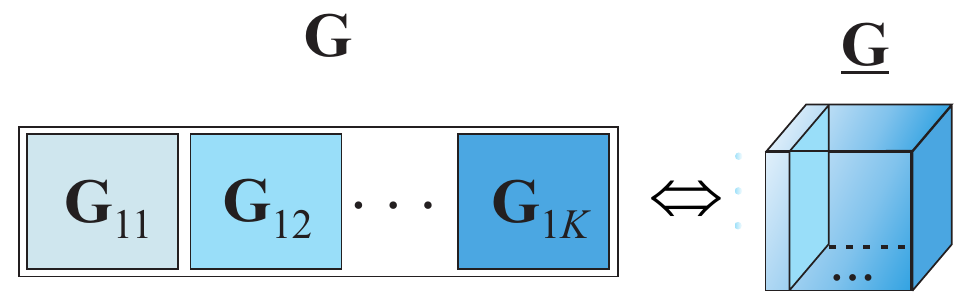}\\
\end{center}
(b)\\
\begin{center}
\includegraphics[width=5.6cm]{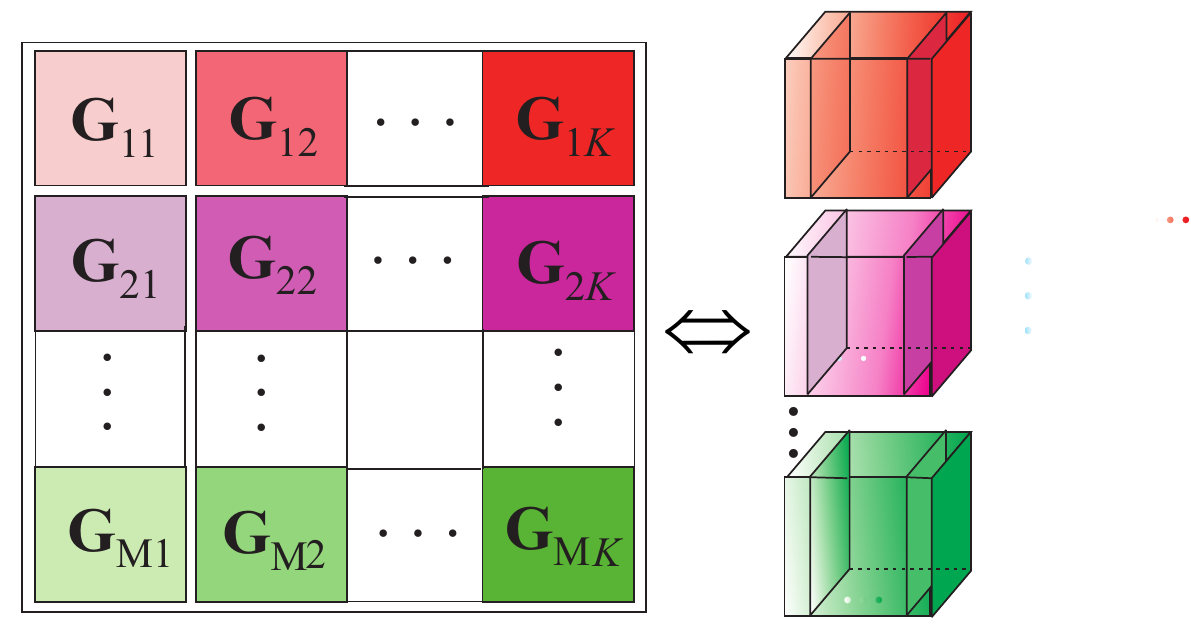}
\end{center}
\caption{Block matrices  and their representations by (a) a 3rd-order tensor and (b) a 4th-order tensor.}
\label{Fig:Tens-block-matr}
\end{figure}

 Tensors are adopted in  diverse branches of science  such as a  data analysis, signal and image processing \cite{NMF-book,Cich-Lath,Cichocki-SICE,Kolda08}, Psychometric, Chemometrics, Biometric, Quantum Physics/Information, and Quantum Chemistry  \cite{Kroonenberg,Smilde,Hackbush2012}.
Modern scientific areas such as bioinformatics or computational neuroscience  generate massive amounts of data collected in various forms of large-scale, sparse tabular,  graphs or networks  with multiple aspects and high dimensionality.

 Tensors, which are multi-dimensional generalizations of matrices (see Fig. \ref{Fig:fibers} and Fig. \ref{Fig:Tens-block-matr}), provide often a useful representation for such data. Tensor decompositions (TDs) decompose  data tensors  in factor matrices, while tensor networks (TNs) represent higher-order tensors by interconnected lower-order tensors.
 %(see Fig. \ref{Fig:fibers} and Fig. \ref{Fig:Tens-block-matr}).
%%%%%%%%
We  show that TDs and TNs provide natural extensions of  blind source separation (BSS) and 2-way (matrix) Component Analysis (2-way CA) to multi-way component analysis (MWCA) methods.
 In addition, TD and TN algorithms are suitable for dimensionality reduction and they can handle missing values, and noisy data.
Moreover, they are potentially useful
for analysis of linked  (coupled) block of tensors with millions and even billions of non-zero entries, using the map-reduce paradigm, as well as divide-and-conquer approaches \cite{Suter13,Wang-out-core05,Phan-CP}.
This all suggest that   multidimensional data can  be represented by linked
multi-block  tensors which can be decomposed into common (or correlated) and distinctive (uncorrelated, indpendent) components \cite{Cichocki-SICE,Zhou-PAMI,Yokota2012}.
Effective analysis of coupled tensors requires the development of new models and associated
algorithms  that can identify the core relations that exist
among the different tensor  modes, and the same tome scale to extremely large datasets.
Our  objective is to develop  suitable models and algorithms for linked low-rank
tensor approximations (TAs), and associated scalable software  to
make such analysis possible.

Review and tutorial papers \cite{Cich-Lath,Kolda08,Comon-ALS09,Lu-2011,Morup11}  and books \cite{Smilde,Kroonenberg,NMF-book} dealing with TDs   already exist,
however, they typically focus  on standard TDs and/or do not  provide explicit links
 to  big data processing topics and do not explore natural connections with  emerging
areas including  multi-block coupled tensor analysis and tensor networks.
This paper extends beyond the standard TD models
and aims to elucidate the power and flexibility of TNs  in the analysis of multi-dimensional, multi-modal, and multi-block  data, together with their role as a mathematical backbone for the discovery of hidden structures in  large-scale data \cite{NMF-book,Cich-Lath,Kolda08}.

{\bf Motivations - Why low-rank tensor approximations?}   A wealth of literature on  (2-way) component analysis (CA)  and BSS exists, especially on  Principal Component Analysis (PCA), Independent Component Analysis (ICA), Sparse Component Analysis (SCA), Nonnegative Matrix Factorizations (NMF), and Morphological Component Analysis (MCA) \cite{Comon-Jutten2010,NMF-book,Torre2012}.
These techniques are maturing, and are promising tools for
 blind source separation (BSS),  dimensionality reduction, feature extraction,  clustering, classification, and visualization \cite{NMF-book,Torre2012}.

The ``flattened view'' provided by  2-way CA and matrix factorizations (PCA/SVD, NMF, SCA, MCA) may be inappropriate  for large classes of real-world data which exhibit multiple couplings and cross-correlations. In this context, higher-order tensor networks   give us the opportunity to develop  more sophisticated models performing distributed computing and capturing  multiple interactions and couplings, instead of standard  pairwise interactions.
In other words,  to discover hidden components within multiway data the analysis tools should account for intrinsic multi-dimensional distributed patterns present in the data.

 \begin{figure}[t!]
\centering
%\psfrag{y}{\color{black}$y$}
%\includegraphics[width=8.8cm,height=4.8cm]{TNsymbols1c.eps}
\includegraphics[width=8.8cm,height=4.8cm]{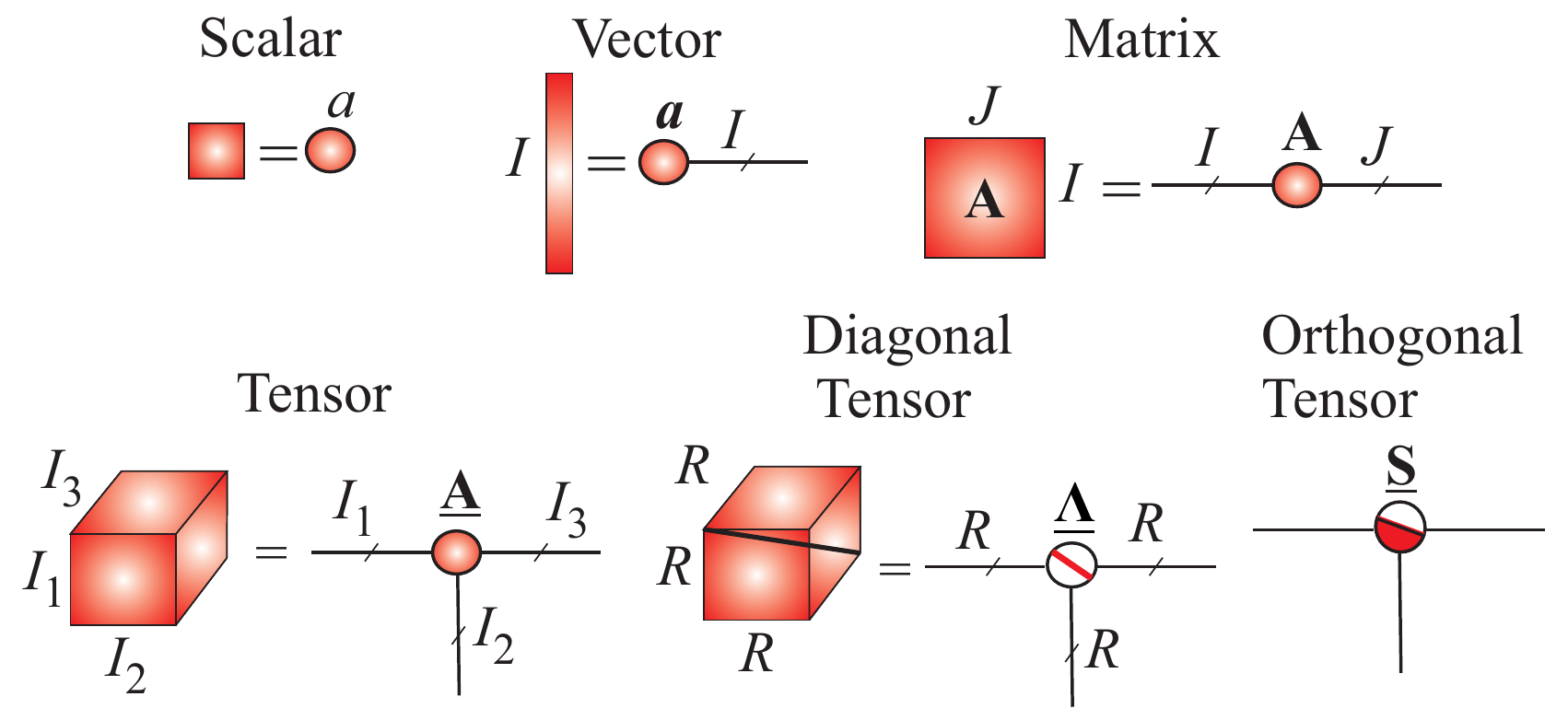}
\caption{Basic symbols for tensor network diagrams.}
\label{Fig:symbols}
\end{figure}

\section{\bf Basic Tensor Operations}

\minrowclearance 2ex
\begin{table}[t!]
\caption{Basic tensor notation and symbols. A tensor are denoted by underline bold capital letters, matrices by uppercase bold letters, vectors by lowercase boldface letters and scalars by lowercase letters.} \centering
 {\small  \shadingbox{
    \begin{tabular*}{\linewidth}[t]{@{\extracolsep{\fill}}ll} \hline
{$\underline \bX \in \Real^{I_1 \times I_2 \times \cdots \times I_N}$} & \minitab[p{.4\linewidth}]{$N$th-order tensor of size $I_1 \times I_2 \times \cdots \times I_N$} \\
{$\underline \bG,  \; \underline \bG_r, \; \underline \bG_{\bX}, \; \underline \bG_{\bY}, \; \underline \bS$}   & {core tensors} \\  [-2ex]
{$\underline {\mbi {\Lambda}} \in \Real^{R \times R \times \cdots \times R}$}   & \minitab[p{.50\linewidth}]{\hspace{-0.6em} $N$th-order diagonal core tensor with nonzero $\lambda_r$ entries on main diagonal} \\  [-2ex]
%{$\tI \in \Real^{I \times I  \times \cdots \times I}$}& \minitab[p{.4\linewidth}]{diagonal tensor of size $I \times I \times \cdots \times I$, with diagonal entries equal to one} \\ [-2ex]
{$\bA = [\ba_1,\ba_2,\ldots,\ba_R] \in \Real^{I \times R} $}& \minitab[p{.50\linewidth}]{matrix with  column vectors $\ba_r \in \Real^I$ and entries $a_{ir}$} \\
{$\bA, \, \bB, \,\bC, \; \bB^{(n)}, \; \bU^{(n)} $}& {component matrices} \\
$\bi = [i_1,i_2, \ldots,i_N]$& vector of indices\\ %[-1ex]
%{$\hat{\underline \bY}$}& {an approximation of $\underline \bY \in \Real^{I_1 \times I_2 \times \cdots \times I_N}$} \\
$\bX_{(n)} \in \Real^{I_n \times I_1 \cdots I_{n-1} I_{n+1} \cdots I_N}$ & \minitab[p{.65\linewidth}]{mode-$n$ unfolding of $\underline \bX$}\\[-3ex]
$\bx_{:,i_2,i_3,\ldots,i_N}$ & \minitab[p{.51\linewidth}]{mode-1 fiber of $\underline \bX$ obtained by fixing all but one index}\\[-3ex]
$\bX_{:,:,i_3,\ldots,i_N}$ & \minitab[p{.50\linewidth}]{tensor slice of $\underline \bX$ obtained by fixing all but two indices}\\[-1ex]
$\underline \bX_{:,:,:,i_4,\ldots,i_N}$ & \minitab[p{.50\linewidth}]{subtensor of $\underline \bX$, in which several indices are fixed}\\[-1ex]
% ${\bA}^{T}$,   ${\bA}^{-1}$, ${\bA}^{\dag}$ & \minitab[p{.4\linewidth}]{transpose, inverse and Moore-Penrose pseudo-inverse of $\bA$} \\[-3ex]
%{$\bC = \bA\otimes\bB$}&\minitab[p{.4\linewidth}]{Kronecker product of $\bA \in \Real^{I\times J}$ and $\bB \in \Real^{L\times M}$, with $c_{L(i-1)+l,M(j-1)+m} = a_{i,j} b_{l,m}$} \\[-3ex]
%{$\bC = \bA\odot \bB$}&\minitab[p{.4\linewidth}]{Khatri-Rao product of $\bA \in \Real^{I\times J}$ and $\bB \in \Real^{K\times J}$ yields $\bC \in \Real^{I K \times J}$, with columns $\bc_j = \ba_j \otimes \bb_j$}\\[-3ex]
%%{$\circledast, \oslash$}&{Hadamard product and division} \\
%{$\underline \bX = \ba \circ \bb \circ \bc \in \Real^{I \times J \times K}$}&\minitab[p{.4\linewidth}]{outer product forms a rank-1tensor with entries $x_{ijk} = a_i b_j c_k$} \\
%
$\bx=\vtr{\underline \bX}$& vectorization of $\underline \bX$  \\
$\diag\{\bullet\}$& diagonal matrix \\
\hline
    \end{tabular*}
    }}
%    \end{center}
\label{table_notation1}
\end{table}
\minrowclearance 0ex
\begin{figure} [h]
(a)\\
\begin{center}
\includegraphics[width=8.6cm]{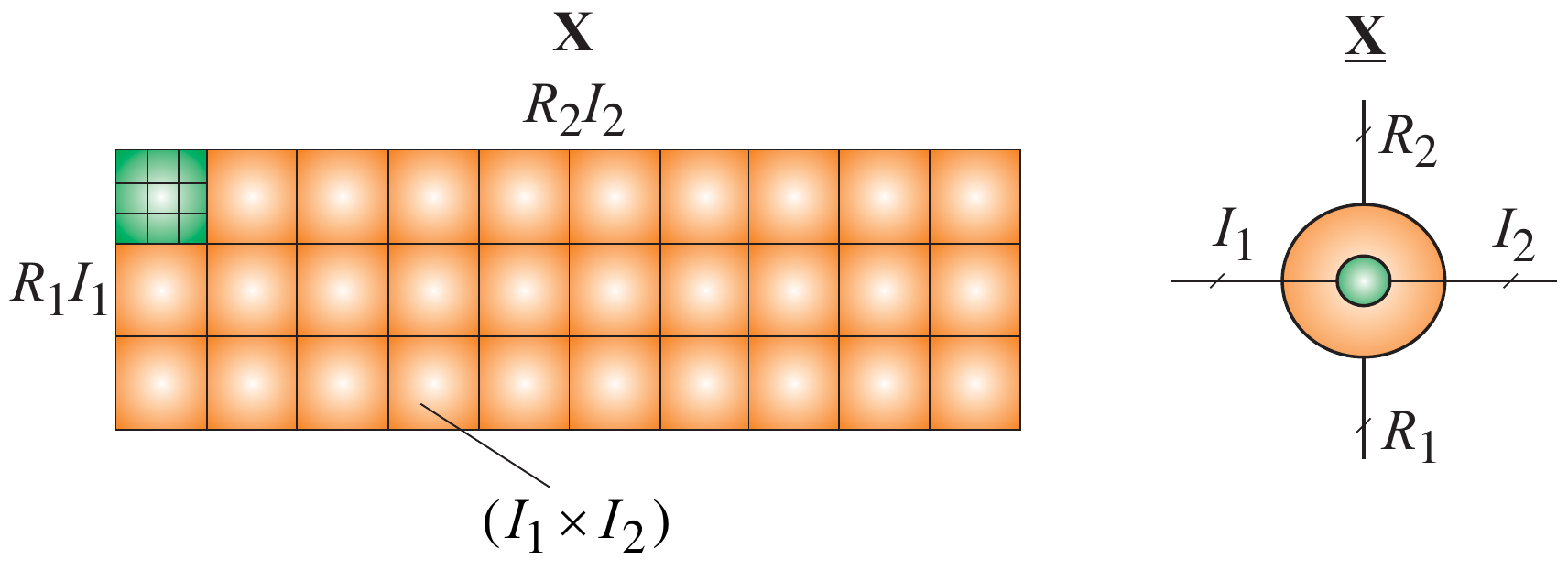}\\
\end{center}
(b)\\
\begin{center}
\includegraphics[width=7.6cm]{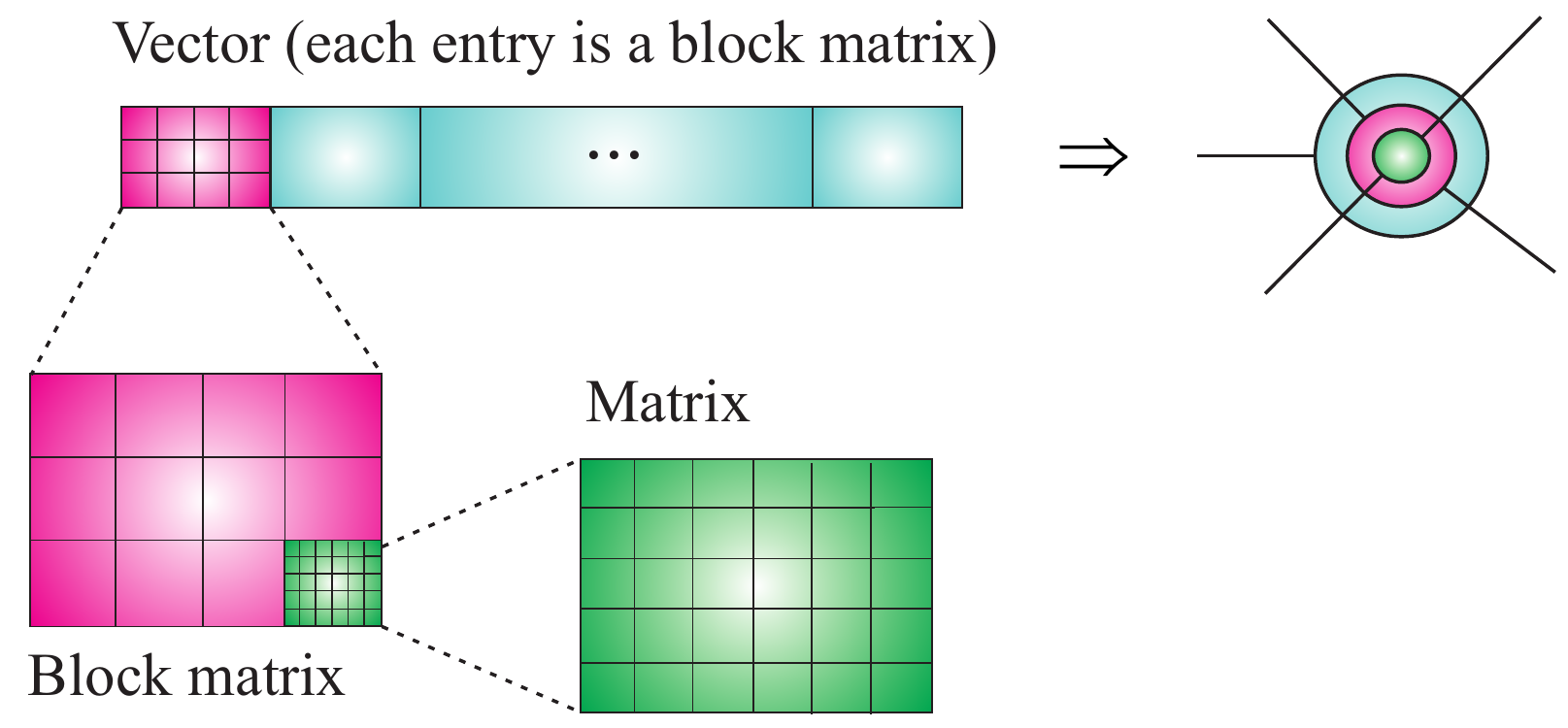}\\
\end{center}
(c)\\
\begin{center}
\includegraphics[width=7.6cm]{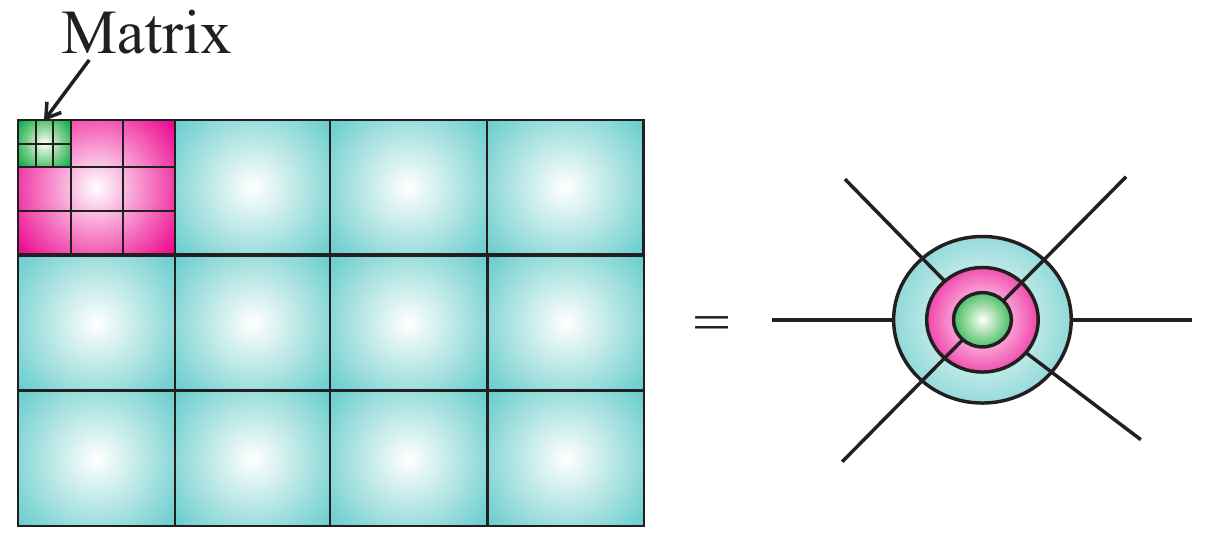}
\end{center}
\caption{Hierarchical block matrices and their representations as  tensors:
(a) a 4th-order tensor for a block matrix $\bX \in \Real^{R_1 I_1 \times R_2 I_2}$, comprising block matrices $\bX_{r_1,r_2} \in \Real^{I_1 \times I_2}$, (b) a 5th-order tensor and (c) a 6th-order tensor.}
\label{Fig:Tens-block5-8}
\end{figure}
\begin{figure}[ht]
\centering
\includegraphics[width=3.8cm]{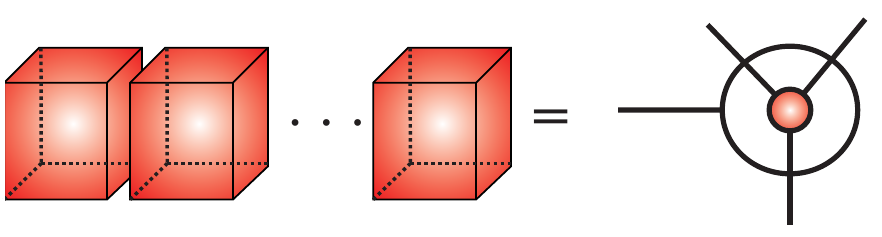}\\
4th-order tensor \\
\vspace{0.3cm}
\includegraphics[width=4.2cm]{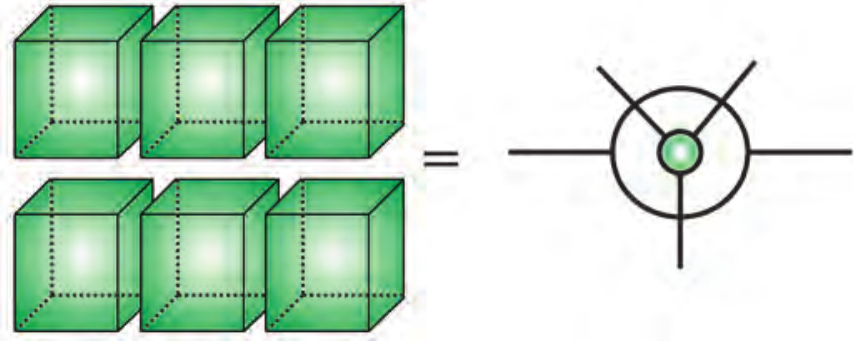}
\hspace{0.1cm}
\includegraphics[width=4.2cm]{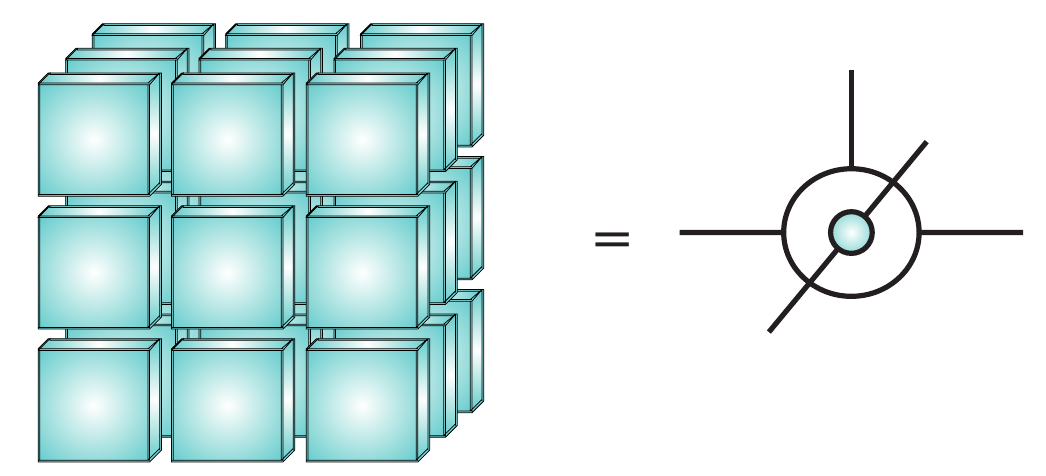}\\
5th-order tensors\\
\vspace{0.4cm}
%4th-order tensor \hspace{2.0cm} 6th-order tensor\\
%{\vspace{0.5cm}
\includegraphics[width=4.0cm]{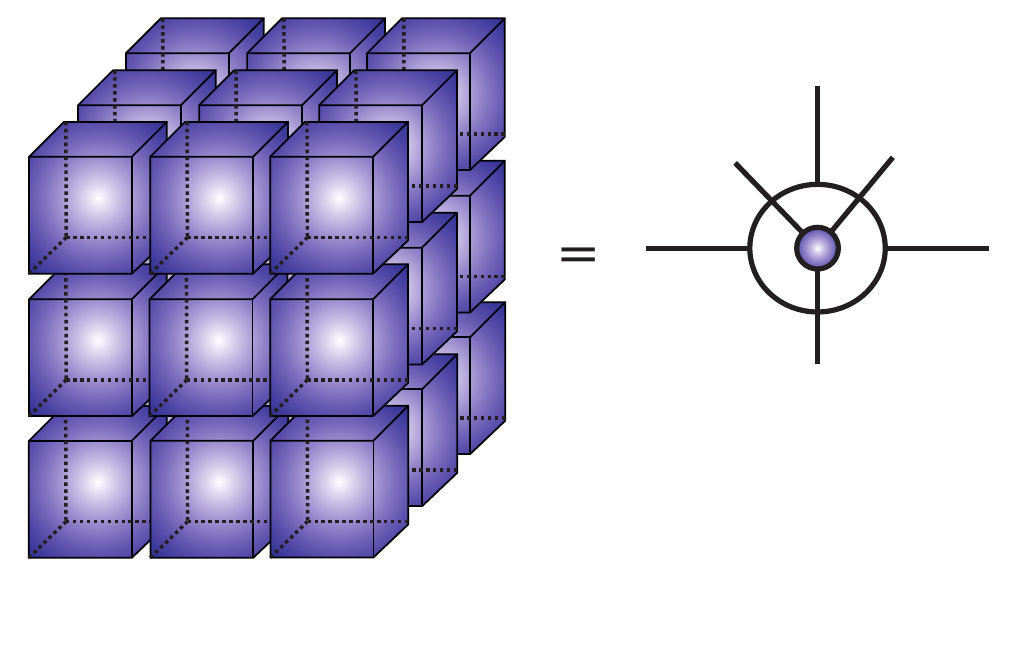}\\
%\vspace{0.1cm}
 6th-order tensor
\caption{Graphical representations and symbols for higher-order block tensors. Each block represents a 3rd-order tensor or 2nd-order tensor. An external circle represent a global structure of the block tensor (e.g., a vector, a matrix, a 3rd-order tensor) and inner circle represents a structure of each element of the block tensor.}
\label{Fig:symbols2}
\end{figure}

\begin{figure}[ht]
%\centering
(a) \\
\includegraphics[width=8.6cm]{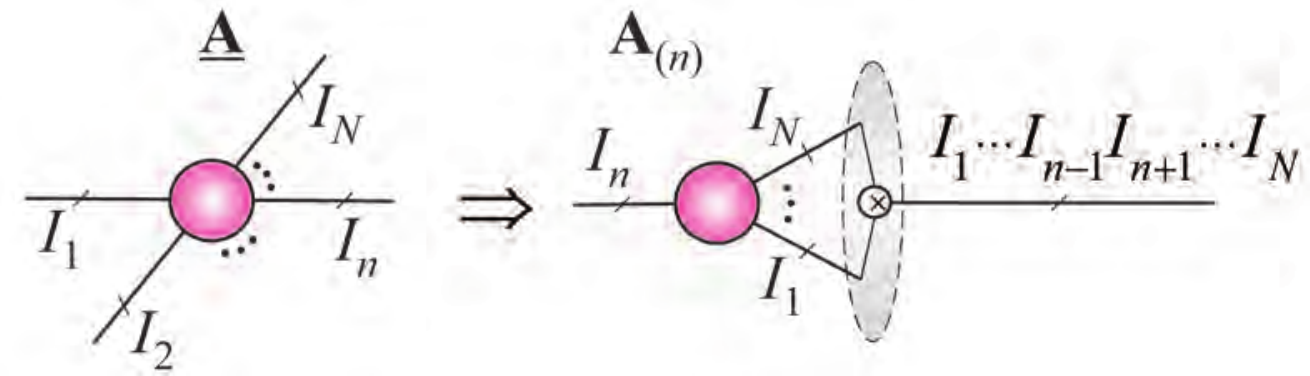}\\
%\includegraphics[width=8.6cm,height=4.6cm]{matrix.eps}\\
%\vspace{0.4cm}
(b)\\
\includegraphics[width=8.6cm]{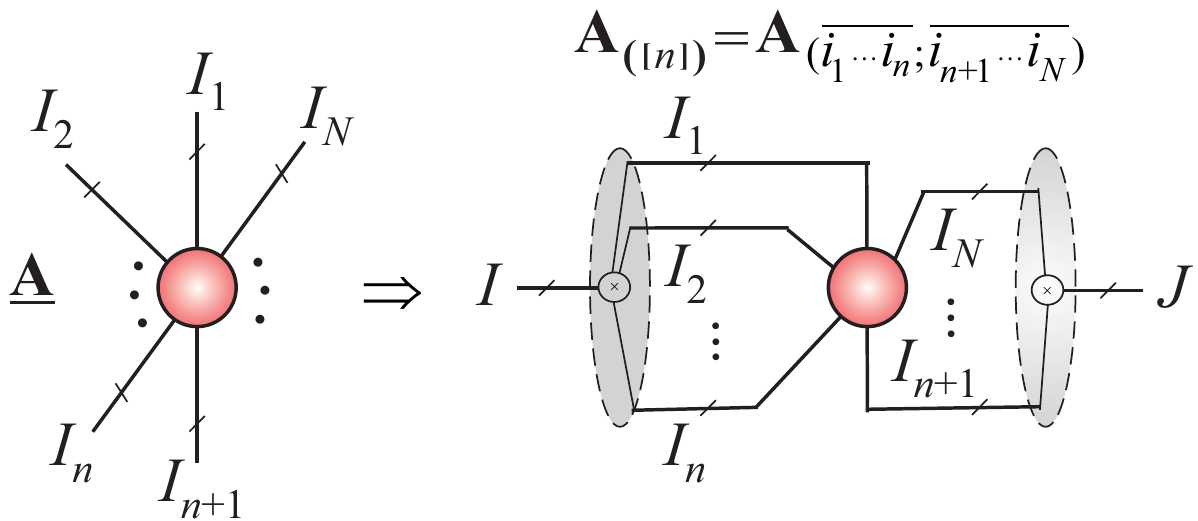}
\caption{Unfoldings in tensor networks: (a) Graphical representation of the basic mode-$n$ unfolding (matricization, flattening) $\bA_{(n)} \in \Real^{I_n \times I_1\cdots I_{n-1} I_{n+1} \cdots I_N}$  for an $N$th-order tensor $\underline \bA \in \Real^{I_1 \times I_2 \times \cdots \times I_N}$. (b) More general  unfolding of the $N$th-order tensor into a matrix $\bA_{([n])} = \bA_{(\overline{i_1, \ldots, i_n}\; ; \; \overline{i_{n+1}, \ldots, i_N})} \in \Real^{I_1 I_2 \cdots I_n \times I_{n+1} \cdots I_N}$. All entries of an unfolded tensor  are arranged in a specific order, e.g., in  lexicographical order. In a more general case, let $\brr=\{m_1,m_2,\ldots,m_R\} \subset \{1,2,\ldots, N\}$ be the row indices and
$\bc=\{n_1,n_2,\ldots, n_C \} \subset \{1,2,\ldots, N\} -\brr$ be the column indices, then the mode-$(\brr,\bc)$ unfolding of $\underline \bA$ is denoted as
$\bA_{(\brr,\bc)} \in \Real^{I_{m_1} I_{m_2} \cdots I_{m_R}  \times I_{n_1} I_{n_2} \cdots I_{n_C}}$.}
\label{Fig:unfolding}
\end{figure}

\minrowclearance 2ex
\begin{table}[ht]
\caption{Basic tensor/matrix operations.} \centering
  {\small \shadingbox{
    \begin{tabular*}{1.04\linewidth}[t]{@{\extracolsep{\fill}}ll} \hline
%$\Real_+$ & nonnegative real number \\
%$\Real^n$ & $n$-dimensional real vector space  \\
%
% tensor-times-matrix
\\[-4em]
$\underline \bC = \underline \bA \times_n \bB$ & \minitab[p{.6\linewidth}]{\\[-3.2ex] mode-$n$ product of $\underline \bA \in \Real^{I_1 \times I_2 \times \cdots \times I_N}$ and $\bB \in \Real^{J_n \times I_n}$ yields $\underline \bC \in  \Real^{I_1 \times \cdots \times I_{n-1} \times J_n \times I_{n+1} \times \cdots \times I_N}$, with entries $c_{i_1, \ldots, i_{n-1}, \, j, \, i_{n+1}, \ldots i_N} = \sum_{i_n=1}^{I_n} a_{i_1, \ldots, i_n, \ldots, i_N} b_{j, \, i_n}$, or equivalently $\bC_{(n)} = \bB \, \bA_{(n)}$} \\
\multicolumn{2}{@{\hspace{-.1ex}}l}{$\underline \bC = \llbracket \underline \bA; \bB^{(1)},  \ldots, \bB^{(N)}\rrbracket = \underline \bA \times_1 \bB^{(1)} \times_2 \bB^{(2)}
\cdots \times_N \bB^{(N)} $}\\
%
%\multicolumn{2}{@{\hspace{-.1ex}}l}{Multilinear product returns tensor $\underline \bC \in \Real^{I_1 \times I_2 \times \cdots \times I_N}$, with} \\
%\multicolumn{2}{@{\hspace{-.1ex}}l}{entries $c_{i_1,i_2,\ldots, i_N}=\sum_{r_1,r_2,\ldots,r_N}^{R_1,R_2,\ldots,R_N} a_{r_1,r_2,\ldots, r_N} b^{(1)}_{i_1,r_1} b^{(2)}_{i_2,r_2} \cdots b^{(N)}_{i_N,r_N}$}\\

%$\underline \bC = \llbracket \underline \bA; \bB^{(1)},  \ldots, \bB^{(N)}\rrbracket =  {\underline \bA} \times_1 \bB^{(1)} \times_2
%\cdots \times_N \bB^{(N)}$ & \minitab[p{.51\linewidth}]{\\[-9.2ex] Multilinear operator returns tensor $\underline \bC$, with entries $c_{i_1,i_2,\ldots, i_N}$}\\

%$\underline \bC = \underline \bA \times^1_1 \underline \bB$ & \minitab[p{.55\linewidth}]{\\[-3.2ex]tensor product of $\underline \bA \in \Real^{I_1 \times I_2 \times \cdots \times I_N}$  and $\underline \bB \in \Real^{I_1 \times J_2 \times \cdots \times J_M}$ with $c_{\bi_{2:N},\bj_{2:M}} = \sum_{i=1}^{I_1} a_{i,\bi_{2:M}} b_{i,\bj_{2:M}}$} \\
%
$\underline \bC = \underline \bA \circ \underline \bB$ & \minitab[p{.61\linewidth}]{\\[-3.2ex] tensor or outer product of $\underline \bA \in \Real^{I_1 \times I_2 \times \cdots \times I_N}$ and $\underline \bB \in \Real^{J_1 \times J_2 \times \cdots \times J_M}$ yields $(N+M)$th-order tensor $\underline \bC$, with entries $c_{i_1, \ldots, i_N, \,j_1, \ldots, j_M} = a_{i_1, \ldots, i_N} b_{j_1, \ldots, j_M}$} \\
%
%$\tC = \tA \times_n^m \tB$ & \minitab[p{.45\linewidth}]{\\[-3.2ex] mode-$n$ product of $\tA \in \Real^{I_1 \times I_2 \times \ldots \times I_N}$ and $\tB \in \Real^{J_1 \times J_2 \times \cdots \times J_M}$,  with $I_n=J_m$ yields $\tC \in \Real^{I_1 \times \cdots I_{n-1} \times I_{n+1} \cdots \times I_N \times J_{1} \times \cdots J_{m-1} \times J_{m+1} \cdots \times J_M}$ with entries $c_{i_1 \cdots i_{n-1} \, i_{n+1} \cdots i_N, \, j_1, \cdots j_{m-1} \, j_{m+1} \cdots j_M} = \sum_{i=1}^{I_n} a_{i_1 \cdots i_{n-1} \; i \; i_{n+1} \cdots i_N} b_{j_1 \cdots j_{m-1} \; i \; j_{m+1} \cdots j_M}$ } \\
%
{$\underline \bX = \ba \circ \bb \circ \bc \in \Real^{I \times J \times K}$}&\minitab[p{.58\linewidth}]{tensor or outer product of vectors  forms a rank-1 tensor, with entries $x_{ijk} = a_i b_j c_k$} \\
%\multicolumn{2}{@{\hspace{-.1ex}}l}{$\bigotimes_{n=1}^{N} \bB^{(n)} = \bB^{(N)} \otimes \cdots \otimes \bB^{(n)} \otimes \cdots \otimes \bB^{(1)} $} \\
%\multicolumn{2}{@{\hspace{-.1ex}}l}{${\bigotimes}_{k\neq n} \bB^{(k)} = \bB^{(N)} \otimes \cdots \otimes \bB^{(n+1)} \otimes \bB^{(n-1)} \, \cdots \otimes \bB^{(1)}$} \\
%%\multicolumn{2}{l}{${\bigcircledast}_{n=1}^{N} \bB^{(n)} = \bB^{(N)} \circledast \cdots \circledast \bB^{(n)} \circledast
%%\cdots \circledast \bB^{(1)} $ }\\
%%\multicolumn{2}{l}{${\bigcircledast}_{k\neq n} \bA^{(k)} = \bA^{(N)} \circledast \cdots \circledast \bB^{(n+1)} \circledast \bB^{(n-1)}  \circledast
%%\cdots \circledast \bA^{(1)} $} \\
%\multicolumn{2}{@{\hspace{-.1ex}}l}{${\bigodot}_{n=1}^{N} \bB^{(n)} = \bB^{(N)} \odot \cdots \odot \bB^{(n)} \odot \cdots \odot \bB^{(1)}$} \\
%\multicolumn{2}{@{\hspace{-.1ex}}l}{${\bigodot}_{k\neq n} \bB^{(k)} = \bB^{(N)} \odot \cdots \odot \bB^{(n+1)} \odot \bB^{(n-1)} \, \cdots \odot \bB^{(1)} $ } \\
 ${\bA}^{T}$,   ${\bA}^{-1}$, ${\bA}^{\dag}$ & \minitab[p{.58\linewidth}]{transpose, inverse and Moore-Penrose pseudo-inverse of $\bA$} \\[-2ex]
%{$\bC = \bA\otimes\bB$}&\minitab[p{.53\linewidth}]{Kronecker product of $\bA \in \Real^{I\times J}$ and $\bB \in \Real^{L\times M}$, with $c_{L(i-1)+l,M(j-1)+m} = a_{i,j} b_{l,m}$} \\[-3ex]
%
{$\underline \bC = \underline \bA \otimes \underline \bB$}&\minitab[p{.6\linewidth}]{Kronecker product of $\underline \bA \in \Real^{I_1 \times I_2 \times \cdots \times I_N}$ and $\underline \bB \in \Real^{J_1 \times J_2 \times \cdots \times J_N}$ yields $ \underline \bC \in \Real^{I_1 J_1 \times \cdots \times I_N J_N}$, with entries
$c_{\overline{i_1, j_1}, \ldots, \overline{i_N,j_N}} = a_{i_1, \ldots,i_N} \: b_{j_1, \ldots,j_N}$, where $\overline{i_n, j_n} = j_n+(i_n-1)J_n$} \\[-2ex]
{$\bC = \bA\odot \bB$}&\minitab[p{.6 \linewidth}]{Khatri-Rao product of $\bA \in \Real^{I\times J}$ and $\bB \in \Real^{K\times J}$ yields $\bC \in \Real^{I K \times J}$, with columns $\bc_j = \ba_j \otimes \bb_j$}\\[-1ex]
%{$\circledast, \oslash$}&{Hadamard product and division} \\
%{$\underline \bX = \ba \circ \bb \circ \bc \in \Real^{I \times J \times K}$}&\minitab[p{.4\linewidth}]{outer product forms a rank-1tensor with entries $x_{ijk} = a_i b_j c_k$} \\
\\ \hline
    \end{tabular*}
    }}
%    \end{center}
\label{table_notation2}
\end{table}
\minrowclearance 0ex

A higher-order tensor can be interpreted as a multiway array, as illustrated graphically in Figs. \ref{Fig:fibers}, \ref{Fig:Tens-block-matr} and \ref{Fig:symbols}. %for a third-order tensor.
 %as shown in Fig. \ref{Fig3waytensor}.
 Our adopted convenience is that tensors are denoted  by  bold underlined capital letters, e.g., $\underline \bX  \in \Real^{I_{1} \times I_{2} \times \cdots \times I_{N}}$, and that all data are real-valued.
%\zgxRemove{Of most models complex variants can be formulated in a straightforward manner.}
The order  of a tensor is the number of its ``modes'', ``ways'' or ``dimensions'', which can include space, time, frequency,  trials, classes, and dictionaries.
 % (see Fig. \ref{Fig3waytensor}).
Matrices (2nd-order tensors) are denoted by boldface capital letters, e.g., $\bX$, and vectors (1st-order tensors)  by boldface lowercase letters; for instance  the columns of the matrix $\bA=[\ba_1,\ba_2, \ldots,\ba_R]  \in \Real^{I \times R}$ are denoted by $\ba_r$ and elements of a matrix (scalars) are denoted by lowercase letters, e.g., $a_{ir}$ (see Table \ref{table_notation1}).

The most common types of tensor multiplications are denoted by: $\otimes$ for the Kronecker, $\odot$ for the Khatri-Rao,  $\*$ for the Hadamard (componentwise),  $\circ$ for the outer  and  $ \times_n$ for the mode-$n$  products (see Table  \ref{table_notation2}).

TNs and TDs can be represented by tensor network diagrams, in which  tensors are represented
graphically by  nodes or any shapes (e.g., circles,  spheres, triangular, squares, ellipses) and each outgoing edge (line)  emerging from a shape represents a mode (a way, dimension, indices)
(see Fig. \ref{Fig:symbols})
Tensor  network diagrams are very useful not only in visualizing tensor
decompositions, but also in their different transformations/reshapings and graphical illustrations of mathematical (multilinear) operations.

It should also be noted that block matrices and hierarchical block matrices can be represented by tensors.
For example,  3rd-order and 4th-order tensors that can be represented by block matrices  as illustrated in Fig. \ref{Fig:Tens-block-matr} and all algebraic operations can be performed on block matrices.
Analogously,  higher-order tensors can be represented as illustrated in Fig. \ref{Fig:Tens-block5-8} and  Fig. \ref{Fig:symbols2}.
Subtensors are formed when a subset of indices is fixed. Of particular interest are  {\it fibers}, defined  by fixing every index but one,  and
{\it matrix slices} which are two-dimensional sections (matrices) of a tensor,
obtained by fixing all the indices but two  (see Fig. \ref{Fig:fibers}).
% Note that fibers are always assumed to be  column vectors \cite{Kolda08}.
A matrix has two modes: rows and columns, while an $N$th-order tensor has $N$ modes.

The process of unfolding (see Fig. \ref{Fig:unfolding}) flattens a tensor into a  matrix. In the simplest scenario, mode-$n$ unfolding (matricization,  flattening) of the tensor  $\underline \bA \in \Real^{I_{1} \times I_{2} \times \cdots \times I_{N}}$ yields
a matrix $\bA_{(n)} \in \Real^{I_{n} \times (I_{1} \cdots I_{n-1} I_{n+1} \cdots I_N)}$, with entries $a_{i_n,(j_1,\ldots,i_{n-1},j_{n+1},\ldots,i_n)}$ such that remaining indices $(i_1,\ldots,i_{n-1},i_{n+1},\ldots,i_N)$ are arranged in a specific order, e.g., in the lexicographical order \cite{Kolda08}. In tensor networks we use, typically a generalized mode-$([n])$ unfolding as illustrated in Fig. \ref{Fig:unfolding} (b).

%By a multi-index $i = \overline{i_1,i_2, \ldots, i_N}$ we denote an index which takes all possible
%combination of values of $i_1, i_2, \ldots, i_n$,  for $i_n = 1,2,\ldots, I_n$ we have $\overline{i_1,i_2, \ldots, i_N}=
%i_1 + (i_2-1)I_1 + \cdots + (i_N-1) I_1 I_2 \cdots I_N$.

By a multi-index $i = \overline{i_1,i_2, \ldots, i_N}$, we denote an index which takes all possible
combinations of values of $i_1, i_2, \ldots, i_n$,  for $i_n = 1,2,\ldots, I_n$
in a specific and consistent orders.  The entries of matrices or tensors in matricized and/or vectorized forms can be ordered in at least two different ways.
%we have $\overline{i_1,i_2, \ldots, i_N}=
%i_1 + (i_2-1)I_1 + \cdots + (i_N-1) I_1 I_2 \cdots I_{N-1}$.

{\bf Remark:}
The multi--index can be defined  using two  different conventions:

1) The little–-endian convention
\be
\overline{ i_1,i_2,\ldots, i_N } &=& i_1 + (i_{2} - 1) I_1+ (i_{3} - 1) I_1 I_2 \notag \\
&\cdots&  + (i_N - 1) I_1 \cdots  I_{N-1}.
\ee

2) The big–-endian
\be
&&\overline{ i_1,i_2,\ldots, i_N } = i_N + (i_{N-1} - 1)I_N + \notag \\
&&+ (i_{N-2} - 1)I_N I_{N-1}+ \cdots  + (i_1 - 1)I_2 \cdots  I_N. %\notag \\
\ee

The little--endian notation is consistent with the Fortran style of indexing,
while the big--endian notation is similar to numbers written in the positional system and corresponds to reverse lexicographic order.
The definition unfolding of tensors and  the Kronecker (tensor) product $\otimes$ should be also consistent with the chosen
convention{\footnote{The  standard and more popular
definition in multilinear algebra assumes the big--endian convention, while for the development of the efficient program
code  for big data usually   the little--endian  convention seems to be more convenient (see for more detail papers of Dolgov and Savostyanov \cite{Dolgov2013alternating,Dolgov2013alternating2}).}}.
In this paper  we will use the big-endian notation, however  it is enough to remember that $\bc = \ba \otimes  \bb$ means that $c_{\overline{i,j}} = a_i b_j$.

The Kronecker product of two tensors: $\underline \bA \in \Real^{I_1 \times I_2 \cdots \times I_N}$ and $ \underline  \bB \in \Real^{J_1 \times J_2 \times \cdots \times J_N}$ yields $\underline  \bC = \underline \bA \otimes \underline \bB \in \Real^{I_1 I_2 \times \cdots \times I_N J_N}$, with entries $c_{\overline{i_1, j_1},\ldots, \overline{i_N, j_N}} = a_{i_1, \ldots,i_N} \: b_{j_1, \ldots,j_N}$, where $\overline{i_n, j_n} = j_n+(i_n-1)J_n$.

\begin{figure}[ht!]
(a)\\
%\centering
\includegraphics[width=8.99cm]{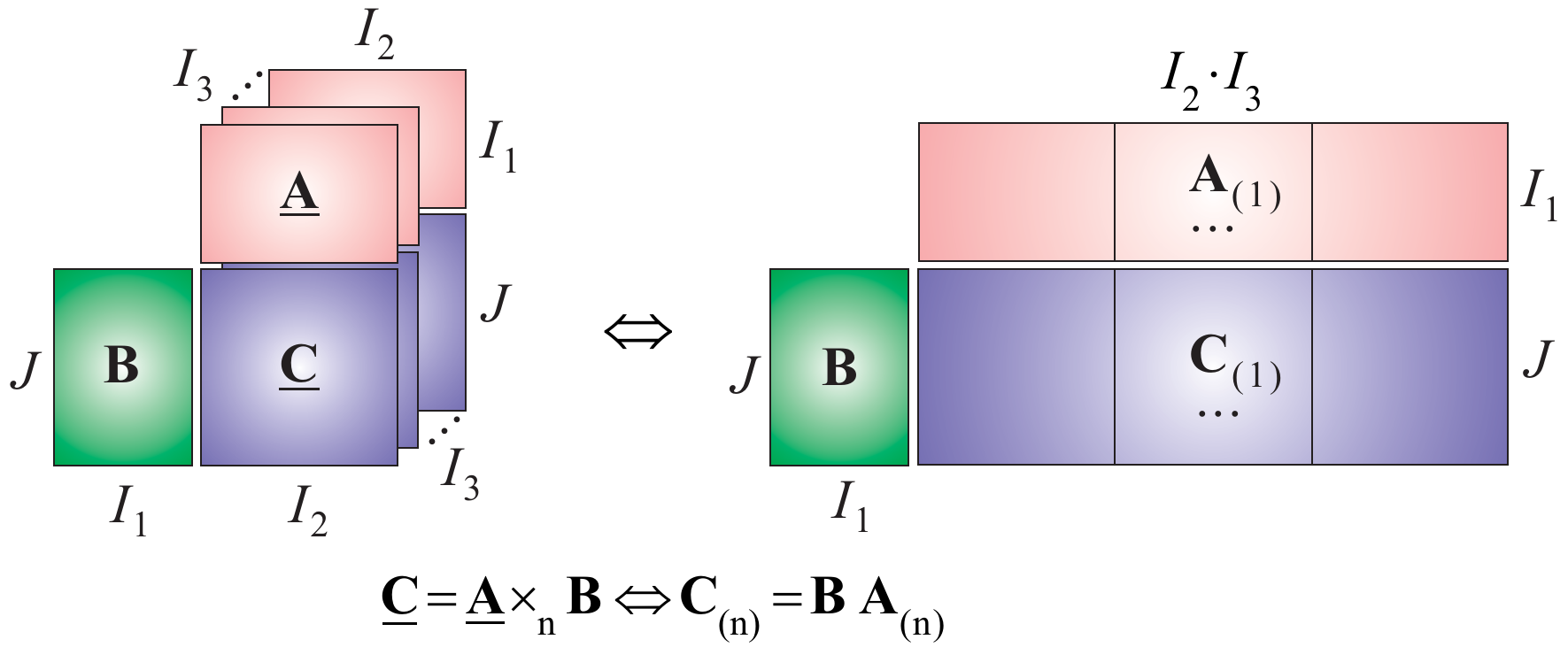}\\
(b)\\
\includegraphics[width=8.9cm,height=2.0cm]{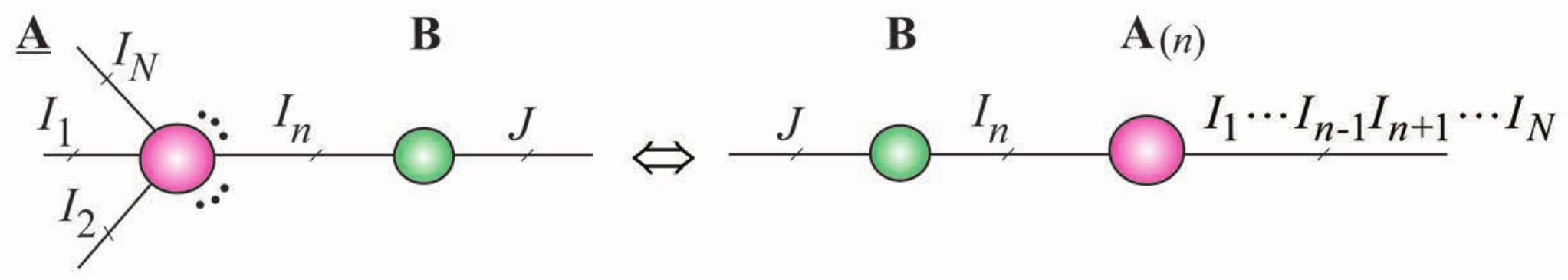}
\caption{From a matrix format to the tensor network format. (a) Multilinear mode-$1$ product of a 3rd-order tensor $\underline \bA \in \Real^{I_1 \times I_2 \times I_3}$ and a factor (component) matrix $\bB \in \Real^{J \times I_1}$  yields a tensor $\underline \bC = \underline \bA \times_1 \bB \in \Real^{J \times I_2 \times I_3}$. This is equivalent to simple matrix multiplication formula $\bC_{(1)}=\bB \bA_{(1)}$. (b) Multilinear mode-$n$ product  an $N$th-order tensor and a factor matrix $\bB \in \Real^{J \times I_n}$.}
\label{Fig:AxnB}
\end{figure}

The  mode-$n$  product of a tensor
$\underline \bA \in \Real^{I_{1} \times \cdots \times I_{N}}$ by a vector  $\bb \in \Real^{I_n}$ is  defined as a tensor
$ \underline \bC = \underline \bA \bar \times_n \bb \in \Real^{I_1 \times \cdots \times I_{n-1}  \times I_{n+1} \times \cdots \times I_N}$, with entries $ c_{i_1,\ldots,i_{n-1},i_{n+1},\ldots, i_{N}} =\sum_{i_n=1}^{I_n} (a_{i_1,i_2,\ldots,i_N})\; b_{i_n}$, while a mode-$n$  product of the tensor
$\underline \bA \in \Real^{I_{1} \times \cdots \times I_{N}}$ by a matrix  $\bB \in \Real^{J \times I_n}$ is the tensor $\underline \bC = \underline \bA \times_n \bB \in \Real^{I_1 \times \cdots \times I_{n-1} \times J \times I_{n+1} \times \cdots \times I_N}$,
with entries $ c_{i_1,i_2,\ldots,i_{n-1},j,i_{n+1},\ldots, i_{N}} =\sum_{i_n=1}^{I_n} a_{i_1,i_2,\ldots,i_N} \; b_{j,i_n}$. This can also be expressed in a matrix form as $\bC_{(n)} =\bB \bA_{(n)}$ (see Fig. \ref{Fig:AxnB}), which allows us to employ fast matrix by vector and matrix by matrix multiplications for very large scale problems.

\begin{figure}[ht]
%\centering
%\psfrag{y}{\color{black}$y$}
(a) \hspace{5.2cm} (b)\\
%\includegraphics[width=4.9cm,height=4.8cm]{TM5.eps}
%\hspace{0.1cm}
% \includegraphics[width=3.3cm]{Tenvec.eps}\\
\includegraphics[width=8.9cm,height=4.8cm]{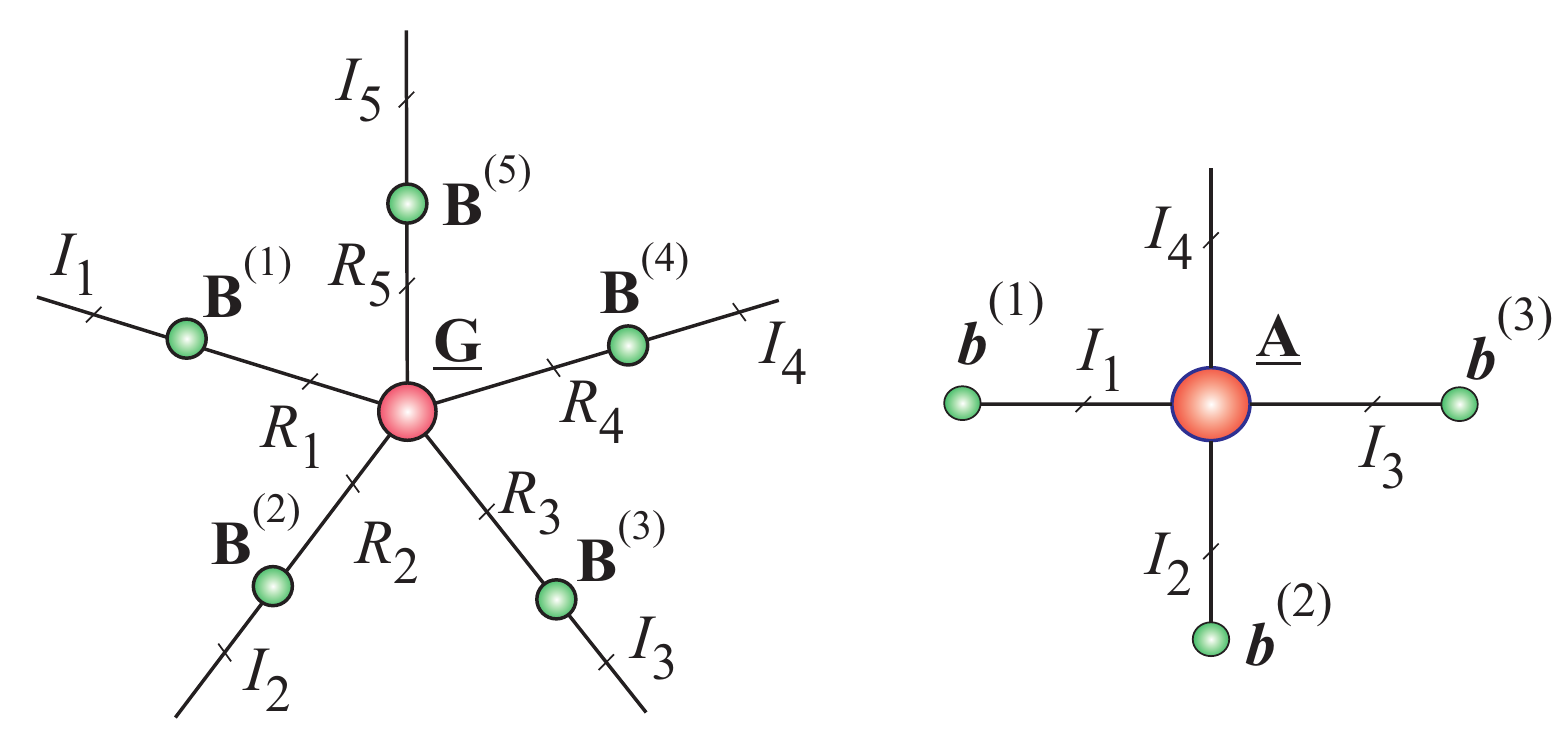}
%\hspace{0.1cm}
% \includegraphics[width=3.3cm]{Tenvec.eps}\\
%
\caption{ Multilinear products via tensor network diagrams. (a) Multilinear full product of tensor (Tucker decomposition) $\underline \bG \in \Real^{R_1 \times R_2 \times  \cdots \times R_5}$ and factor (component) matrices $\bB^{(n)} \in \Real^{I_n \times R_n}$ ($n=1,2,\ldots,5$) yields the Tucker tensor decomposition $\underline \bC = \underline \bG \times_1 \bB^{(1)}  \times_2 \bB^{(2)}  \cdots   \times_5 \bB^{(5)} \in \Real^{I_1 \times I_2 \times \cdots \times I_5}$.
 %with entries $ c_{i_1,i_2,\ldots, i_5}=\sum_{r_1,r_2,\ldots,r_5=1}^{R_1,R_2,\ldots,R_5} a_{r_1,r_2,\ldots, r_5} b^{(1)}_{i_1,r_1} b^{(2)}_{i_2,r_2} \cdots b^{(5)}_{i_5,r_5}$ and (
 b) Multilinear product of tensor $\underline \bA \in \Real^{I_1 \times I_2 \times \cdots \times I_4}$  and  vectors $\bb_n \in \Real^{I_n}$ $(n=1,2,3)$ yields a vector $\bc= \underline \bA \bar\times_1 \bb^{(1)}  \bar\times_2 \bb^{(2)} \bar\times_3  \; \bb^{(3)} \in \Real^{I_4}$.
% with entries $c_{i_4}=\sum_{i_1,i_2,i_3=1}^{I_1,I_2,I_3} a_{i_1,i_2,i_3,i_4} b^{(1)}_{i_1}  b^{(2)}_{i_2}  b^{(3)}_{i_3}$.
}
\label{Fig:TO1}
\end{figure}

If we take all the modes, then we have a full multilinear product
 of a tensor and a set of matrices, which  is compactly written as \cite{Kolda08} (see Fig. \ref{Fig:TO1} (a))):
\be
\underline \bC &=& \underline \bG \times_1 \bB^{(1)}  \times_2 \bB^{(2)}  \cdots  \times_N \bB^{(N)} \notag \\
&=& \llbracket \underline \bG; \bB^{(1)}, \bB^{(2)}, \ldots, \bB^{(N)} \rrbracket.
\label{mprod}
\ee

\begin{figure}[h]
\centering
%%
%(a) \hspace{3.7cm} (b)\\
%\includegraphics[width=4.4cm,height=2.4cm]{nmprod.eps}
%\hspace{0.01cm}
% \includegraphics[width=3.8cm,height=2.1cm]{inprod.eps}\\ \\
% %
% (c) \hspace{3.7cm} (d)\\
%\includegraphics[width=8.6cm]{Tcontr44-2.eps}
 \includegraphics[width=8.6cm,height=6.1cm]{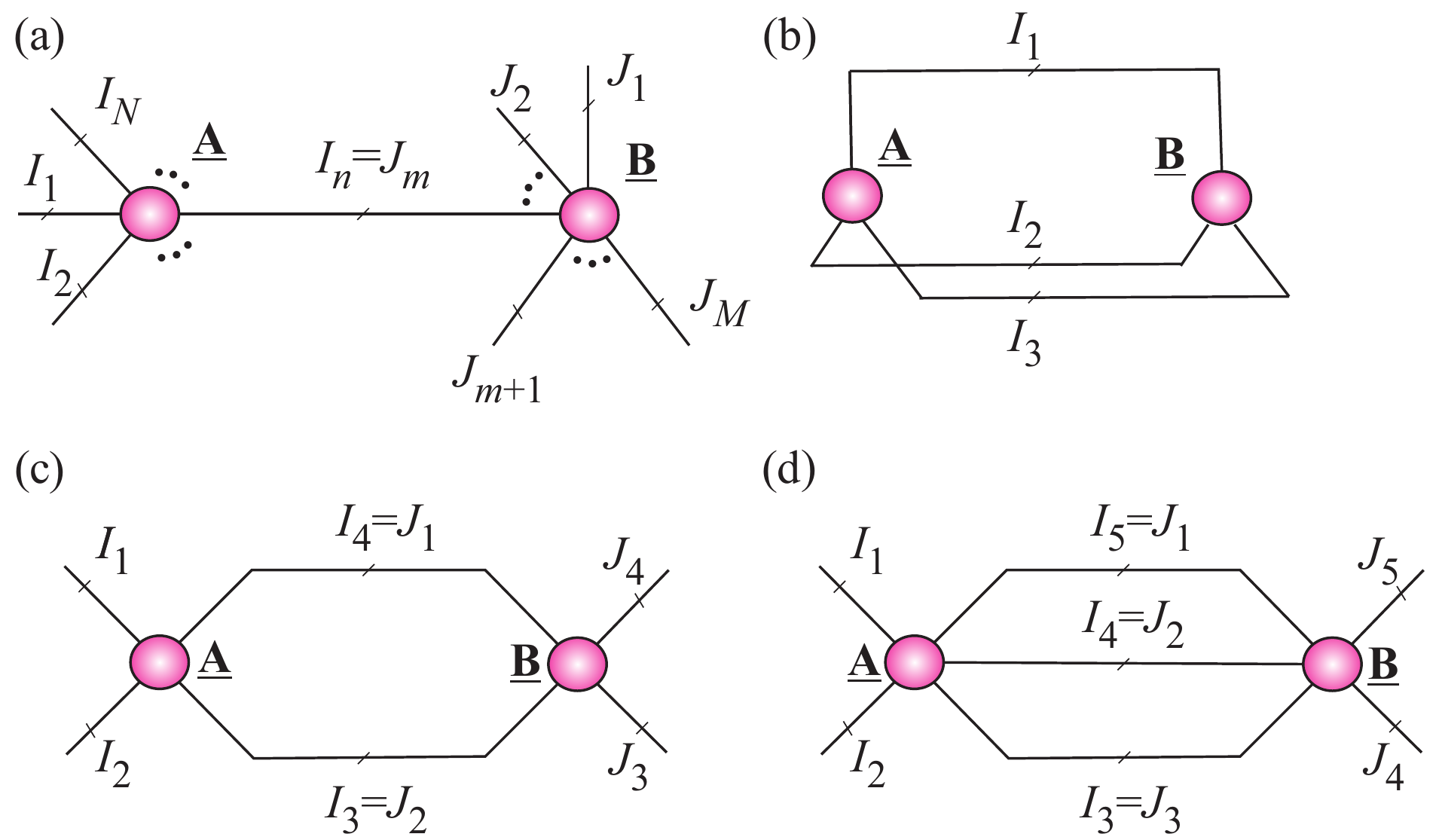}
\caption{Examples of tensor contractions: (a) Multilinear product of two tensors is denoted by  $\underline \bC = \underline \bA \; \times_n^m \; \underline \bB$. (b) Inner product of two 3rd-order tensors yields a scalar $c=\langle\underline \bA, \underline \bB\rangle=\underline \bA \; \times_{1,2,3}^{1,2,3} \; \underline \bB  =\underline \bA \; \times\; \underline \bB=\sum_{i_1,i_2,i_3} \; a_{i_1,i_2,i_3} \; b_{i_1,i_2,i_3}$.  (c) Tensor contraction of two 4th-order tensors yields  $\underline \bC = \underline \bA \; \times_{4, 3}^{1, 2} \; \underline \bB \in \Real^{I_1 \times I_2 \times J_3 \times J_4}$, with entries $c_{i_1,i_2,j_3,j_4}= \sum_{i_3,i_4} \; a_{i_1,i_2,i_3,i_4}  \; b_{i_4,i_3,j_3,j_4}$. (d) Tensor contraction of two 5th-order tensors yields 4th-order tensor $\underline \bC = \underline \bA \; \times_{3, 4, 5}^{1, 2, 3} \; \underline \bB \in \Real^{I_1 \times I_2 \times J_4 \times J_5}$, with entries $c_{i_1,i_2,j_4,j_5}= \sum_{i_3,i_4,i_5}\; a_{i_1,i_2,i_3,i_4,i_5}  \; b_{i_5,i_4,i_3,j_4,j_5}$.}
\label{Fig:TO2}
\end{figure}

In a similar way to mode-$n$ multilinear product, we can define  the mode-$ (_n^m)$ product of two tensors (tensor contraction) $\underline \bA \in \Real^{I_1 \times I_2 \times \ldots \times I_N}$ and $\underline \bB \in \Real^{J_1 \times J_2 \times \cdots \times J_M}$,  with  common modes $I_n=J_m$ that yields  an $(N+M-2)$-order tensor $ \underline \bC  \in \Real^{I_1 \times \cdots I_{n-1} \times I_{n+1} \times  \cdots \times I_N \times J_{1} \times \cdots J_{m-1} \times J_{m+1} \times \cdots \times J_M}$: % (see Fig. \ref{Fig:TO2} (a)):
 \be \underline \bC =  \underline \bA \; {\times}_n^m \; \underline \bB,
 \ee
  with entries $c_{i_1, \ldots i_{n-1}, \, i_{n+1}, \ldots i_N, \, j_1, \ldots j_{m-1}, \, j_{m+1}, \ldots j_M} = \sum_{i=1}^{I_n} a_{i_1, \ldots i_{n-1}, \; i \; i_{n+1}, \ldots i_N} b_{j_1, \ldots j_{m-1}, \; i, \; j_{m+1}, \ldots j_M}$ (see Fig. \ref{Fig:TO2} (a)).
  This operation can be considered as  a contraction of two tensors in single common mode. Tensors can be contracted in several modes or even in all modes as illustrated in Fig. \ref{Fig:TO2}.

If not confusing a super- or sub-index $m,n$ can be neglected. For example, the multilinear product of the tensors $\underline \bA \in  \Real^{I_{1} \times I_{2} \times \cdots \times I_{N}}$ and $\underline \bB \in  \Real^{J_{1} \times J_{2} \times \cdots \times J_{M}}$,  with a common  modes $I_N=J_1$ can be written   as
\be
\underline \bC = \underline \bA \; \times_N^1 \; \underline \bB = \underline \bA \times_N \underline \bB = \underline  \bA \bullet \underline \bB \notag \\  \in  \; \Real^{I_{1} \times I_2 \times  \times I_{N-1} \times  \times J_2 \times \cdots \times J_{M}},
\ee
with entries:
 $c_{i_2,i_3,\ldots,i_N,j_1,j_3,\ldots,j_M}= \sum_{i=1}^{I_1} a_{i,i_2,\ldots,i_N} \; b_{j_1,i,j_3,\ldots,j_M}$.
%$c_{\bi_{2:N},\bj_{2:M}}= \sum_{i=1}^{I_1} a_{i, \bi_{2:N}} b_{i,\bj_{2:M}}$ by using the MATLAB notation $\bi_{p:q}=\{i_p,i_{p+1},\ldots,i_{q-1},i_q\}$.
Furthermore, note that for multiplications of matrices and vectors this notation implies that $\bA \times^1_2 \bB =\bA \bB$, $\;\bA \times^2_2 \bB =\bA \bB^T$,  $\;\bA \times^{1,2}_{1,2} \bB = \langle\bA, \bB\rangle$, and $\bA \times^1_2 \bx =\bA \times_2 \bx =\bA \bx$.

{\bf Remark:} If we use contraction for more than two tensors  the order has to be specified (defined) as follows:\\ $\underline \bA \times^b_a
\underline \bB \times_c^d \underline \bC=\underline \bA \times^b_a
(\underline \bB \times_c^d \underline \bC)$ for $b < c$.

The outer or tensor product $\underline \bC =\underline \bA \circ \underline \bB$ of the tensors $\underline \bA \in \Real^{I_1 \times \cdots \times I_N}$ and $\underline \bB \in \Real^{J_1 \times \cdots \times J_M}$ is the tensor $\underline \bC \in \Real^{I_1 \times \cdots \times I_N \times J_1 \times \cdots \times J_M }$, with entries $c_{i_1,\ldots,i_N,j_1,\ldots,j_M} = a_{i_1,\ldots,i_N} \; b_{j_1,\ldots,j_M}$.
Specifically, the outer product of two nonzero vectors
$\ba \in \Real^I, \; \bb \in \Real^J$ produces a rank-1 matrix
$\bX=\ba \circ \bb = \ba \bb^T \in \Real^{I \times J}$
and the outer product of  three nonzero vectors: $\ba \in \Real^I, \; \bb \in \Real^J$ and  $\bc \in \Real^K$ produces a 3rd-order rank-1  tensor:
$\underline \bX= \ba \circ \bb \circ \bc \in \Real^{I \times J  \times K}$,
whose entries are $x_{ijk} =a_i \; b_j \; c_k$.
A tensor $\underline \bX \in \Real^{I_1 \times I_2 \times \cdots \times I_N}$ is said to be rank-1 if  it can be expressed exactly as $\underline \bX = \bb_1 \circ \bb_2 \circ \cdots  \circ \bb_N$, with entries $x_{i_1,i_2,\ldots,i_N} = b_{i_1} b_{i_2} \cdots b_{i_N}$, where $\bb_n \in \Real^{I_n}$ are nonzero vectors.
%
%The superscripts: $(\cdot)^T$,  $(\cdot)^{-1}$,$(\cdot)^{\dagger}$ denote respectively for the transpose,  matrix inverse, and Moore-Penrose pseudo-inverse operators.
%%The superscripts: $(\cdot)^T$, $(\cdot)^H$, $(\cdot)^{-1}$,$(\cdot)^{\dagger}$ are used for the transpose, Hermitian conjugate, matrix inverse and Moore-Penrose pseudo-inverse operations.
%
%General basic operations, e.g., $vec(\cdot)$, $\diag\{\cdot\}$, are defined as in  MATLAB.
 We refer to \cite{NMF-book,Kolda08} for more detail regarding the basic  notations and tensor operations.

\section{\bf Tensor Networks}

\begin{figure}[t!]
\centering
%\psfrag{y}{\color{black}$y$}
%\includegraphics[width=8.6cm]{TN1m.eps}
%\includegraphics[width=4.1cm,height=3.0cm]{SPM1c.eps}\\
\includegraphics[width=8.99cm]{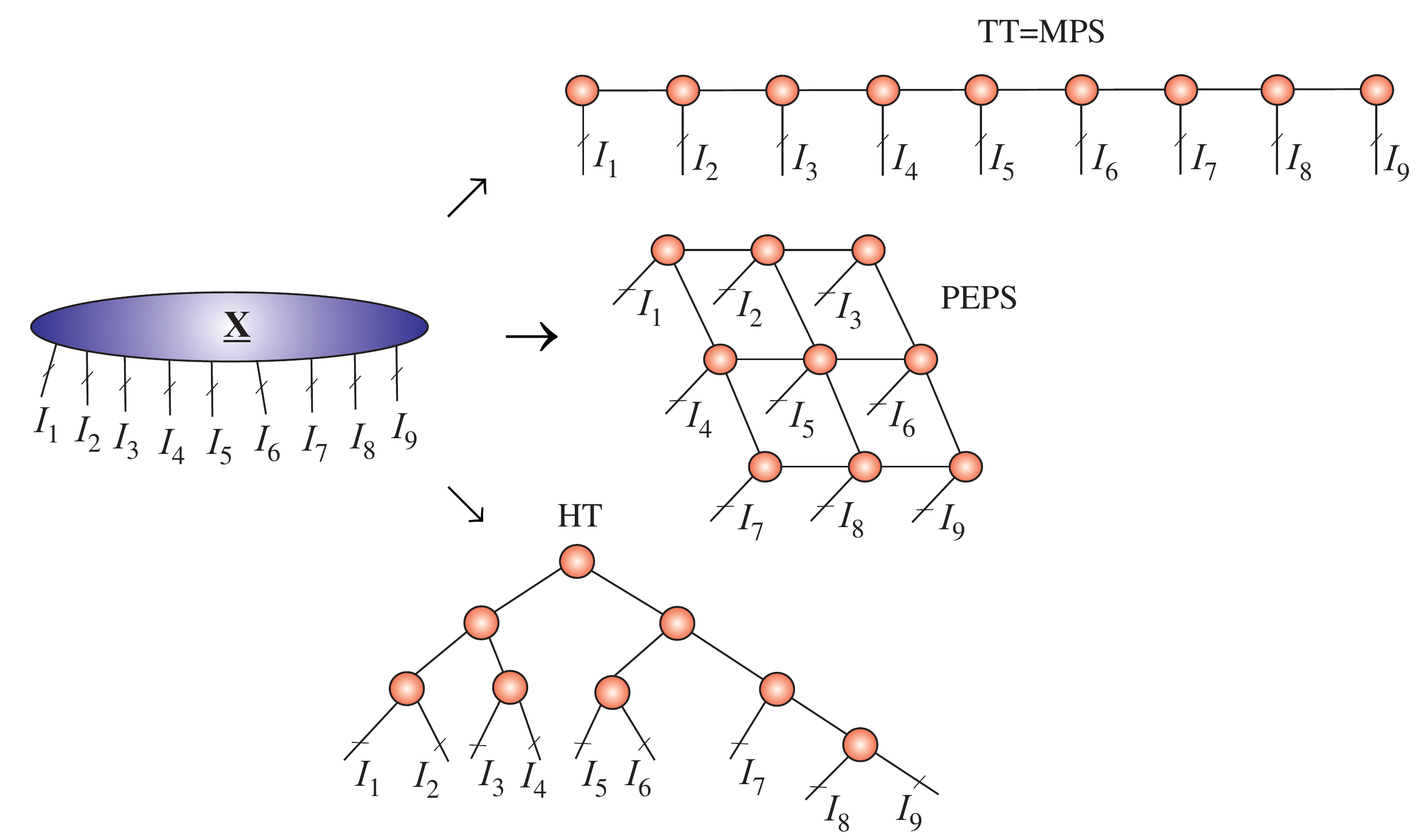}
\caption{Examples of tensor networks. Illustration of representation of 9th-order tensor $\underline \bX \in \Real^{I_1 \times I_2 \times \cdots \times I_9}$ by different kinds of tensor networks (TNs): Tensor Train (TT) which is equivalent to the Matrix Product State (MPS) (with open boundary conditions (OBC)), the Projected Entangled-Pair State (PEPS), called also Tensor Product States (TPS,  and Hierarchical Tucker (HT) decomposition, which is equivalent to the Tree-Tensor Network State (TTNS).  The objective is to decompose a high-order tensor into sparsely connected low-order and low-rank tensors, typically 3rd-order and/or 4th-order tensors, called cores.}
\label{Fig:TN1}
\end{figure}

A tensor network  aims to represent or decompose a higher-order tensor  into a set of  lower-order tensors (typically,   2nd (matrices) and 3rd-order tensors called cores or components) which are  sparsely interconnected.
In other words, in  contrast to TDs, TNs represent decompositions of the data tensors into a set of sparsely (weakly) interconnected  lower-order tensors.
Recently, the curse of dimensionality for higher-order tensors has been considerably alleviated or even completely avoided through the concept of tensor networks  (TN) \cite{hTucker,OseledetsTT09}.
A TN can be represented by a set of  nodes interconnected by lines. The lines (leads, branches, edges)  connecting tensors between each
other correspond to contracted modes, whereas lines that do not go from one tensor to another
correspond to open (physical) modes  in the TN (see Fig. \ref{Fig:TN1}).

An edge connecting two nodes indicates a
contraction of the respective tensors in the associated pair of modes as illustrated in Fig. \ref{Fig:TO2}.
%In contrast to  classical graphs in tensor network diagrams  do not need  connect two nodes, but may be connected to only one node.
Each  free (dangling) edge corresponds to a mode, that is not contracted and, hence, the order of the entire  tensor network is given by the
number of free edges (called often physical indices).
A tensor network may not contain any loops,
i.e., any edges connecting a node with itself. Some examples of tensor network diagrams
are given in Fig. \ref{Fig:TN1}.
%All the tensor decompositions
% (i.e., CP, Tucker, HT, TT decompositions) can be represented as tensor networks.
%

\begin{figure}[p]
(a)\\
  %  \includegraphics[width=8.6cm]{HT8s.eps} \\
%    (b)\\
% \includegraphics[width=8.6cm,height=15.6cm]{HTFull.eps}
  \includegraphics[width=8.6cm]{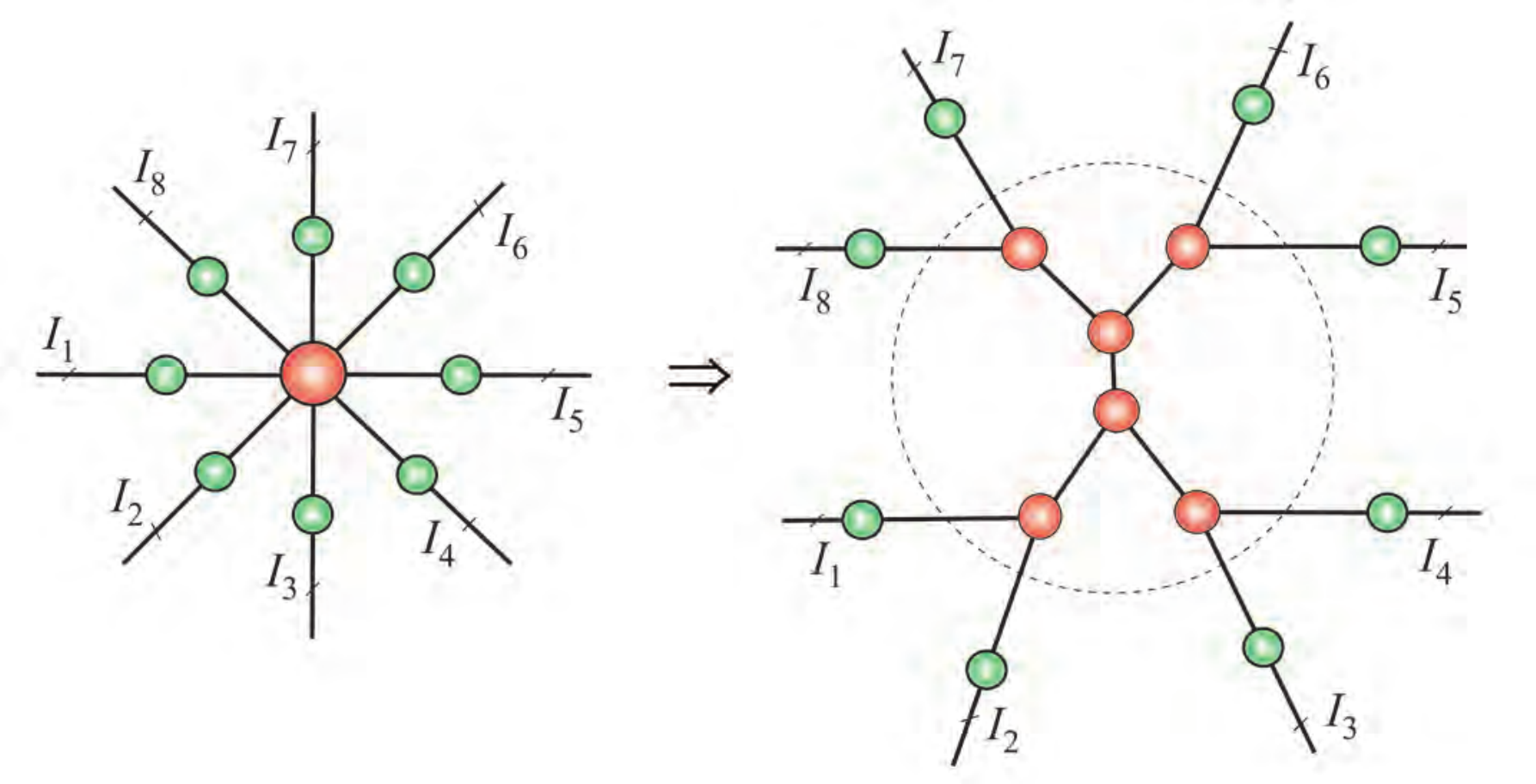} \\
%  \vspace{0.4cm}
    (b)\\
 \includegraphics[width=8.6cm,height=15.6cm]{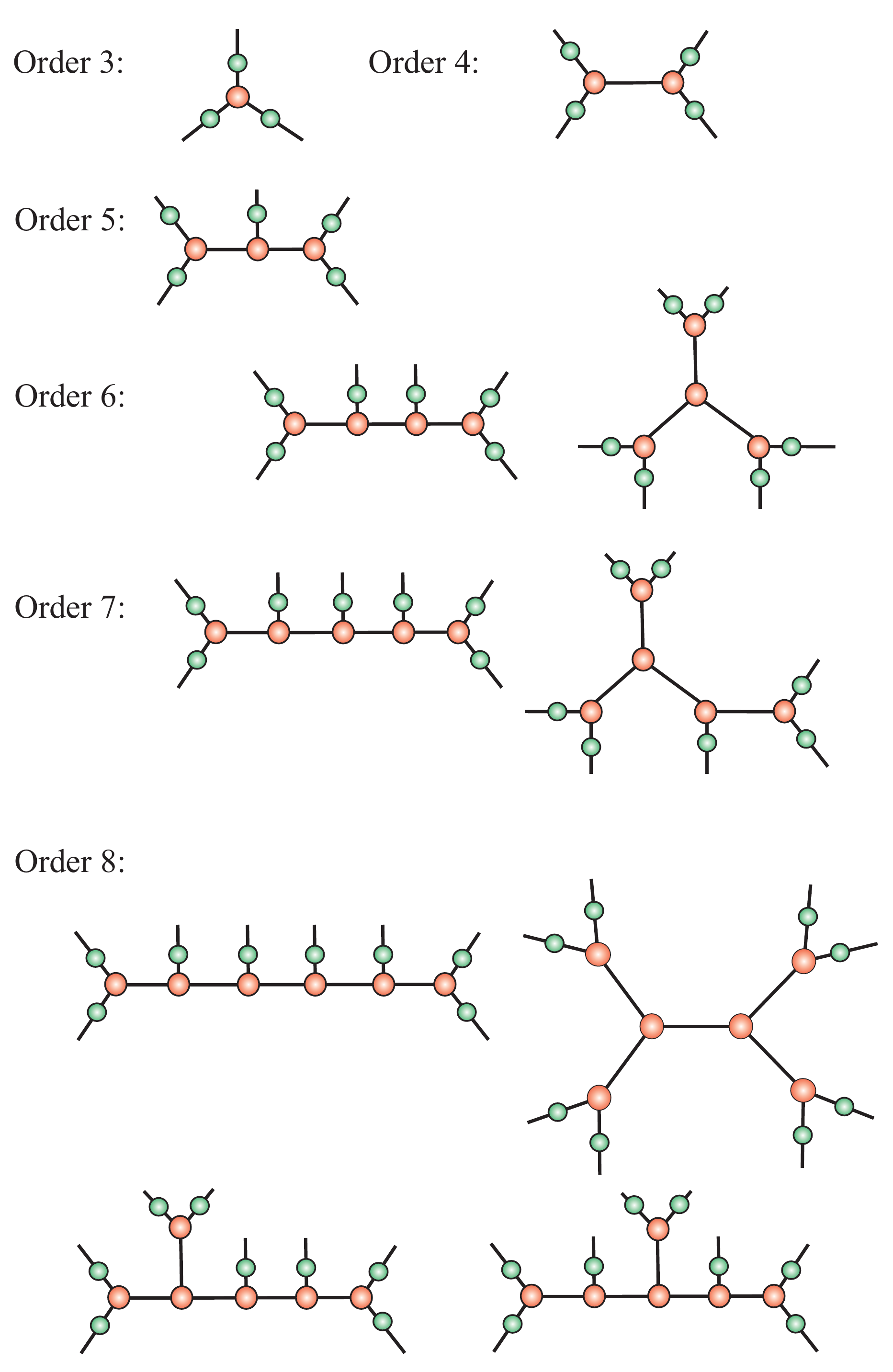}
\caption{(a) The standard Tucker decomposition and its transformation into Hierarchical Tucker (HT) model for an 8th-order tensor using  interconnected 3rd-order core tensors. (b) Various exemplary  structure HT/TT models for different order of data tensors. Green circles indicate factor matrices while red circles indicate cores.}
\label{Fig:HT8}
\end{figure}

If a tensor network is a tree, i.e.,  it does not contain any cycle, each of its edges
splits the modes of the data tensor  into two  groups, which is  related to the
suitable matricization of the tensor.
If, in such a tree tensor network, all nodes have degree 3 or less, it corresponds
to an Hierarchical Tucker (HT) decomposition  shown in Fig. \ref{Fig:HT8} (a). The HT  decomposition has been first introduced in scientific computing  by Hackbusch and K\"uhn and further developed by Grasedyck, Kressner, Tobler and others \cite{Uschmajew-Vander2013,Hackbush2012,Grasedyck-Hrank,Toblerthesis,KressnerTobler14,Grasedyck-rev}.
%
%Exemplary underlying tensor networks for Tucker decompositions in HT  format of order 3 to 8 are shown
%in Fig. \ref{Fig:HT3-8}.
 Note that for 6th-order tensor, there are two such tensor networks (see Fig. \ref{Fig:HT8} (b)), and for 10th-order  there are 11 possible HT decompositions \cite{Toblerthesis,KressnerTobler14}.

%\begin{figure}[ht]
%%\centering
%(a)\\
% \includegraphics[width=7.8cm]{HT8s.eps} \\ \\
% (b)\\
%\includegraphics[width=8.6cm]{HTFull.eps}
%\caption{(a) Transformation of the standard  Tucker decomposition into Hierarchical Tucker (HT) model for 8th-order tensor using a interconnected 3rd-order core tensors. (b) Representation of large-scale  higher-order Tucker decomposition  via HT decompositions in which core tensors  are represented by only 3rd-order tensors.}
%   %(for simplicity, we assumed that all core tensors
%   %$\underline \bG_r \in \Real^{R_1 \times R_2 \times R_3}$ have the same dimensions).}
%\label{Fig:HT3-8}
%\end{figure}

%\begin{figure}[ht!]
%\centering
%\includegraphics[width=8.6cm]{Dimtree7.eps}
%\caption{Tree graph illustrating construction of a Tree Tensor Network (TNN) for a 7th-order tensor.}
%   %(for simplicity, we assumed that all core tensors
%   %$\underline \bG_r \in \Real^{R_1 \times R_2 \times R_3}$ have the same dimensions).}
%\label{Fig:TTN}
%\end{figure}

\begin{figure}[t]
\centering
\includegraphics[width=8.8cm]{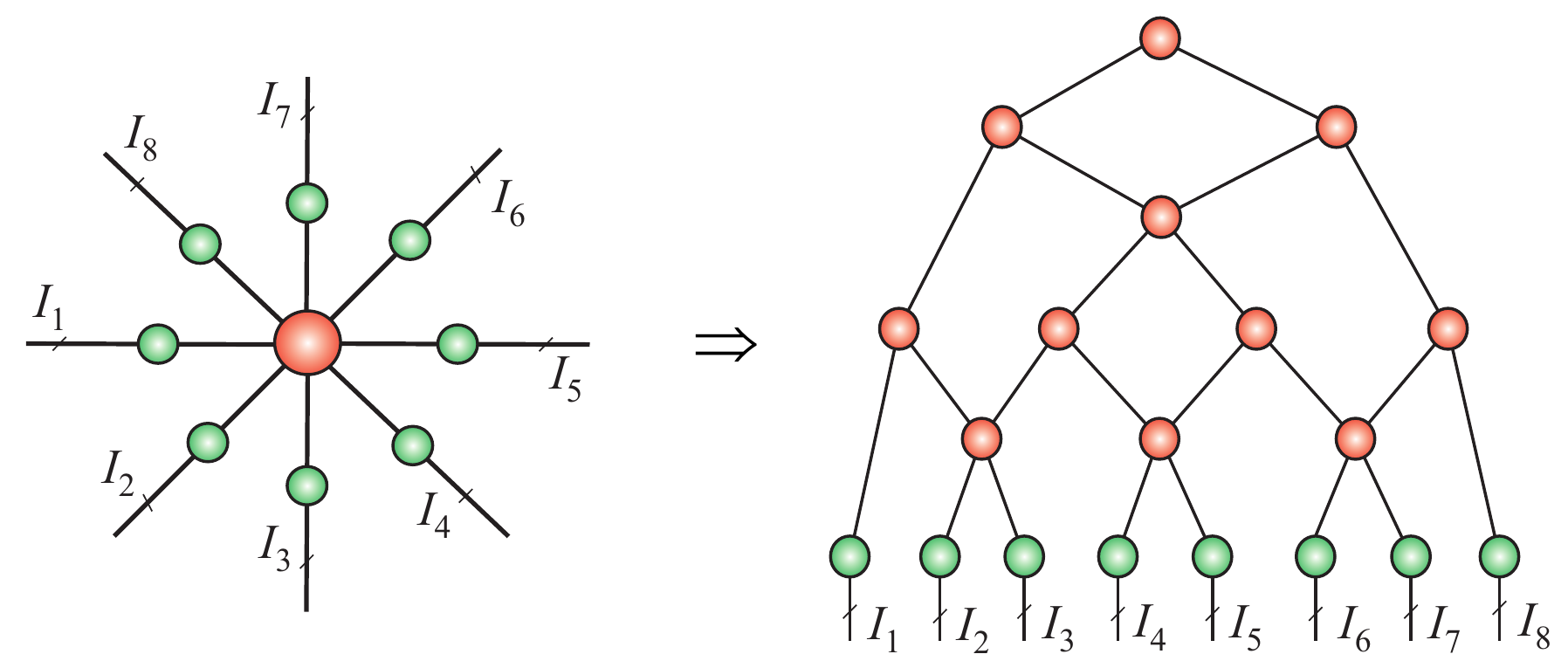}
\caption{Alternative distributed representation of 8th-order  Tucker decomposition where a core tensor is replaced by MERA (Multi-scale Entanglement Renormalization Ansatz) tensor network which employs 3rd-order and 4th-order core tensors. For some data-sets, the advantage of such model is relatively low size (dimensions) of the distributed cores.}
\label{Fig:MERA}
\end{figure}

A simple approach to reduce the size  of core tensors is to apply distributed tensor networks (DTNs), which consists in
two kinds of cores (nodes): Internal cores (nodes) which have no free edges and external cores which have free edges representing physical indices of a data tensor as illustrated in Figs. \ref{Fig:HT8} and \ref{Fig:MERA}.

 The idea  in the case of the Tucker model, is that a core tensor is replaced by distributed sparsely interconnected cores of lower-order, resulting in a Hierarchical Tucker (HT) network  in  which  only some cores are connected (associated) directly with factor matrices
\cite{Hackbush2012,Grasedyck-Hrank,Grasedyck-rev,Uschmajew-Vander2013}.

For some very high-order data tensors  it has been observed
that the ranks $R_n$ (internal dimensions of cores)  increase rapidly with the order of the tensor and/or with an increasing accuracy of approximation for any choice of tensor network, that is, a tree (including TT and  HT decompositions) \cite{KressnerTobler14}.
%In such case we may employ MERA as illustrated in Fig. \ref{Fig:MERA}.
%
For such cases, the Projected Entangled-Pair State (PEPS) or the Multi-scale Entanglement Renormalization Ansatz (MERA) tensor networks can be used. These   contain
cycles, but have  hierarchical structures (see Fig. \ref{Fig:MERA}). For the PEPS and MERA TNs the ranks
can be kept considerably smaller, at the cost of  employing 5th and 4th-order cores  and consequently a higher computational complexity w.r.t. tensor contractions.
The  main advantage of PEPS and MERA is that  the size of
each  core tensor in the internal tensor network structure is usually much smaller than the  cores in TT/HT decompositions, so consequently the total number of parameters can be  reduced.
However, it should be noted that the contraction of the resulting tensor network becomes more difficult
when compared to the basic tree structures represented by TT and  HT models.
This is due to the fact that the PEPS and MERA  tensor networks contain loops.
%To retain numerical accessibility, either approximate treatments have to
%be applied (as in contraction schemes introduced in the context of PEPS [19]) or the tree-like
%structure has to be maintained, e.g. by limiting the tree width of the concatenated tensor network %(as in [38]).
%

\section{\bf Basic Tensor Decompositions and their Representation via Tensor Networks Diagrams}

The main objective of a standard tensor decomposition is to factorize a data tensor into physically interpretable or meaningful  factor matrices and   a single core tensor which indicates the links between components (vectors of factor matrices) in different modes.

\subsection{\bf Constrained Matrix Factorizations and Decompositions -- Two-Way Component Analysis}

Two-way  Component Analysis (2-way CA)  exploits  {\it a priori} knowledge about different characteristics, features or morphology of components  (or source signals) \cite{NMF-book,Cichocki-GCA} to  find the hidden components thorough constrained  matrix factorizations of the form
\begin{equation}
 \label{2CA}
 \bX  = \bA \bB^T + \bE = \sum_{r=1}^R \ba_r \circ \bb_r +\bE=\sum_{r=1}^R \ba_r  \bb^T_r +\bE,
\end{equation}
where the constraints imposed on factor matrices $\bA$ and/or $\bB$ include orthogonality, sparsity, statistical independence, nonnegativity or smoothness.
The  CA can be considered as a bilinear (2-way) factorization, where $\bX \in \Real^{I \times J}$ is a known matrix of observed data, $\bE  \in \Real^{I \times J}$ represents residuals or noise,    $\bA = [ \ba_1, \ba_2, \ldots, \ba_R] \in \Real^{I \times R}$ is the unknown (usually, full column rank $R$) mixing matrix with $R$ basis vectors $\ba_{r} \in  \Real^I $,
 and   $\bB=[\bb_1,\bb_2,$ $\ldots,\bb_R]$ $ \in \Real^{J  \times R}$
is the matrix of unknown components (factors, latent variables, sources).
%Figure \ref{Fig:CPD1}  shows that Eq.(\ref{2CA}) can be considered as a 2-way (bilinear) factorization where $\bX \in \Real^{I \times J}$ is a known matrix of observed data, $\bE  \in \Real^{I \times J}$ represents residuals or noise,    $\bA = [ \ba_1, \ba_2, \ldots, \ba_R] \in \Real^{I \times R}$ is the unknown (usually full column rank) mixing matrix with $R$ basis vectors $\ba_{r} \in  \Real^I $,
% and   $\bB=[\bb_1,\bb_2,$ $\ldots,\bb_R]$ $ \in \Real^{J  \times R}$
%is the matrix of unknown components (factors, latent variables, sources).

Two-way component analysis (CA) refers to a class of signal processing  techniques that decompose or encode  superimposed or mixed signals into  components with certain constraints or properties.
 %(e.g., orthogonality, sparseness, statistically independence, non-negativity, smoothness).
%
The CA  methods  exploit  {\it a priori} knowledge about the true
nature or diversities  of latent  variables.
By diversity, we refer to  different characteristics, features or morphology of sources  or hidden latent variables \cite{Cichocki-GCA}.

For example, the columns of the matrix $\bB$ that represent different data sources should be:  as statistically
 independent as possible for ICA;  as sparse as possible for SCA;  take only nonnegative values for
(NMF) \cite{NMF-book,Cichocki-GCA,Comon-Jutten2010}.

{\bf Remark:} Note that  matrix factorizations have an inherent symmetry, Eq. (\ref{2CA}) could be written as $\bX^T \approx \bB \bA^T$,  thus interchanging the roles of  sources and mixing process.

Singular value decomposition (SVD) of the data matrix $\bX \in \Real^{I \times J}$ is a special case of the factorization in Eq. (\ref{2CA}). It  is exact and  provides an explicit notion of the range and null space of the matrix $\bX$ (key issues in low-rank approximation), and is given  by
\begin{equation}
\bX= \bU \mbi \Sigma \bV^T =\sum_{r=1}^R\sigma_r \; \bu_r \bv_r^T =\sum_{r=1}^R \sigma_r \; \bu_r \circ \bv_r,
%\bX= \bU \mbi \Sigma \bV^T = \mbi \Sigma  \times_1 \bU  \times_2 \bV
% =\sum_{r=1}^R\sigma_r \bu_r \bb_r^T, \notag
\label{SVD1}
\end{equation}
where $\bU $ and $\bV $ are column-wise orthonormal
matrices and $\mbi \Sigma$ is a diagonal matrix  containing only nonnegative singular values $\sigma_r$.

%Several applications lead to the computation of simultaneous matrix factorizations,
%\be
%\bX_k \approx \bA_k \bB_k^T, \qquad (k=1,2,\ldots,K),
%\label{GBSS1}
%\ee
%subject to various constraints, including constraints that couple the decompositions. An example is the analysis of multiple-subject, multiple-task datasets in brain function studies. In several cases where the matrices $\bX_k$ have equal size

Another virtue of component analysis comes from a representation of multiple-subject, multiple-task datasets
by a set of data matrices $\bX_k$, allowing us
 to perform  simultaneous matrix factorizations:
\be
\bX_k \approx \bA_k \bB_k^T, \qquad (k=1,2,\ldots,K),
\label{GBSS1}
\ee
subject to various  constraints.
%(e.g., $\bA_k=\bA, \; \forall k $ and their columns are mutually independent and/or sparse).
In the case of statistical  independence constraints, the problem can be related to  models of group ICA through suitable pre-processing,  dimensionality reduction and post-processing procedures \cite{GICA-rev}. %\cite{GICA-rev,groupICA2008guo,Langers-GICA,LinkedICA}.

The field of CA is  maturing  and has generated efficient  algorithms  for  2-way component analysis (especially, for sparse/functional PCA/SVD, ICA, NMF and SCA) \cite{Comon-Jutten2010,NMF-book,Zhou-Cichocki-SP}. The rapidly emerging field of tensor decompositions is the next important step that naturally generalizes 2-way CA/BSS algorithms and paradigms.
%
%Tensor decompositions and tensor networks are therefore the next key step that generalizes the existing 2-way component analysis paradigms to  enable flexible and powerful processing of such data via decomposition them into simpler interconnected factors.
%
%We needs to develop a similar infrastructure that can be strongly supported by
%2-way CA algorithms and paradigms.
%
We proceed to show how  constrained matrix factorizations and
component analysis (CA) models can be naturally  generalized to multilinear
models using constrained tensor decompositions,  such as the Canonical Polyadic Decomposition (CPD) and Tucker  models, as
illustrated in Figs. \ref{Fig:CPD1} and \ref{Fig:Tucker}.

%\section{Tensor Decompositions}

%Usually tensor is decomposed into several parts or terms as illustrated in Fig \ref{Fig:model}.
%The main objective of tensor decomposition to decompose tensor to easily physically interpretable or meaningful  factor matrices and a core tensor which indicate links between components (vectors of factor matrices) in different modes.
%%%%
%\begin{figure}[ht]
%\centering
%\includegraphics[width=7.8cm]{model.eps}
%%\includegraphics[width=7.8cm]{CPDVec4ss.eps}
%\caption{
%Illustration of decomposition of data tensor $\underline \bX$ into three fundamental terms.}
%\label{Fig:model}
%\end{figure}

\subsection{\bf The Canonical Polyadic Decomposition (CPD)} \label{sect:CPD}

% LDL: check: include {Acar-Morup11}?
% remove/replace Comon, other tales?
% other norm: \cite{vorobyov}

%\subsection{Definition}

\begin{figure}[t!]
%\centering
(a) Standard block diagram\\
\vspace{0.3cm}
\begin{center}
 \includegraphics[width=8.6cm]{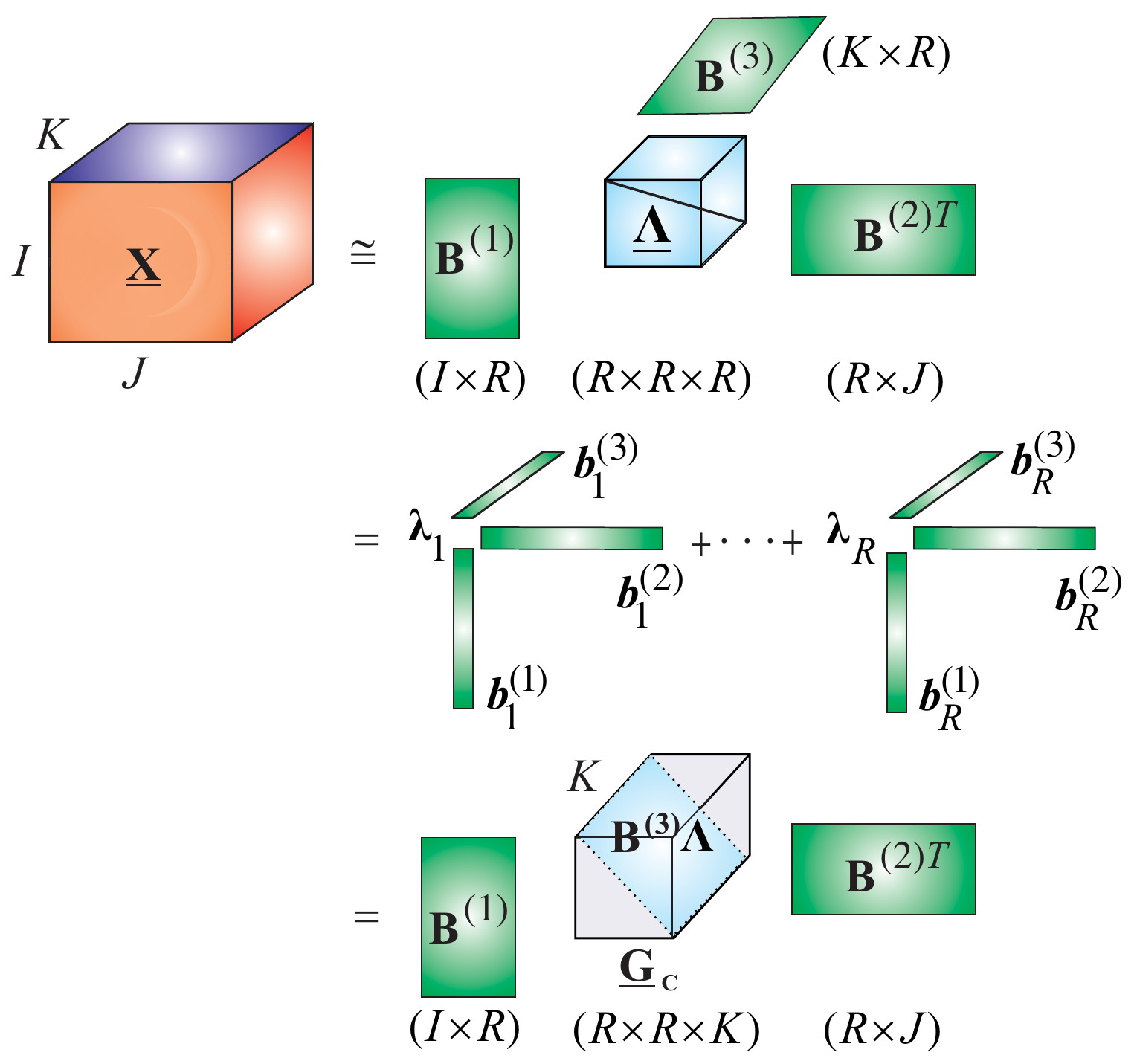}\\
 \end{center}
\vspace{0.2cm}
(b)  CPD in tensor network notation
\begin{center}
\includegraphics[width=6.6cm]{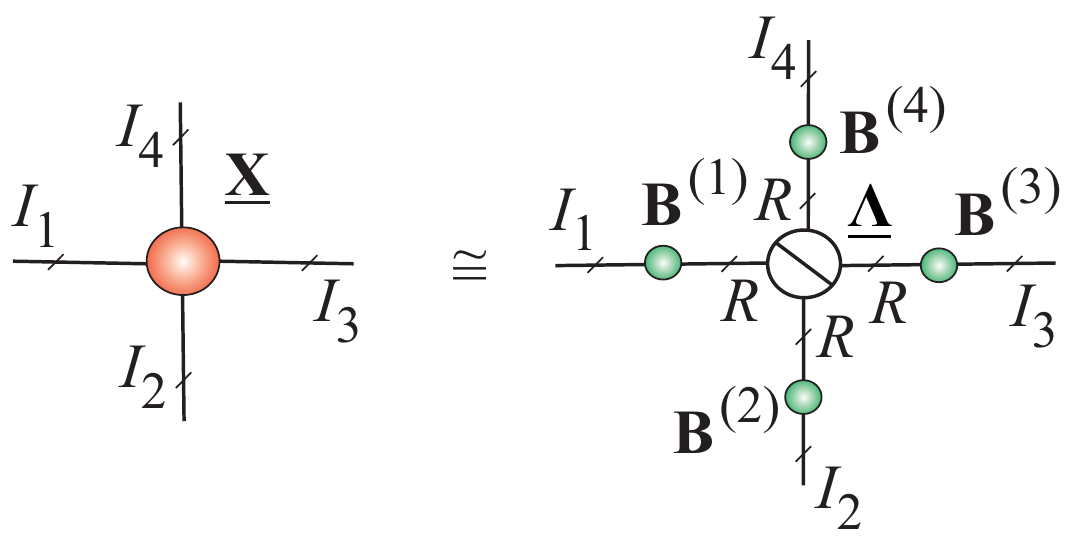}
\end{center}
\caption{Representation of the CPD. (a) The Canonical Polyadic Decomposition (CPD)  of a 3rd-order tensor as: $\underline \bX \cong \underline {\mbi \Lambda} \times_1 \bB^{(1)} \times_2 \bB^{(2)} \times_3 \bB^{(3)} = \sum_{r=1}^R \lambda_r \; \bb^{(1)}_r \circ \bb^{(2)}_r \circ \bb^{(3)}_r = \underline \bG_c \times_1 \bB^{(1)} \times_2 \bB^{(2)}$ with $\underline \bG = \underline {\mbi \Lambda}$ and $\underline \bG_c = \underline {\mbi \Lambda} \times_3 \bB^{(3)}$.  (b) The CPD for a 4th-order tensor as: $\underline \bX \cong  \underline {\mbi \Lambda} \times_1 \bB^{(1)} \times_2 \bB^{(2)} \times_3 \bB^{(3)} \times_4 \bB^{(4)} =\sum_{r=1}^R  \bb^{(1)}_r \circ \bb^{(2)}_r \circ \bb^{(3)}_r \circ \bb^{(4)}_r$. The objective of the CPD is to estimate the factor matrices $\bB^{(n)}$ and a rank of tensor $R$, that is, the number of components $R$.}
\label{Fig:CPD1}
\end{figure}

The CPD  (called also PARAFAC or CANDECOMP) factorizes an $N$th-order tensor $\underline \bX \in \Real^{I_1 \times I_2 \times  \cdots \times I_N}$ into a linear combination of terms $\bb^{(1)}_r \circ  \bb^{(2)}_r  \circ \cdots \circ \bb^{(N)}_r$, which are rank-1 tensors, and is given by \cite{Hitchcock1927,Harshman,PARAFAC1970Carroll}%(see Fig. \ref{Fig:CPD1}):
\begin{equation}
\label{CPDN}
\begin{aligned}
 \underline \bX & \cong  \sum_{r=1}^R  \lambda_r \; \bb^{(1)}_r \circ  \bb^{(2)}_r  \circ \cdots \circ \bb^{(N)}_r \\
 &= \underline {\mbi \Lambda} \times_1 \bB^{(1)} \times_2 \bB^{(2)} \cdots
\times_N \bB^{(N)}\\
 &= \llbracket  \underline {\mbi \Lambda}; \matn[1]{B},  \matn[2]{B}, \ldots, \matn[N]{B}\rrbracket,
  \end{aligned}
\end{equation}
where the only non-zero entries $\lambda_r$  of the diagonal core tensor $\underline \bG =\underline {\mbi {\Lambda}}
\in \Real^{R \times R \times \cdots \times R}$ are located on the main diagonal (see Fig. \ref{Fig:CPD1} for a 3rd-order and 4th-order tensors).

Via the Khatri-Rao products the CPD can also be  expressed in a matrix/vector form as:
\begin{equation}
\label{CPD-KR}
\bX_{(n)} \cong  \bB^{(n)} \mbi \Lambda (\bB^{(1)} \odot \cdots \odot \bB^{(n-1)} \odot \bB^{(n+1)} \odot \cdots \odot\bB^{(N)})^T \notag \\
\end{equation}
\be
&\text{vec}(\underline \bX) \cong [\bB^{(1)} \odot \bB^{(2)} \odot \cdots \odot\bB^{(N)}] \; \mbi \lambda,
\ee
%\be
%\label{CPD-KR}
%&\bX_{(n)} \cong  \bB^{(n)} \mbi \Lambda (\bB^{(N)} \odot \cdots \odot \bB^{(n+1)} \odot \bB^{(n-1)} \odot \cdots \odot\bB^{(1)})^T \notag \\
%&\text{vec}(\underline \bX) \cong [\bB^{(N)} \odot \bB^{(N-1)} \odot \cdots \odot\bB^{(1)}] \; \mbi \lambda,
%%\label{CPD-KR}
%\ee
where  $\bB^{(n)} = [\bb_1^{(n)},\bb_2^{(n)},\ldots,\bb_R^{(n)}] \in \Real^{I_n \times R}$, $\mbi \lambda =[\lambda_1,\lambda_2, \ldots,\lambda_R]^T$ and $\mbi \Lambda = \diag\{\mbi \lambda\}$ is a diagonal matrix.

The rank of tensor $\underline \bX$ is  defined as the smallest $R$ for which CPD (\ref{CPDN}) holds  exactly.

{\bf Algorithms to compute CPD.}  In the presence of noise in real world applications the  CPD is rarely exact and has to be  estimated by  minimizing a suitable  cost function, typically of the Least-Squares (LS) type in the form of the Frobenius norm $||\underline \bX - \llbracket  \underline {\mbi \Lambda}; \matn[1]{B},  \matn[2]{B}, \ldots, \matn[N]{B}\rrbracket ||_F$, or using Least Absolute Error (LAE) criteria \cite{Vorobyov}.
The Alternating Least Squares (ALS) algorithms \cite{Comon-ALS09,Harshman,NMF-book,HALS2009}  minimize the LS cost function by optimizing individually  each component matrix, while keeping the other component matrices fixed.
%For the LS cost function, such a conditional update boils down to   a standard linear LS problem.
%
For instance, assume that the diagonal matrix $\mbi \Lambda$   has been absorbed in one of the component matrices; then, by taking advantage of the Khatri-Rao structure the component matrices $\bB^{(n)}$ can be updated sequentially  as \cite{Kolda08}
\be
\bB^{(n)} \leftarrow \bX_{(n)} \left( \bigodot_{k\neq n} \bB^{(k)} \right) \left( \*_{k\neq n} (\bB^{(k)\;T} \bB^{(k)}) \right)^{\dagger},
\label{ALS-CP}
\ee
 which requires the computation of the pseudo-inverse  of  small $(R \times R)$ matrices.
 %instead of the large $R \times \prod_{k\neq n} I_k$-dimensional matrix $(\bigodot_{k \neq n} \bB^{(k)})^T$ in (\ref{CPD-KR}).

 The ALS is attractive for its simplicity and for well defined problems (not too many, well separated, not collinear components) and high SNR, the performance of ALS  algorithms is often satisfactory.
For  ill-conditioned problems, more advanced algorithms exist, which typically exploit the rank-1 structure of the terms within CPD to perform efficient computation and storage of the Jacobian and Hessian of the cost function \cite{Phan2012-Hess,Sorber2012b}.

%
%\subsection{Constraints}
%
{\bf Constraints.} The CPD is usually unique by itself, and does not require constraints to impose uniqueness \cite{Sidiropoulos2000}.  However, if  components in one or more modes are known to be  e.g.,   nonnegative, orthogonal, statistically independent or sparse, these constraints should be incorporated to relax  uniqueness conditions. More importantly, constraints may increase the accuracy and stability of the CPD algorithms and facilitate better physical interpretability of components \cite{sorensen,Zhou2012-SPL}.

\subsection{\bf The Tucker Decomposition} \label{sect:Tucker}

\begin{figure}[t]
%\centering
(a) Standard block diagram of TD\\
\begin{center}
\includegraphics[width=8.99cm]{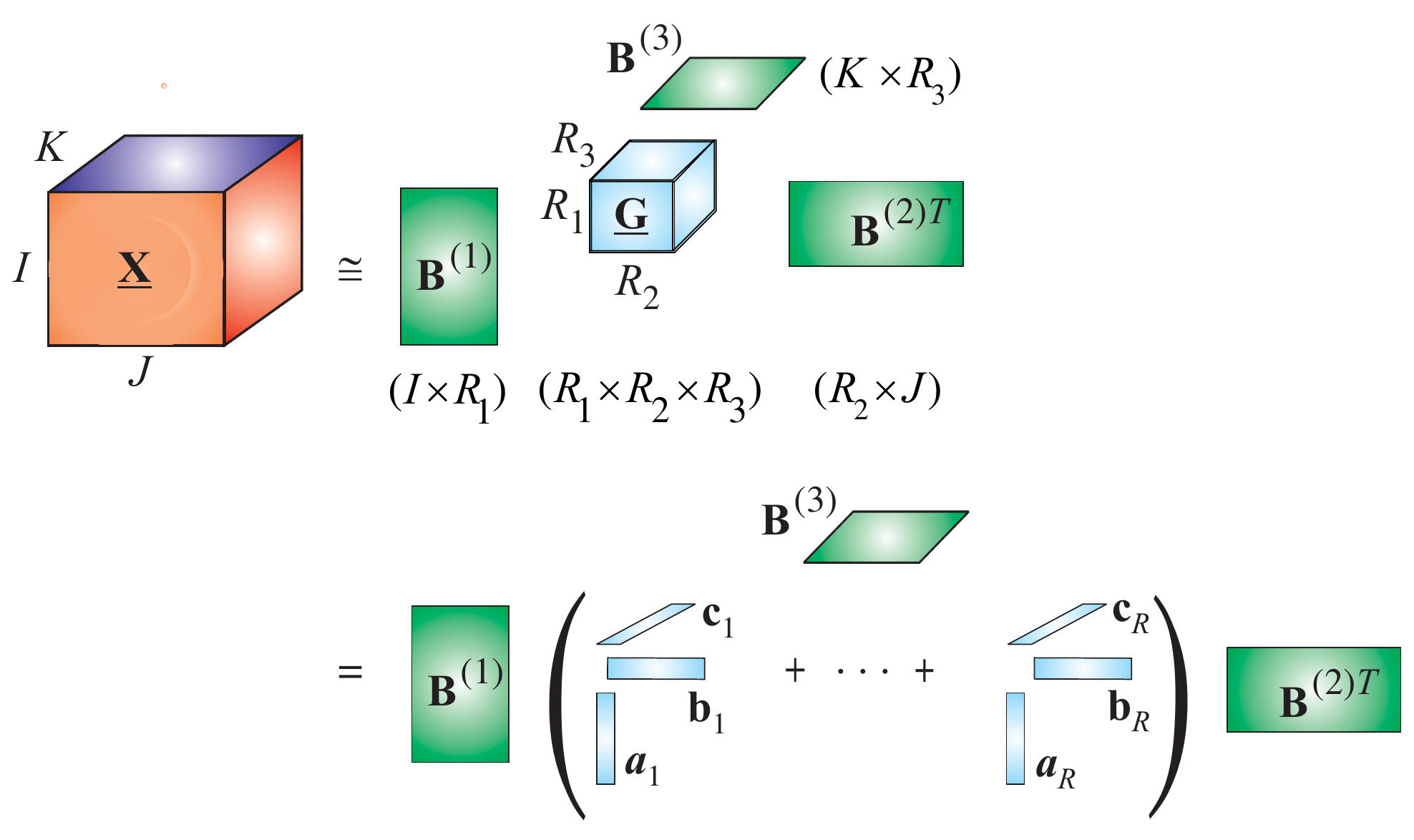}\\
\end{center}
% \includegraphics[width=8.6cm]{TuckCPD2.eps}\\
%\includegraphics[width=8.99cm]{CPD1m.eps}\\
%\includegraphics[width=8.99cm]{Fig15a.eps}\\ \\
%(b) \includegraphics[width=7.8cm]{Tucker4.eps}
(b) TD in tensor network notations \\
\begin{center}
 \includegraphics[width=8.7cm,height=4.2cm]{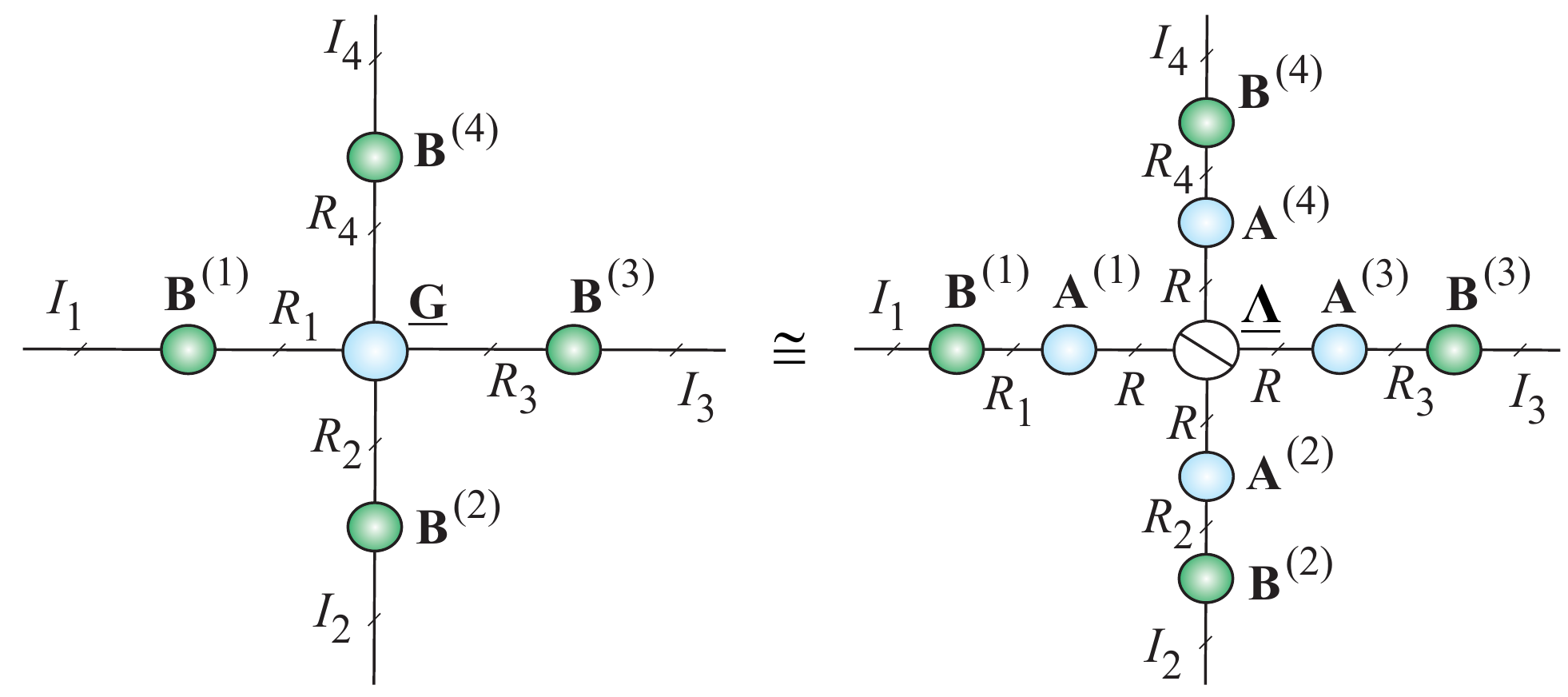}
 \end{center}
\caption{Representation of the Tucker Decomposition (TD). (a) TD of a 3rd-order tensor $\underline \bX \cong \underline \bG \times_1 \bB^{(1)} \times_2 \bB^{(2)} \times_3 \bB^{(3)}$. The objective
is to compute factor matrices $\bB^{(n)}$ and core tensor $\underline \bG$. In some applications, in the second stage, the core tensor is approximately factorized using the CPD as $\underline \bG \cong \sum_{r=1}^R \ba_r \circ \bb_r \circ \bc_r$.  (b) Graphical representation of the Tucker  and CP decompositions in two-stage procedure for a 4th-order tensor as: $\underline \bX \cong \underline \bG \times_1 \bB^{(1)} \times_2 \bB^{(2)} \cdots  \times_4 \bB^{(4)} =\llbracket \underline \bG; \bB^{(1)},  \bB^{(2)}, \bB^{(3)}, \bB^{(4)} \rrbracket \cong (\underline {\mbi \Lambda} \times_1 \bA^{(1)} \times_2 \bA^{(2)} \cdots  \times_4 \bA^{(4)})  \times_1 \bB^{(1)} \times_2 \bB^{(2)} \cdots  \times_4 \bB^{(4)} = \llbracket \underline {\mbi \Lambda}; \; \bB^{(1)} \bA^{(1)}, \; \bB^{(2)} \bA^{(2)}, \; \bB^{(3)} \bA^{(3)}, \; \bB^{(4)} \bA^{(4)} \rrbracket$.}
\label{Fig:Tucker}
\end{figure}

The Tucker decomposition can be expressed  as follows \cite{Tucker1966}:
%a  multilinear product of a core tensor with $N$ factor matrices, possibly with a different number of components $R_n$ in each mode, so that \cite{Tucker1966}:
%%
\be
%\label{GeneralTDModel}
  \underline \bX
 & \cong & \sum\limits_{r_1 = 1}^{R_1} {
 {\cdots
 \sum\limits_{r_N = 1}^{R_N}{g_{r_1 r_2 \cdots r_N} \, \left(\; \bb^{(1)}_{r_1} \circ \bb^{(2)}_{r_2} \circ \cdots \circ \bb^{(N)}_{r_N} \right)}}} \notag \\
& = & \underline \bG \times_1 \bB^{(1)} \times_2 \bB^{(2)} \cdots \times_N \bB^{(N)}  \notag \\
& = & \llbracket\underline \bG; \bB^{(1)},\bB^{(2)},\ldots,\bB^{(N)}\rrbracket.
 \label{GeneralTDModel}
 \ee
 where $\underline \bX \in \Real^{I_1\times I_2 \times \cdots \times I_N}$ is the given data tensor,  $\underline \bG  \in \Real^{R_1 \times R_2 \times \cdots \times R_N}$  is the core tensor and $\bB^{(n)} =[\bb_1^{(n)},\bb_2^{(n)},\ldots,\bb_{R_n}^{(n)}] \in \Real^{I_n\times R_n}$ are the mode-$n$ component  matrices, $n=1,2,\ldots,N$ (see Fig. \ref{Fig:Tucker}).
% %

 Using Kronecker products the decomposition in (\ref{GeneralTDModel}) can be expressed in a matrix and vector form as follows:
%
%\begin{eqnarray}
\begin{equation}
%\bX_{(n)} &\cong& \bB^{(n)} \bG_{(n)} (\bigotimes_{k \neq n} \bB^{(k)})^T \\
\bX_{(n)} \cong \bB^{(n)} \bG_{(n)} (\bB^{(1)} \cdots \otimes \bB^{(n-1)} \otimes \bB^{(n+1)} \cdots \otimes \bB^{(N)})^T \notag
\end{equation}
\be
\text{vec}(\underline \bX) \cong [\bB^{(1)} \otimes \bB^{(2)} \cdots \otimes \bB^{(N)}] \text{vec}(\underline \bG).
\label{Tucker-kron}
\ee

\begin{figure}[ht]
(a) \\
\vspace{0.2cm}
\begin{center}
\includegraphics[width=4.8cm]{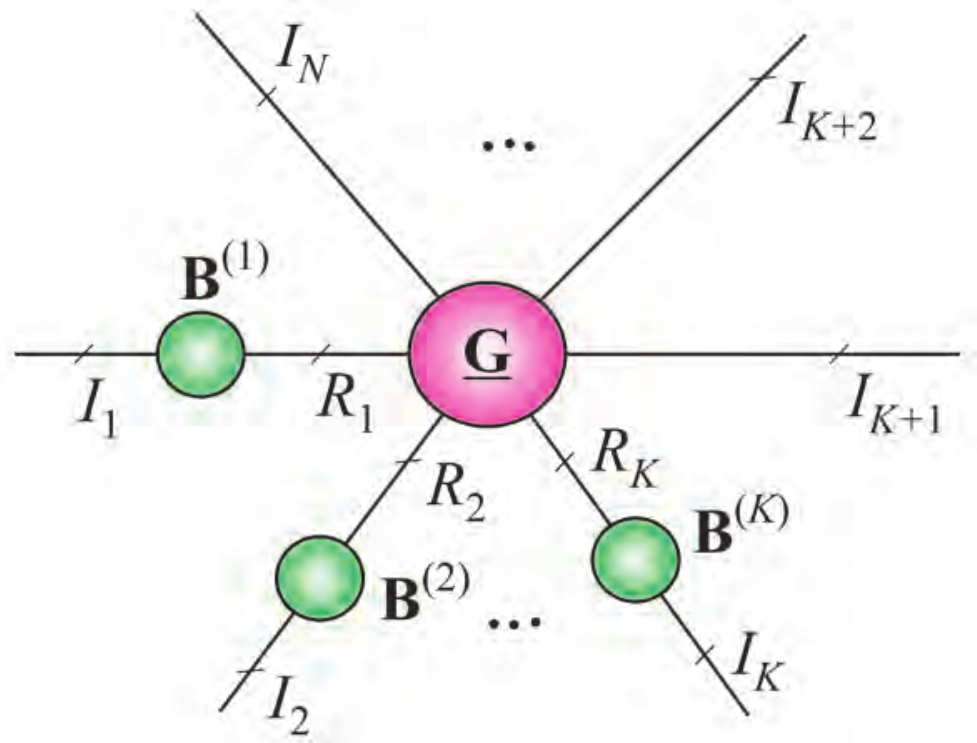}\\
\end{center}
(b)\\
\begin{center}
 \includegraphics[width=7.8cm]{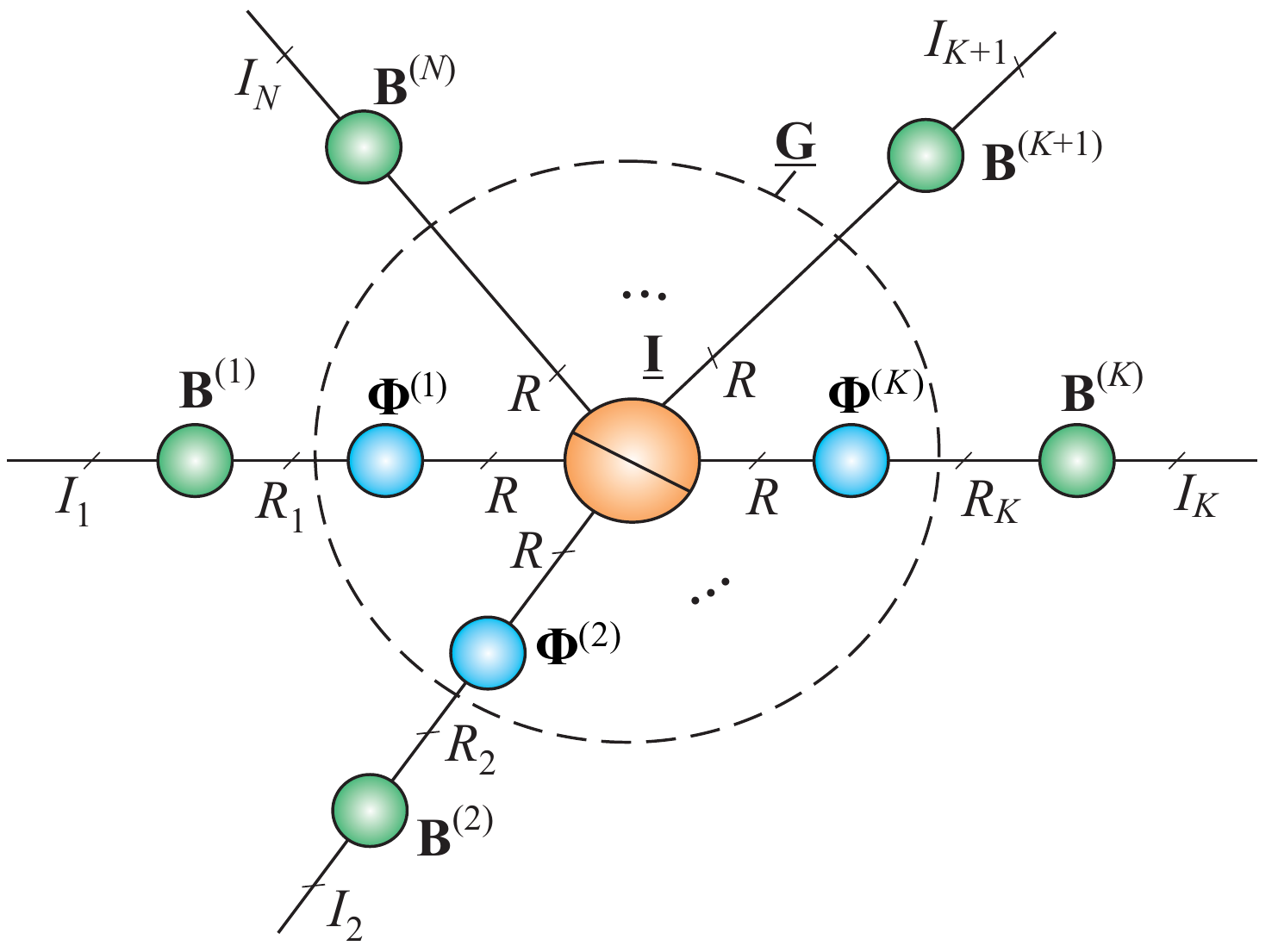}
 \end{center}
\caption{ Graphical illustration of  constrained Tucker and  CPD models: (a) Tucker-$(K,N)$  decomposition of a $N$th-order tensor, with $N \geq K$, $\underline \bX \cong \underline \bG \times_1 \bB^{(1)} \times_2 \bB^{(2)} \cdots \times_K \bB^{(K)} \times_{K+1} \bI \times_{K+2} \cdots \times_N \bI = \llbracket\underline \bG;\bB^{(1)},\bB^{(2)}, \ldots,\bB^{(K)} \rrbracket $,  (b) Constrained CPD  model, called PARALIND/CONFAC-$(K,N)$   $\underline \bX \cong \underline \bG \times_1 \bB^{(1)} \times_2 \bB^{(2)} \cdots \times_N \bB^{(N)} = \llbracket\underline \bI;\bB^{(1)} \Phi^{(1)},\bB^{(2)}\Phi^{(2)}, \ldots,\bB^{(K)} \Phi^{(K)},\bB^{(K+1)}, \ldots,\bB^{(N)} \rrbracket $,
where core tensor $\bG= \underline \bI \times_1 \Phi^{(1)} \times_2 \Phi^{(2)} \cdots \times_K \Phi^{(K)} $ with  $K \leq N$.}
\label{Fig:Paralind}
\end{figure}

The core tensor (typically,  $R_n < I_n$) models a potentially complex pattern of mutual interaction between the vectors (components) in  different modes.

{\bf Multilinear rank.}  The $N$-tuple $(R_1,R_2,\ldots,R_N)$ is called the multilinear-rank of $\underline \bX$, if the Tucker decomposition holds exactly.

%An essential difference between CPD and Tucker is that while for CPD  the core tensor is diagonal and the factor matrices $\bB^{(n)}$ do not necessarily have full column rank.
%but admit  rank-1  decomposition suitable for a wide class of signal separation.
Note that the CPD can be  considered as a special case of the Tucker decomposition, in which the core tensor has nonzero elements only on main diagonal. In contrast to the CPD the Tucker decomposition, in general, is non unique. However, constraints imposed on all factor matrices and/or core tensor can reduce the indeterminacies  to only column-wise permutation and scaling \cite{Zhou-Cichocki-MBSS}.

In Tucker model some selected factor matrices can be identity  matrices, this leads to Tucker-$(K,N)$ model, which is graphically illustrated in Fig. \ref{Fig:Paralind} (a). In a such model $(N-K)$ factor matrices are equal to identity matrices. In the simplest scenario for 3rd-order tensor $\underline \bX \in \Real^{I_1 \times I_2 \times I_3} $ the Tucker-(2,3) model, called simply Tucker-2, can be described as
\be
\underline \bX \cong \underline \bG \times_1 \bB^{(1)} \times_2 \bB^{(2)}.
\ee
Similarly, we can define PARALIND/CONFAC-$(K,N)$ models{\footnote{PARALIND is abbreviation of PARAllel with LINear Dependencies, while CONFAC means CONstrained FACtor model (for more detail see  \cite{Favier-deAlmeida14} and references therein.)}}described as \cite{Favier-deAlmeida14}
\be
\underline \bX &\cong& \underline \bG \times_1 \bB^{(1)} \times_2 \bB^{(2)} \cdots \times_N \bB^{(N)} \\
 &= & \llbracket\underline \bI;\bB^{(1)} \Phi^{(1)}, \ldots,\bB^{(K)} \Phi^{(K)},\bB^{(K+1)} \ldots,\bB^{(N)} \rrbracket,\notag
\ee
where the core tensor, called constrained tensor or interaction tensor, is expressed as
\be
\bG= \underline \bI \times_1 \Phi^{(1)} \times_2 \Phi^{(2)} \cdots \times_K \Phi^{(K)},
\ee
 with  $K \leq N$. The factor matrices $\Phi^{(n)} \in \Real^{R_n \times R}$, with $R \geq \max(R_n)$ are constrained matrices, called often interaction matrices (see Fig. \ref{Fig:Paralind} (b).

 Another important, more complex  constrained CPD model, which can be represented
 graphically as nested Tucker-$(K,N)$ model is the PARATUCK-$(K,N)$ model (see review paper of Favier and de Almeida \cite{Favier-deAlmeida14} and references therein).

%There  are important special cases (see Fig. \ref{Fig:Paralind} (c) and (d):
% the PARATUCK-2 model can be described as
%  \be
%  \underline \bX \cong \underline \bG \times_1 \bB^{(1)} \times_2 \bB^{(2)},
%  \ee
%with the core tensor
%\be \bG= \underline \bI \times_1  \Psi^{(1)} \times_2 \Psi^{(2)} \times_3 \hat \bC
%\ee
% and ) PARATUCK-(2,4) model"
%\be
%\underline \bX \cong \underline \bG \times_1 \bB^{(1)} \times_2 \bB^{(2)},
%with core tensor
%\be
%\bG= \underline \bI \times_1 \Psi^{(1)} \times_2 \Psi^{(2)} \times_3 \bar \bF \times \bar \bD.

\subsection{\bf Multiway Component Analysis Using Constrained  Tucker Decompositions}
%\label{sec:MBSS}

A great success of 2-way component analysis  (PCA, ICA, NMF, SCA) is
largely due to the various constraints we can impose.
Without constraints matrix factorization loses its most
sense as the components are rather arbitrary and they do not have any physical meaning. There are various
constraints that lead to all kinds of component analysis
methods which are able give unique components with
some desired physical meaning and properties and hence
serve for different application purposes. Just similar to matrix
factorization, unconstrained Tucker decompositions
generally can only be served as multiway data compression
as their results lack physical meaning. In the most
practical applications we need to consider constrained
Tucker decompositions which can provide multiple sets
of essential unique components with desired physical
interpretation and meaning. This is direct extension
of 2-way component analysis and is referred to as
multiway component analysis (MWCA) \cite{Cich-Lath}.

The MWCA based on Tucker-$N$ model can be considered as a natural and simple extension of multilinear SVD and/or multilinear ICA,  in which we apply any efficient  CA/BSS algorithms to each mode, which provides essential uniqueness \cite{Zhou-Cichocki-MBSS}.
%
%{\bf Application 2. Multiway Component Analysis.} Latent (hidden) components in data
%exhibit a variety of complex properties, diversities  and features - true unknown hidden components
%are seldom all e.g. orthogonal or statistically independent, and the use of a single criterion, like SVD or ICA will not extract all the latent components \cite{NMF-book,Cichocki-GCA}.
%
%Tensor naturally accommodate  multiple criteria or diversities in component matrices in various modes, making MWCA more versatile than multilinear SVD and/or multilinear ICA \cite{Zhou-Cichocki-MBSS}.
%
%Since multiway array data can be always interpreted in many different ways, some {\it a priori}
%knowledge is needed to determine which diversities, characteristics, features or properties represent  true latent (hidden) components with physical meaning.

%{\bf Remark} Note that the MWCA (Multiway Component Analysis) can be considered as  a constrained tensor decomposition  into component matrices, whose vectors or components have specific diversities or morphological properties.

%Notice that in order to  estimate the components (vectors of component matrices) we need to apply a suitable BSS or CA method to unfolding matrices of the data tensor in each individual mode.  The unfolded tensor  in each individual mode represents a linear mixture of the components (vectors of component matrices), so that components with specific desired diversities and properties can be extracted directly by applying CA/BSS.

 There are two different models to  interpret and implement  constrained Tucker decompositions for  MWCA. (1) the columns of the component matrices $\bB^{(n)}$ represent the desired  latent variables,   the core tensor $\underline \bG$ has a role of  ``mixing process'', modeling the links among the components from different modes, while the data tensor $\underline \bX$  represents a collection of 1-D or 2-D mixing signals; (2) the core tensor represents the desired (but  hidden)  $N$-dimensional signal (e.g., 3D MRI image or 4D video), while the component   matrices represent  mixing or filtering processes through e.g., time-frequency transformations or wavelet dictionaries \cite{Cichocki-SICE}.

The  MWCA based on the Tucker-$N$ model can be computed directly
in two steps:
(1)  for $n=1,2, \ldots, N$ perform model reduction and unfolding of data tensors sequentially
and apply a  suitable set of  CA/BSS algorithms to reduced unfolding matrices $\tilde \bX_{(n)}$,  - in each mode we can  apply different  constraints and algorithms;
(2)  compute the core tensor using e.g., the inversion formula: %$ {\widehat{{\underline \bG}}}
 $ \hat{\underline \bG} = \underline \bX  \times_{1}\matn[1]{B}{}^\dagger \times_{2}\matn[2]{B}{}^\dagger
  \cdots \times_{N}\matn[N]{B}{}^\dagger$ \cite{Zhou-Cichocki-MBSS}. This step is quite important because core tensors illuminate complex links among the multiple components
  in different modes \cite{NMF-book}.

 \begin{figure}[t!]
%\centering
%\psfrag{y}{\color{black}$y$}
(a) \\
\includegraphics[width=8.6cm]{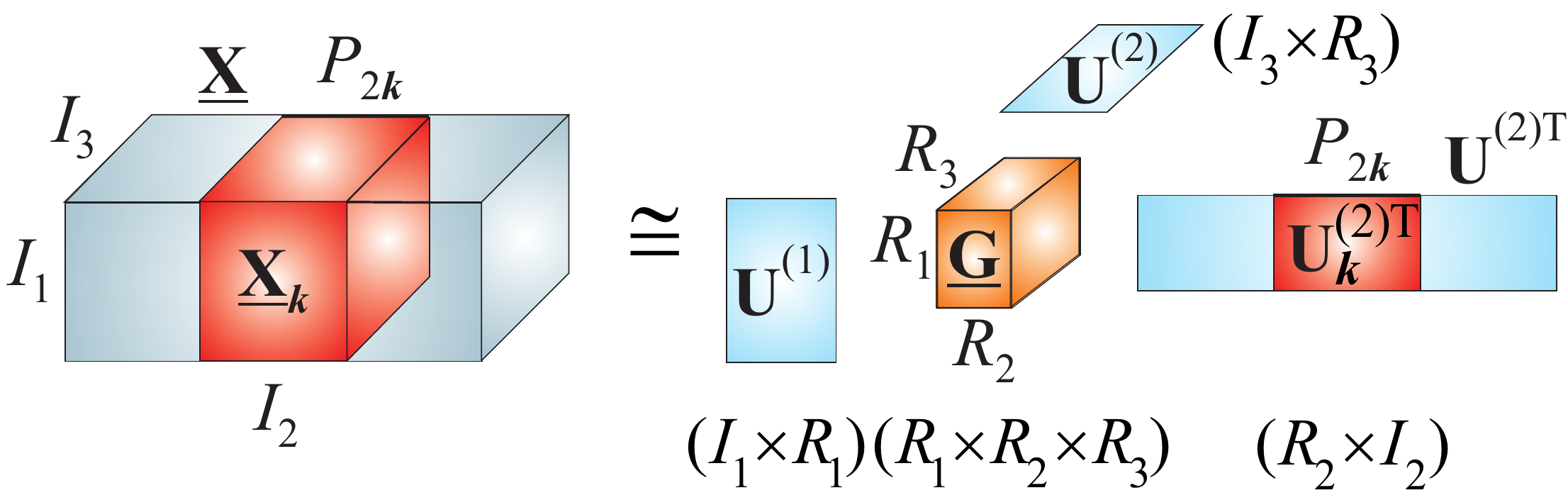}\\ \\
(b) \\
\includegraphics[width=8.8cm]{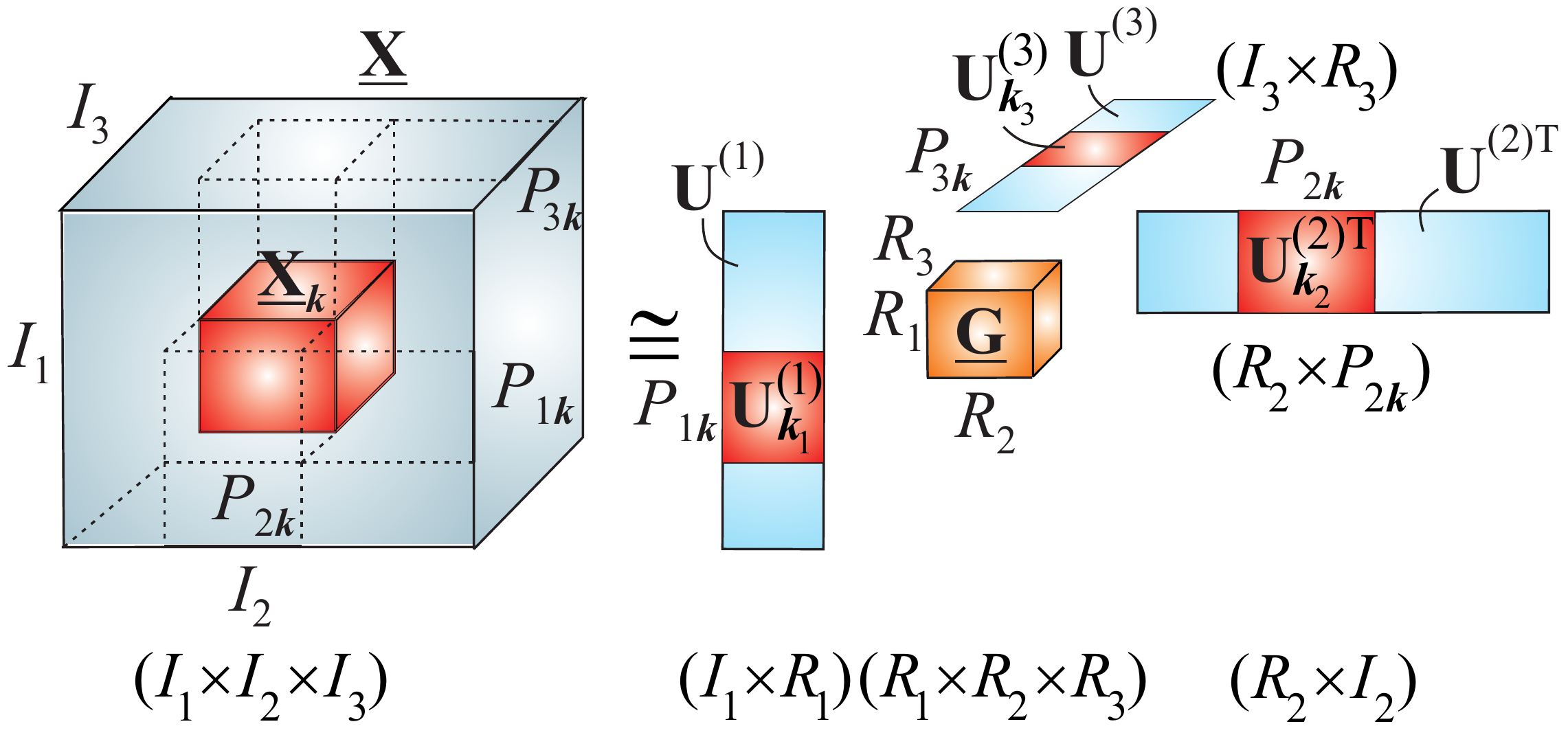}\
%(c)
%\includegraphics[width=8.6cm,height=4.6cm]{Grid-Tensor1.eps}
%\includegraphics[width=7.2cm]{6order33.eps}\\
%\includegraphics[width=7.8cm]{4ordten.eps}
\caption{Conceptual models for performing  the  Tucker decomposition (HOSVD) for  large-scale 3rd-order data tensors by dividing the  tensors into blocks (a) along one largest dimension mode, with blocks $\underline \bX_k \cong \underline \bG \times_1 \bU^{(1)} \times_2 \bU_k^{(2)} \times_3 \bU^{(3)}$, $(k=1,2\ldots,K)$, and (b) along all modes  with blocks $\underline \bX_k \cong \underline \bG \times_1 \bU_{k_1}^{(1)} \times_2 \bU_{k_2}^{(2)} \times_3 \bU_{k_3}^{(3)}$. The models can be used for an
anomaly detection by fixing a core tensor and some factor matrices and by monitoring the
changes along one or more specific modes. First, we compute tensor decompositions for sampled
(pre-selected) small blocks and in the next step we analyze changes in specific factor matrices  $\bU^{(n)}$.}
\label{Fig:blocks}
\end{figure}

\section{\bf Block-wise Tensor Decompositions for Very Large-Scale Data}

Large-scale tensors cannot be processed by commonly used computers,
since not only their size exceeds available working memory but also processing of huge data is very slow.
The basic idea is to perform partition of a big data tensor into smaller blocks and then perform tensor related operations block-wise using a suitable tensor format (see Fig. \ref{Fig:blocks}).
A data management system that divides the data tensor into blocks is  important approach to both  process and to save large datasets.
The method is based on a decomposition of the original tensor dataset into small block tensors, which are approximated via TDs.
Each block is approximated using low-rank reduced
tensor decomposition, e.g., CPD or a Tucker decomposition.

There are three important steps for such approach before we would be able to generate an
output: First, an effective tensor representation
should be chosen for the input dataset; second,
the resulting tensor needs to be partitioned into sufficiently small
blocks stored on a distributed memory system, so that each block can fit into the
main memory of a single machine; third, a suitable algorithm for TD needs to be
adapted so that it can take the blocks of the original tensor, and still
output the correct approximation as if the tensor for the original
dataset had not been partitioned \cite{Wang-out-core05,Phan-CP,Suter13}.

Converting the input data tensors from its original
format into this block-structured tensor format is straightforward,
and needs to be performed as a preprocessing step. The resulting
blocks should be saved into separate files on hard disks to allow
efficient random or sequential access to all of blocks, which is required by most TD and TN algorithms.

 We have successfully applied such
techniques  to CPD  \cite{Phan-CP}.  Experimental results indicate that our
algorithms cannot only process out-of-core data, but also achieve
high computation speed and good performance.

 \begin{figure}[t!]
%\centering
%\psfrag{y}{\color{black}$y$}
(a) \\
\includegraphics[width=8.7cm,height=6.0cm]{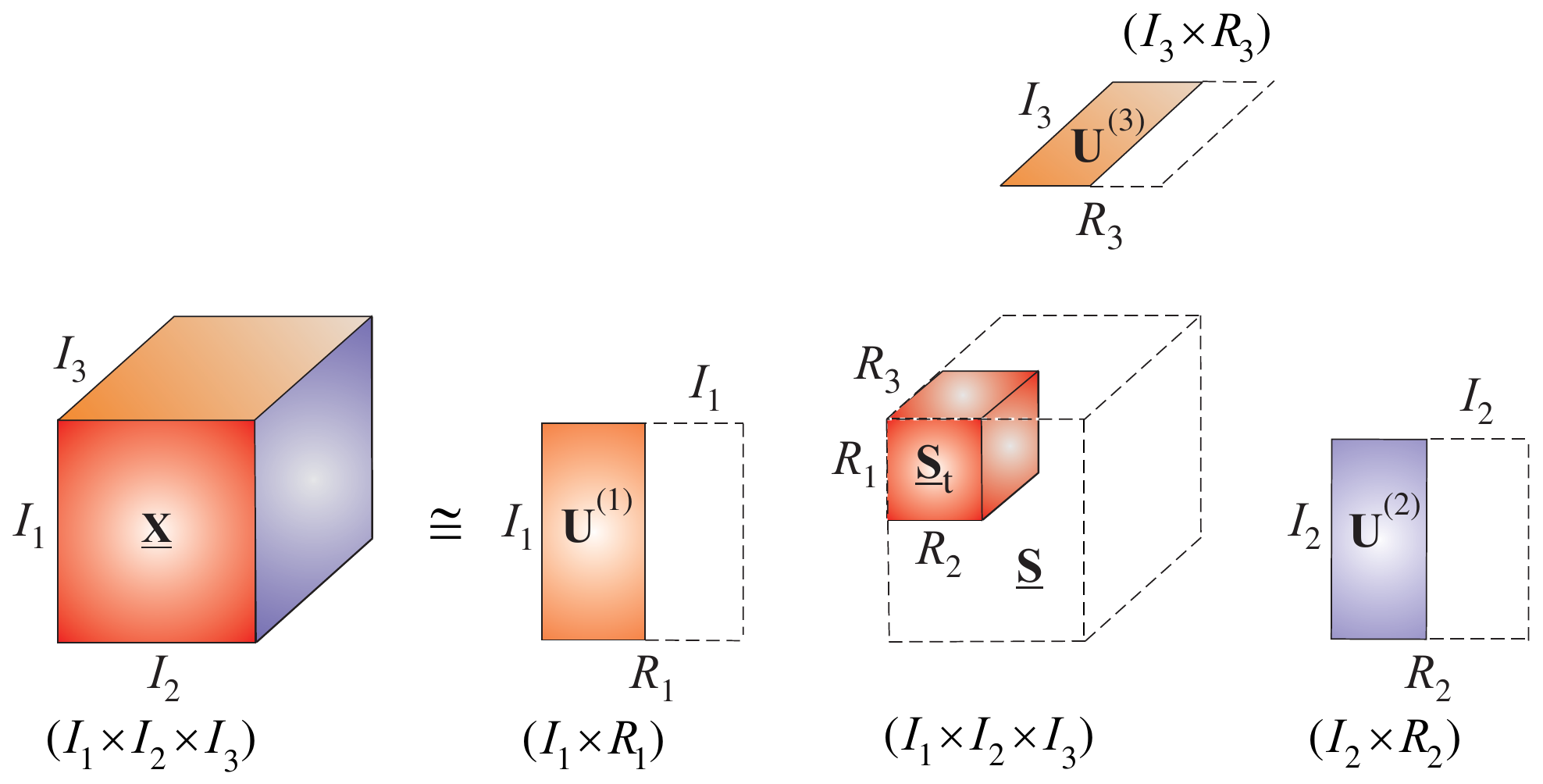}
(b) \\
\includegraphics[width=8.6cm]{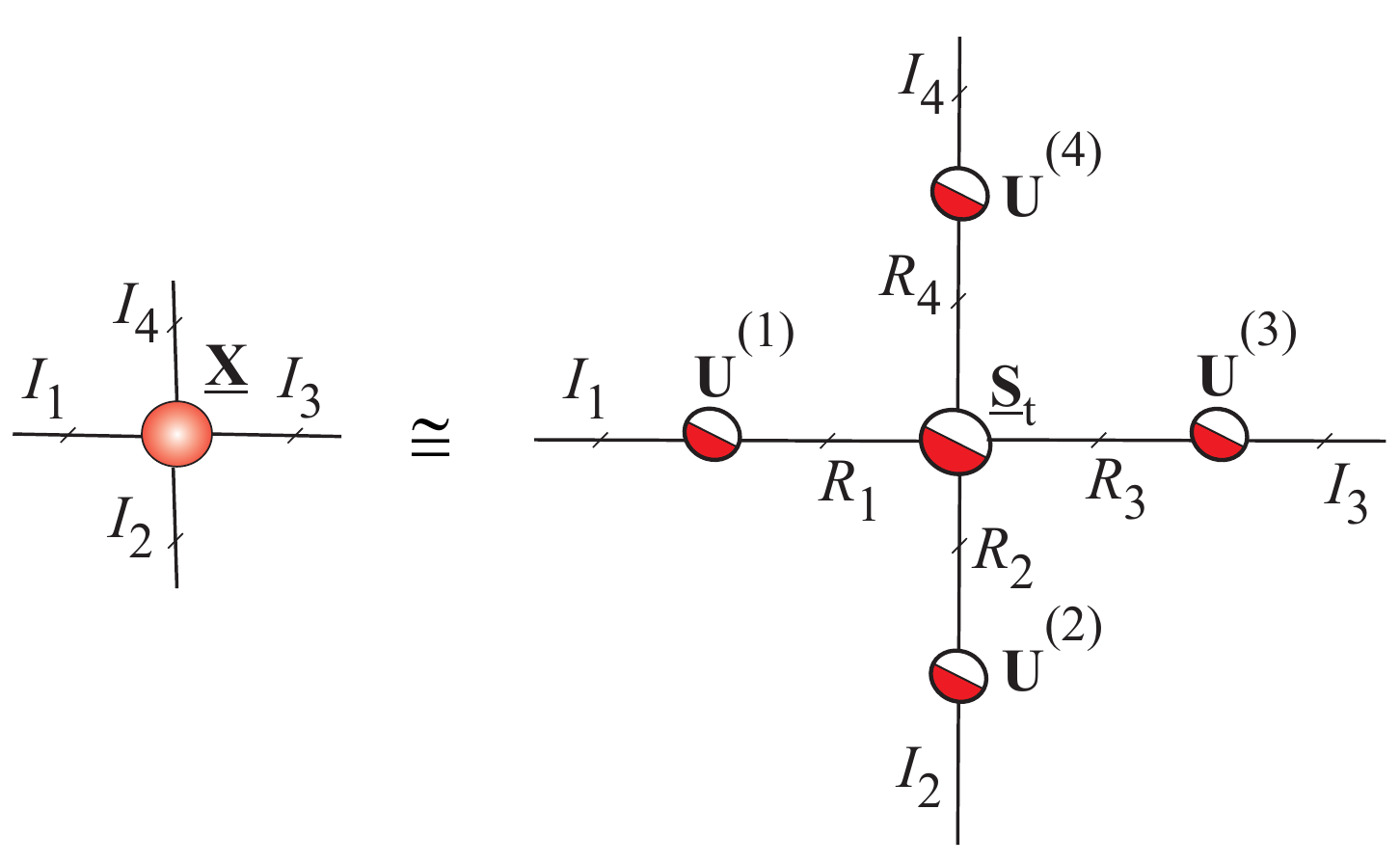}
%\centering
%\includegraphics[width=7.8cm]{HOSVD4.eps}
\caption{Graphical illustration of the HOSVD. (a) The exact HOSVD and truncated (approximative) HOSVD a for 3rd-order tensor as:
  $\underline \bX \cong \underline \bS_t \times_1 \bU^{(1)} \times_2 \bU^{(2)}  \times_3 \bU^{(3)}$ using a truncated SVD.  (b) Tensor network notation for the HOSVD for a 4th-order tensor $\underline \bX \cong \underline \bS_t \times_1 \bU^{(1)} \times_2 \bU^{(2)}  \times_3 \bU^{(3)} \times_4 \bU^{(4)}$. All factor matrices $\bU^{(n)}$ and the core tensor $\underline \bS_t$ are orthogonal; due to orthogonality of the core tensor the HOSVD is unique for a specific multilinear rank.}
\label{Fig:HOSVD}
\end{figure}

\section{\bf Multilinear SVD (MLSVD) for Large Scale Problems}
%{\bf Multilinear SVD (MLSVD).}

   MultiLinear Singular Value Decomposition (MLSVD),
   called also higher-order SVD (HOSVD) can be considered as a
   special form of the Tucker decomposition \cite{Lathauwer01,HOOI:Lathauwer:2000},
in which all factor matrices $\bB^{(n)}=\bU^{(n)} \in \Real^{I_n \times I_m}$ are orthogonal and the
 core tensor
$\underline \bG = \underline \bS \in \Real^{I_1 \times I_2 \times \cdots \times I_N}$ is all-orthogonal (see Fig. \ref{Fig:HOSVD}).

We say that the core tensor is all-orthogonal if it satisfies the following conditions:

(1) All orthogonality: Slices in each mode are mutually orthogonal, e.g., for a 3rd-order tensor
\be
\langle\bS_{:,k,:} \bS_{:,l,:}\rangle=0, \quad \mbox{for} \quad k \neq l,
\ee

(2) Pseudo-diagonality: Frobenius norms of slices in each mode are decreasing with the increase  of
the running index
\be
||\bS_{:,k,:}||_F \geq  ||\bS_{:,l,:}||_F, \quad k \geq l.
\ee
These norms play a role similar to that of the singular values in the matrix SVD.

The orthogonal matrices $\bU^{(n)}$ can be in practice computed by the standard SVD or truncated SVD of unfolded mode-$n$ matrices $\bX_{(n)} = \bU^{(n)} \mbi \Sigma_n \bV^{(n) T}\in \Real^{I_n \times I_1 \cdots I_{n-1} I_{n+1} \cdots I_N}$.
After obtaining  the  orthogonal matrices $\bU^{(n)}$  of left singular vectors of $\bX_{(n)}$, for each $n$, we can compute the core tensor $\underline \bG =\underline \bS$ as
\be
\underline \bS = \underline \bX  \times_1 \bU^{(1)\;T} \times_2 \bU^{(2)\;T} \cdots  \times_N \bU^{(N)\;T},
\label{HOSVD:core}
\ee
such that
\be
\underline \bX = \underline \bS  \times_1 \bU^{(1)}  \times_2 \bU^{(2)}   \cdots  \times_N \bU^{(N)}.
\label{HOSVD}
\ee

Due to orthogonality of  the core tensor $\underline \bS$ its slices are mutually orthogonal,
this reduces to the diagonality in the matrix case.

In some applications we may use a modified HOSVD in which the SVD can be performed not on the unfolding mode-$n$ matrices $\bX_{(n)}$ but on their transposes, i.e., $\bX_{(n)}^T \cong \bV^{(n)} \mbi \Sigma_n \bU^{(n) T}$. This leads to the modified HOSVD corresponding to Grassmann manifolds \cite{lui2010action}, that requires the computation of  very large (tall-and-skinny) factor orthogonal matrices $\bV^{(n)} \in \Real^{ I_{\bar n} \times I_n}$,  where $I_{\bar n} = \prod_{k\neq n} I_k=I_1 \cdots I_{n-1} I_{n+1} \cdots I_N$,  and the  core tensor $\underline {\tilde \bS}
\in \Real^{I_{\bar 1} \times  I_{\bar 2} \times \cdots \times  I_{\bar N}}$ based on the following model:
\be
\underline \bX = \underline {\tilde \bS}  \times_1 \bV^{(1)\;T}  \times_2 \bV^{(2)\;T}   \cdots  \times_N \bV^{(N) \; T}.
\label{MHOSVD}
\ee

%\subsection{\bf Tucker Decompositions and Multilinear SVD for Large-Scale Problems}

% {\bf Multilinear SVD for large-scale problems.}

 \begin{figure}[t!]
%\centering
%\psfrag{y}{\color{black}$y$}
(a) Sequential computation\\
\includegraphics[width=8.8cm]{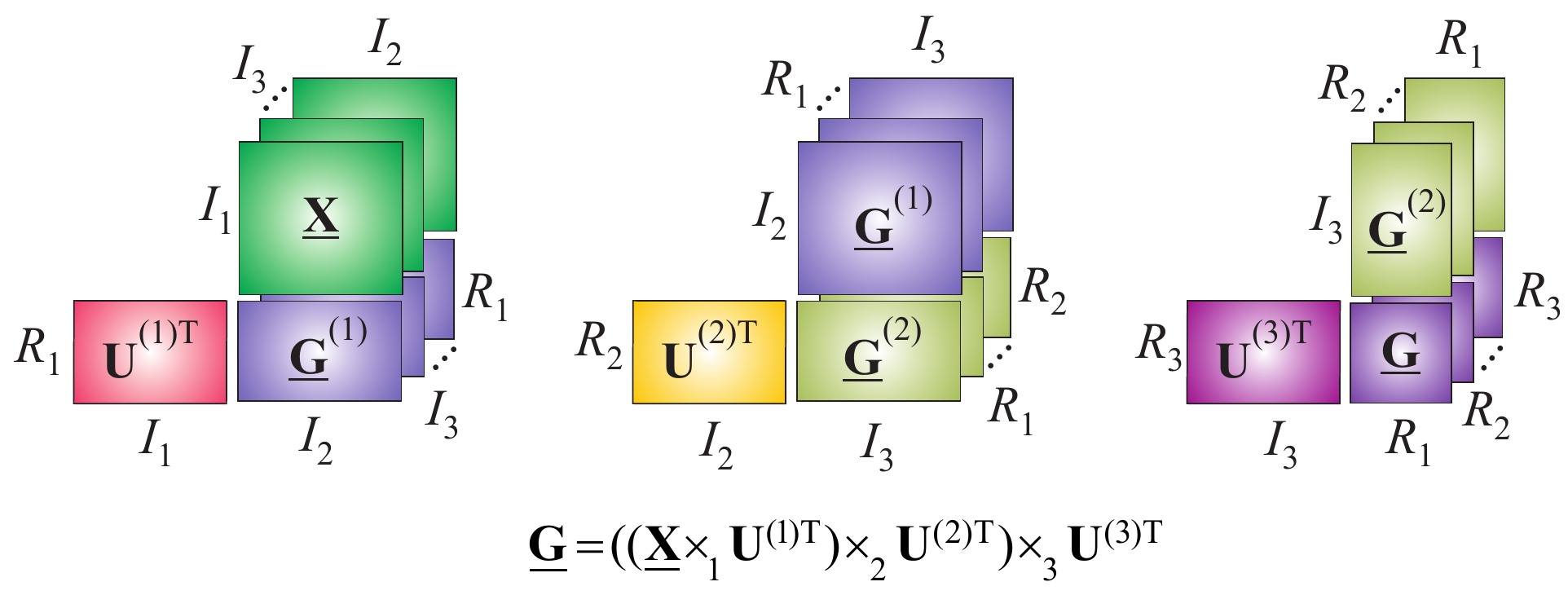} \\
(b) Distributed computation \\
\includegraphics[width=8.6cm]{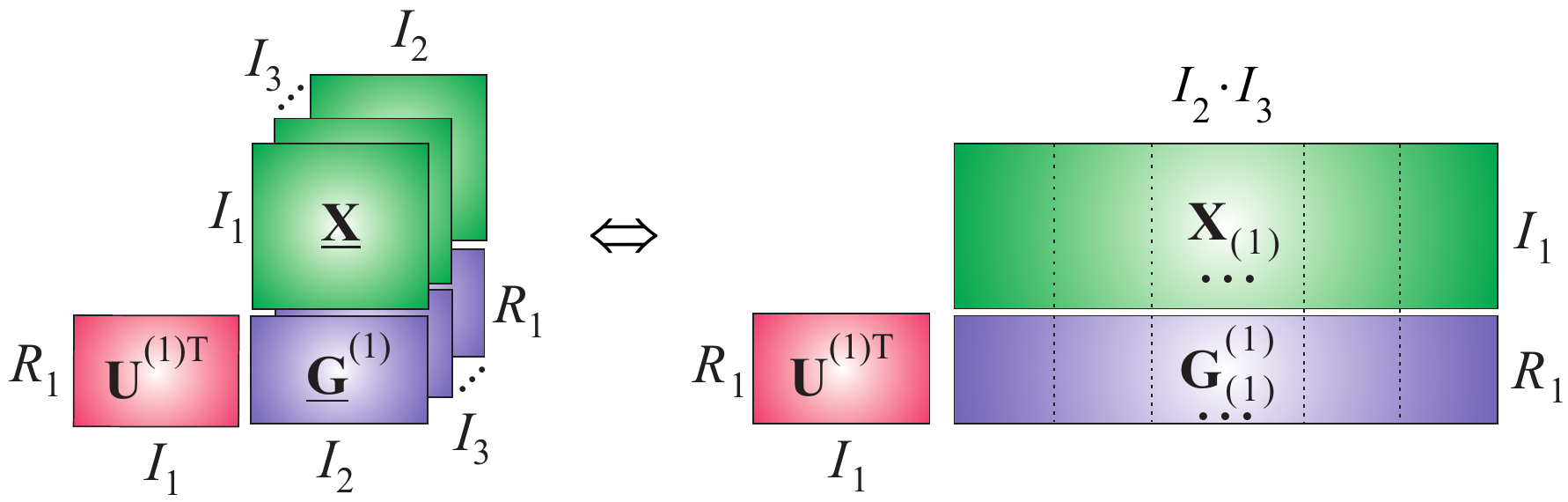}\\ \\
(c) Divide-and-conquer approach \\
\includegraphics[width=8.8cm]{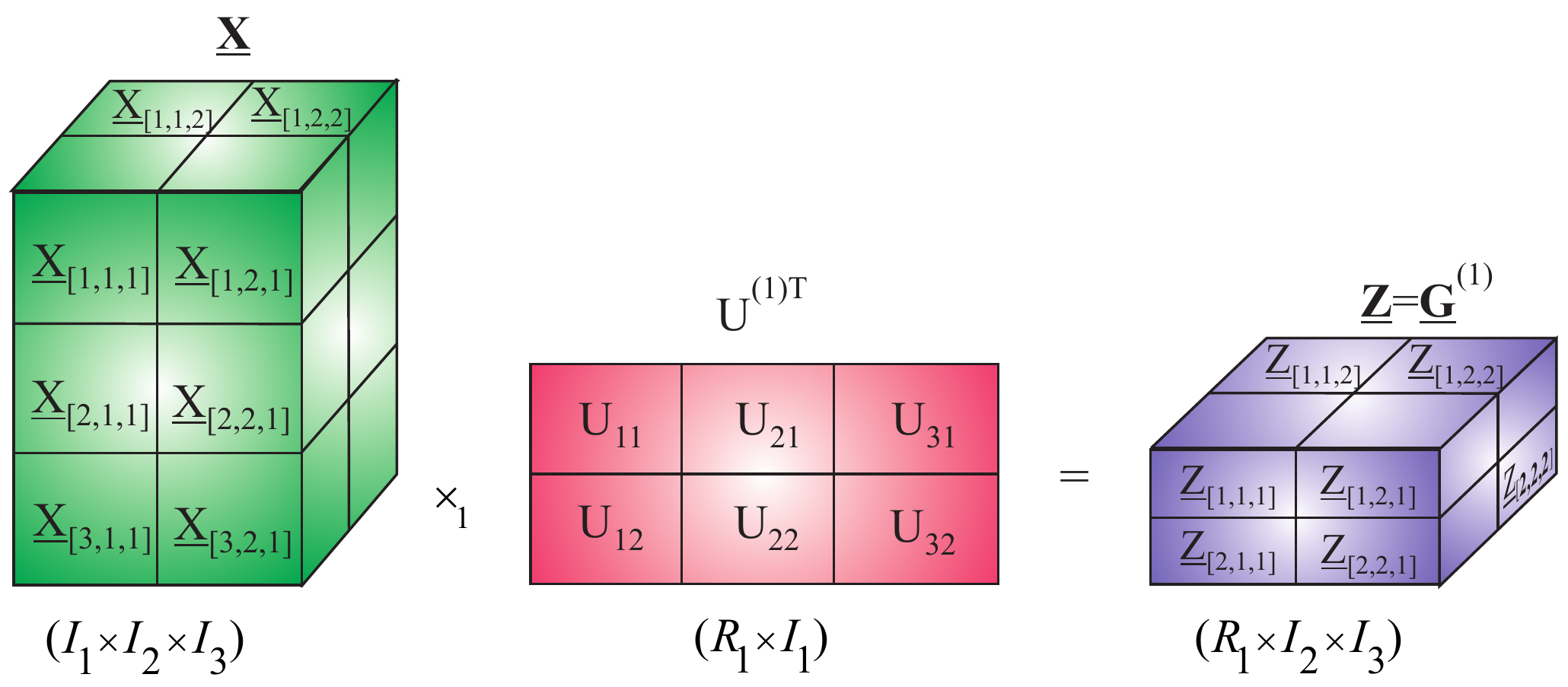}\\
%(c)
%\includegraphics[width=8.6cm,height=4.6cm]{Grid-Tensor1.eps}
%\includegraphics[width=7.2cm]{6order33.eps}\\
%\includegraphics[width=7.8cm]{4ordten.eps}
\caption{Computation of a core tensor for a large-scale  HOSVD:  (a) using sequential computing of multilinear products $\underline \bG = \underline \bS= (((\underline \bX \times_1 \bU^{(1)T}) \times_2  \bU^{(2)T}) \times_3 \bU^{(3)T})$, and (b) by applying fast and distributed implementation of matrix by matrix multiplications; (c) alternative method for very large-scale problems by applying divide and conquer approach, in which a data tensor $\underline \bX$ and  factor matrices $\bU^{(n)T}$  are partitioned into suitable small blocks: Subtensors $\underline \bX_{[k_1,k_2,k_3]} $ and blocks matrices $\bU^{(1)T}_{[k_1,p_1]}$, respectively. We compute the blocks of tensor $\underline \bZ=\underline \bG^{(1)}=\underline \bX \times_1 \bU^{(1)T}$ as follows $\underline \bZ_{[q_1,k_2,k_3]}= \sum_{k_1=1}^{K_1} \bX_{[k_1,k_2,k_3]} \times_1 \bU^{(1)T}_{[k_1,q_1]}$ (see Eq. (\ref{outcore-prod}) for a general case.)}
% as follows $\underline \bZ_{[k_1,k_2,k_3]}= \sum_{k_1=1}^{K_1} \bX_{[k_1,k_2,k_3]} \times_1 \bU^{(1)T}_{[k_1,p_1]} $ for $k_1=1,2,3; \; k_2=1,2; \; k_3=1,2$ and $p_1=1,2$. }
\label{Fig:outcore}
\end{figure}

%\begin{figure}[ht]
%\begin{center}
%%\includegraphics[width=4.8cm]{HOSVDCRT.eps}\\
%%\includegraphics[width=8.6cm]{SVD-blocks.eps}
%\includegraphics[width=8.6cm]{Fig19.eps}
%\end{center}
%\caption{Computation of the SVD using block sub-matrices $\bX_C \in \Real^{I \times P}$ and $\bX_R \in \Real^{P \times J}$ for huge low-rank matrix $\bX \in \Real^{I \times J}$ with the rank $R \ll \{I,J\}$. We select such rows and columns of the matrix $\bX$ that the sub-matrix $\bW \in \Real^{P \times P}$ with $P \geq R$ has the full rank $R$.}
%\label{Fig:HOSVD-subtensors}
%\end{figure}

\begin{figure}[ht]
%\centering
%\psfrag{y}{\color{black}$y$}
%(a)
\begin{center}
\includegraphics[width=8.8cm,height=5.0cm]{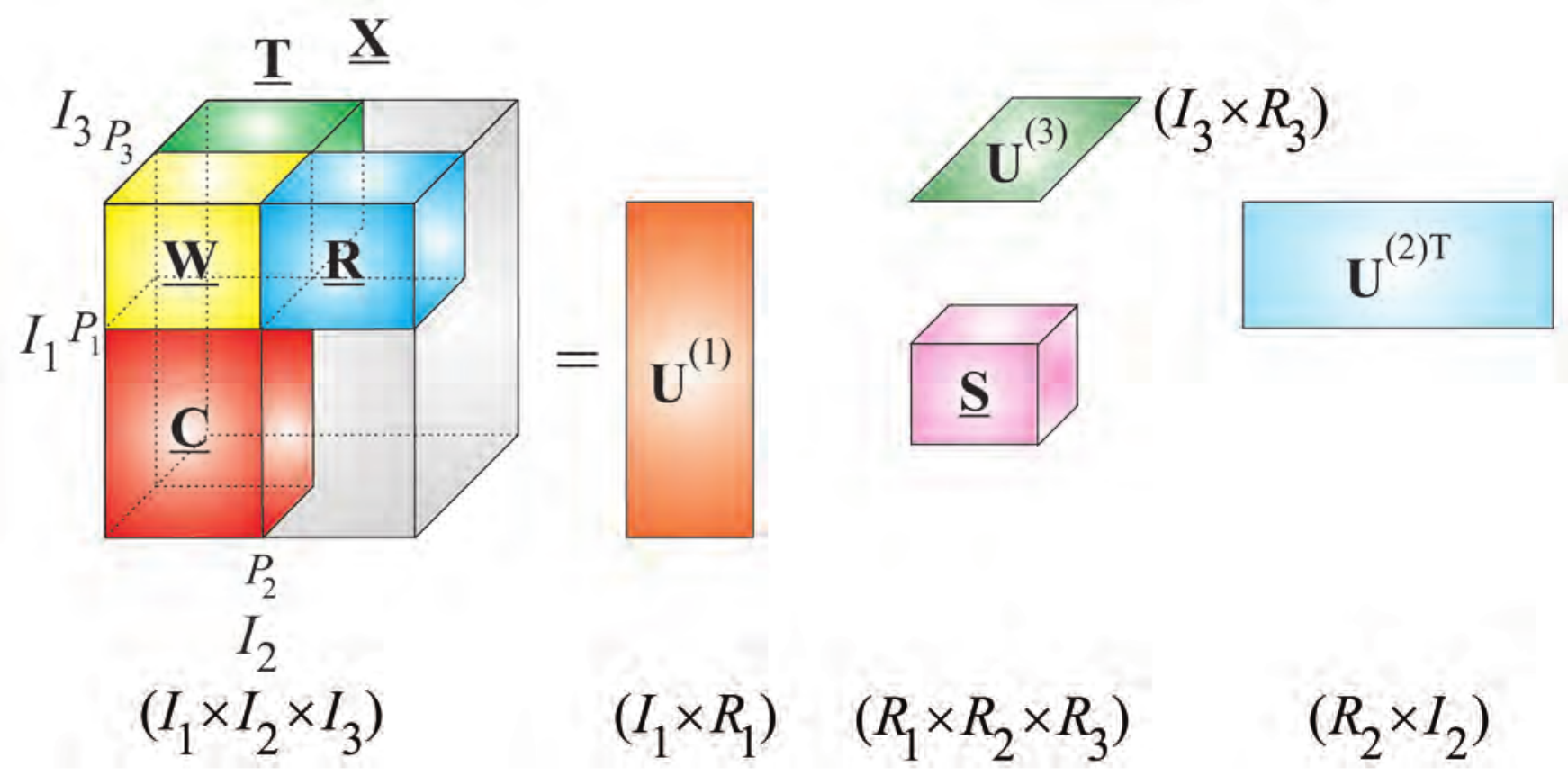}\\
\end{center}
%(b) \\
%\includegraphics[width=8.99cm,height=13.6cm]{HOSVD34.eps}
%\includegraphics[width=8.99cm,height=13.6cm]{Fig20b.eps}
%(c)
%\includegraphics[width=8.6cm,height=4.6cm]{Grid-Tensor1.eps}
%\includegraphics[width=7.2cm]{6order33.eps}\\
%\includegraphics[width=7.8cm]{4ordten.eps}
\caption{Alternative approach to computation of the  HOSVD for very large data tensor, by exploiting multilinear low-rank approximation.  The objective is to select such fibers (up to permutation of fibers) that the subtensor $\underline \bW \in \Real^{P_1 \times P_2 \times P_3}$ with $P_n  \geq R_n$ ($n=1,2,3$) has the same multilinear rank  $\{R_1,R_2,R_3\}$ as the whole huge data tensor $\underline \bX$, with $R_n \ll I_n$. Instead of unfolding of the whole data tensor $\underline \bX$ we need to perform unfolding (and applying the standard SVD) for typically much smaller subtensors $\underline  \bX^{(1)}=\underline \bC \in \Real^{I_1 \times P_2 \times P_3}, \; \underline  \bX^{(2)} = \underline \bR \in \Real^{P_1 \times I_2 \times P_3}, \; \underline  \bX^{(3)} = \underline \bT \in \Real^{P_1 \times P_2 \times I_3}$, each in a single mode-$n$, ($n=1,2,3$). This approach can be applied if data tensor admits low multilinear rank approximation. For simplicity of illustration, we assumed that fibers are permuted in a such way that the first $P-1, P_2,P_3$ fibers  were selected.}
\label{Fig:HOSVD-subtensors}
\end{figure}

 In  practical applications the dimensions of unfolding matrices $\bX_{(n)} \in \Real^{I_n \times I_{\bar n}}$ may be prohibitively large (with $I_{\bar n} \gg I_n$), easily exceeding memory of  standard computers.
  A truncated SVD of a large-scale unfolding matrix $\mathbf{X}_{(n)}=\bU^{(n)} \mathbf{\Sigma}_n
  \bV^{(n) T}$ is performed by  partitioning it into $Q$
slices, as $\mathbf{X}_{(n)} = [
  \mathbf{X}_{1,n}, \mathbf{X}_{2,n}, \ldots , \mathbf{X}_{Q,n}]=\bU^{(n)} \mathbf{\Sigma}_n[
  \mathbf{V}_{1,n}^T , \mathbf{V}_{2,n}^T , \ldots, \mathbf{V}_{Q,n}^T
]$. Next, the orthogonal matrices $\bU^{(n)}$ and the diagonal matrices $\mathbf{\Sigma}_n$ are obtained from eigenvalue decompositions $\mathbf{X}_{(n)}\mathbf{X}_{(n)}^T=
\bU^{(n)} \mathbf{\Sigma}_{n}^2 \bU^{(n) T} =\sum_q\mathbf{X}_{q,n}\mathbf{X}_{q,n}^T \in \Real^{I_n\times I_n}$, allowing for the terms $\mathbf{V}_{q,n}=\mathbf{X}_{q,n}^T \bU^{(n)} \mathbf{\Sigma}_n^{-1}$ to be computed separately. This allows us to optimize the  size of the $q$-th slice $\bX_{q,n} \in \Real^{I_n \times (I_{\bar n}/Q)}$ so as  to match the available computer memory.                                                                                              Such a simple approach to compute matrices $\bU^{(n)}$ and/or $\bV^{(n)}$ does not require loading the entire unfolding matrices  at once into  computer memory, instead the  access to the dataset is sequential. Depending on the size of computer  memory,  the dimension $I_n$ is typically less than 10,000, while  there is no limit on the dimension $I_{\bar n} = \prod_{k\neq n} I_k$. % \cite{byang,Zipunnikov-2011}.
More sophisticated approaches which also exploit partition of matrices or tensors into blocks for  QR/SVD, PCA, NMF/NTF  and ICA   can be found in                                      \cite{TaSNMF14,TaSQR13,Vasilescu,Phan-CP,Zhou-Cichocki-SP}.

When a data tensor $\underline \bX$ is very large and cannot be stored in  computer memory, then
another challenge  is to compute a core tensor $\underline \bG =\underline \bS$ by  directly using the formula:
\be
\bG = \underline \bX \times_1 \bU^{(1)T} \times_2  \bU^{(2)T} \cdots \times_n \bU^{(n)T},
\ee
which is generally performed sequentially as illustrated in Fig. \ref{Fig:outcore} (a)  and (b) \cite{Suter13,Wang-out-core05}.

 For very large tensors it is useful to divide the data tensor $\underline \bX$ into small blocks $\underline \bX_{[k_1,k_2,\ldots, k_N]} $  and in order to store them on hard disks or distributed memory. In similar way, we can divide the orthogonal factor matrices $\bU^{(n)T}$ into corresponding blocks of matrices $\bU^{(n)T}_{[k_n,p_n]}$ as illustrated in Fig. \ref{Fig:outcore} (c) for  3rd-order tensors \cite{Wang-out-core05}. In a general case, we can compute blocks within the resulting tensor $\bG^{(n)}$  sequentially or in parallel way as follows:
\be
\underline \bG^{(n)}_{[k_1,k_2,\ldots,q_n,\ldots,k_N]}= \sum_{k_n=1}^{K_n} \bX_{[k_1,k_2,\ldots,k_n,\ldots,k_N]} \times_n \bU^{(n) \;T}_{[k_n,q_n]}.
\label{outcore-prod}
\ee
If  a data tensor has  low-multilinear rank,
so that its  multilinear rank $\{R_1,R_2,\ldots,R_N\}$ with $R_n \ll I_n,\;\; \forall n$,
we can further alleviate the problem of dimensionality  by  first identifying a subtensor $\underline \bW \in \Real^{P_1 \times P_2 \times \cdots P_N}$ for which $P_n \geq R_n$, using efficient CUR tensor decompositions  \cite{Caiafa-Cichocki-CUR}. Then the HOSVD can be computed from subtensors as illustrated in Fig. \ref{Fig:HOSVD-subtensors} for a 3rd-order tensor.
This feature can be formulated in more general form as the following Proposition.

{\bf Proposition 1}: If a tensor $\underline \bX \in \Real^{I_1 \times I_2 \times \cdots \times I_N}$ has low multilinear rank $\{R_1,R_2,\ldots,R_N\}$, with $R_n \leq I_n, \;\; \forall n$, then it can be fully reconstructed via the HOSVD using only $N$ subtensors $\underline \bX^{(n)} \in \Real^{P_1 \times \cdots \times P_{n-1} \times  I_n \times P_{n+1} \times \cdots \times P_N}, \;\ (n=1,2,\ldots, N)$, under the condition that subtensor $\underline \bW \in \Real^{P_1 \times P_2  \times \cdots \times P_N}$, with $P_n \geq R_n,\;\; \forall n$  has the multilinear rank $\{R_1,R_2,\ldots,R_N\}$.

In practice, we can compute the HOSVD for low-rank, large-scale data tensors in several steps.  In the first step, we can apply the CUR FSTD decomposition \cite{Caiafa-Cichocki-CUR} to identify  close to optimal a subtensor $\underline \bW \in \Real^{R_1 \times R_2 \times \cdots \times R_N}$  (see the next Section),
In the next step, we can use the standard SVD for unfolding matrices $\bX^{(n)}_{(n)}$  of subtensors $\underline \bX^{(n)}$ to compute the left orthogonal matrices $\widetilde \bU^{(n)} \in \Real^{I_n \times R_n}$. Hence,
we compute an auxiliary core tensor $\underline \bG = \underline \bW \times_1 \bB^{(1)} \cdots \times_N \bB^{(N)}$, where $\bB^{(n)} \in \Real^{R_n \times R_n}$ are inverses of the sub-matrices consisting the first $R_n$ rows of the matrices $\widetilde \bU^{(n)}$. In the last step, we perform HOSVD decomposition of
the relatively small core tensor as  $\underline \bG = \underline \bS \times_1 \bQ^{(1)} \cdots \times_N \bQ^{(N)}$, with $\bQ^{(n)} \in \Real^{R_n \times R_n}$ and then desired orthogonal matrices are computed as $\bU^{(n)} = \widetilde \bU^{(n)} \bQ^{(n)}$.

%\subsection{Fiber Sampling Tucker Decomposition (FSTD) for Dimensionality Reduction and Compression of Large-Scale Tensor Data}

\section{\bf CUR Tucker Decomposition for Dimensionality Reduction and Compression of Tensor Data}

Note that instead of using the full tensor, we may  compute an approximative tensor decomposition model from a limited number of entries (e.g., selected fibers, slices or subtensors). Such completion-type strategies have been developed for low-rank and low-multilinear-rank \cite{Acar-Morup11,Gandy2011}.
A simple approach  would be  to apply CUR decomposition or Cross-Approximation by  sampled fibers for the columns of factor matrices  in a Tucker approximation  \cite{mahoney2008tensor,Caiafa-Cichocki-CUR}.
%Furthermore, since large-scale tensors cannot be loaded explicitly in main memory, they  usually  reside in distributed storage by splitting tensors to smaller blocks.
Another approach is to apply tensor networks to represent big data by high-order tensors  not explicitly but in compressed tensor formats (see next sections).
Dimensionality reduction methods are based on the fundamental assumption that  large datasets are highly redundant and can be approximated by low-rank matrices and cores, allowing for a significant  reduction in computational complexity and to discover meaningful components while exhibiting marginal loss of information.

For very large-scale matrices, the so called CUR matrix decompositions  can be employed for  dimensionality reduction \cite{Goreinov:1997,Goreinov:1997b,mahoney2008tensor,Caiafa-Cichocki-CUR,oseledets2010tt}. Assuming a  sufficiently
precise low-rank approximation, which implies that data has some internal structure or smoothness, the idea is to provide data representation through a linear combination of a few ``meaningful'' components, which are exact replicas of columns and rows of the original data matrix \cite{Mahoney:2009}.

\begin{figure}[t]
\centering
\includegraphics[width=8.6 cm]{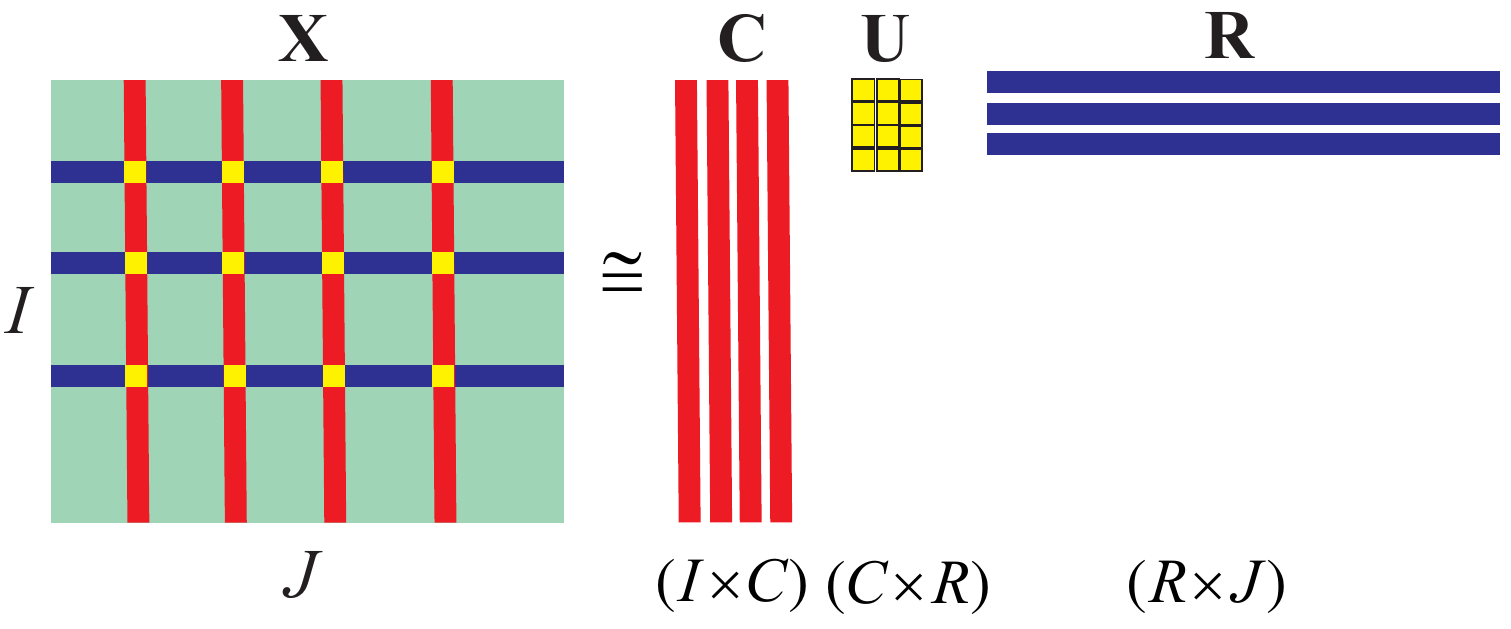}
\caption{CUR decomposition for a huge matrix.}
\label{Fig:CUR}
\end{figure}

\begin{figure}[p!]
%\centering
(a)\\
\includegraphics[width=8.6 cm]{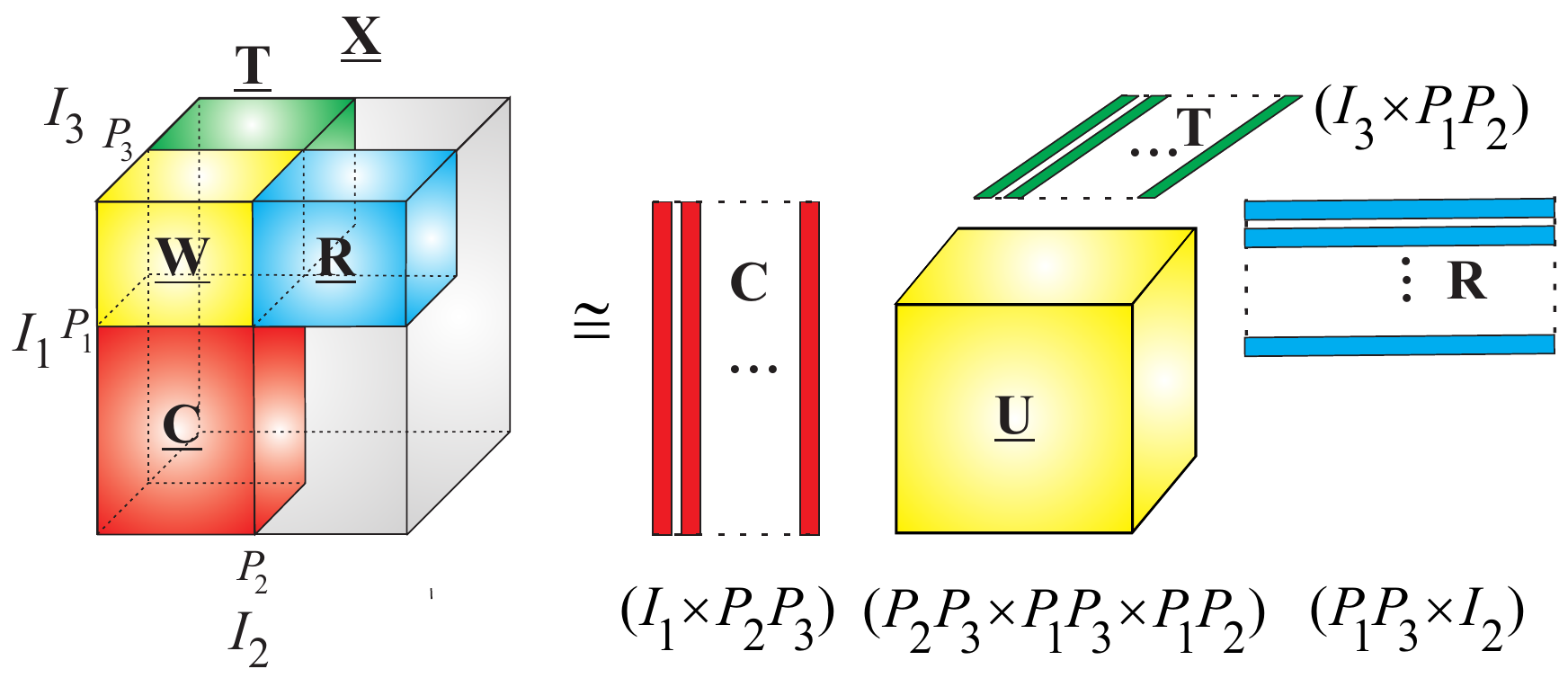}\\ \\
%\includegraphics[width=8.6 cm]{FigS21a.eps}\\ \\
%\vspace{0.1cm}
(b)\\
\includegraphics[width=8.6 cm]{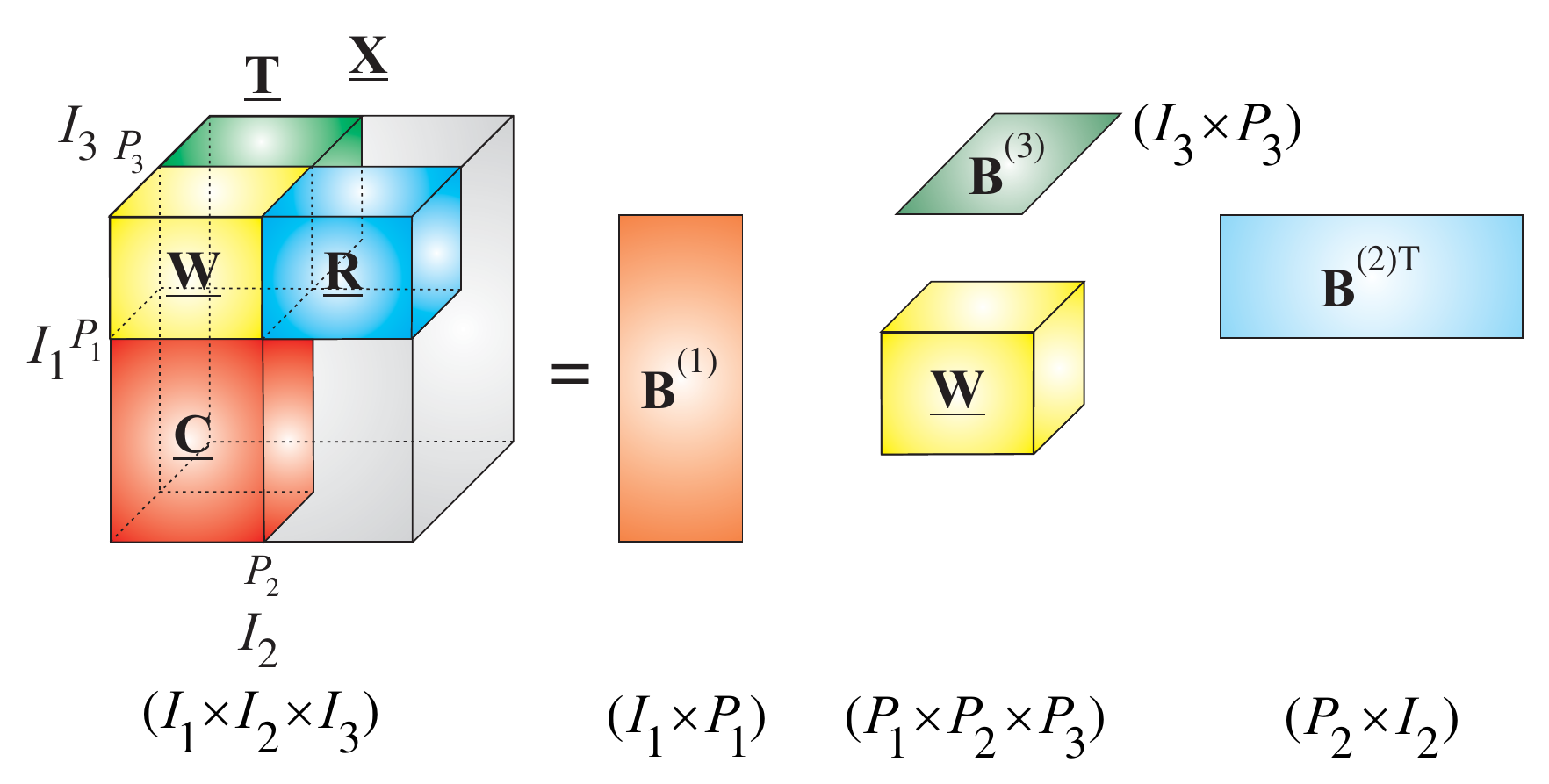}\\ \\
(c)\\
\includegraphics[width=8.1 cm]{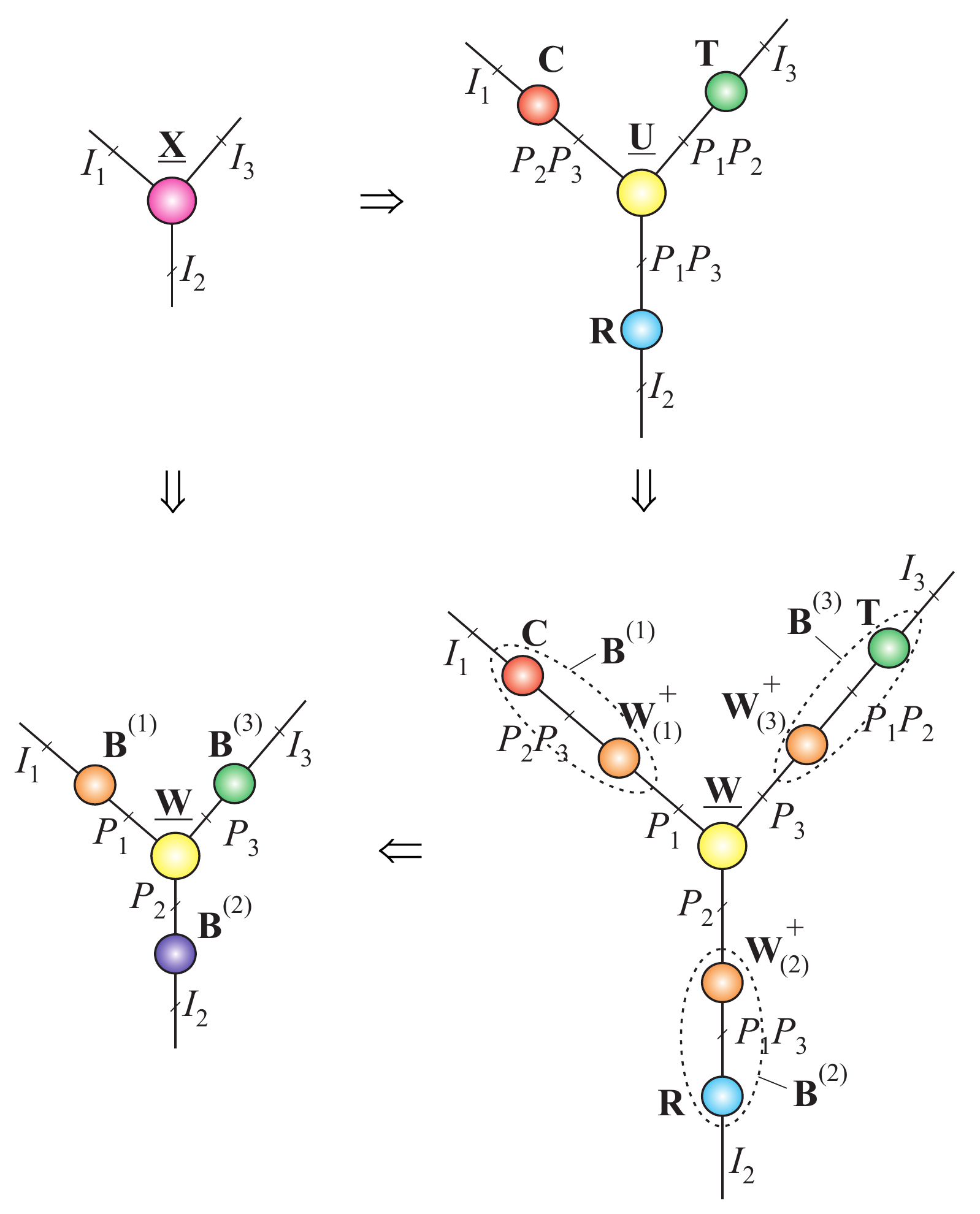}
\caption{(a) CUR decomposition of a large  3rd-order tensor (for simplicity of illustration up to permutation of fibers) $\underline \bX \cong \underline \bU \times_1 \bC \times_2 \bR \times_3 \bT
=\llbracket \underline \bU; \bC, \bR, \bT \rrbracket$, where $\underline \bU = \underline \bW \times_1\mathbf{W}_{(1)}^{\dagger} \times_2 \mathbf{W}_{(2)}^{\dagger} \times_3 \mathbf{W}_{(3)}^{\dagger} =\llbracket \underline \bW;\mathbf{W}_{(1)}^{\dagger}, \mathbf{W}_{(2)}^{\dagger},  \mathbf{W}_{(3)}^{\dagger}\rrbracket$. (b) equivalent decomposition expressed via subtensor $\underline \bW$, (c) Tensor network diagram illustrating transformation from CUR Tucker format (a) to form (b) as: $\underline \bX \cong \underline \bW \times_1 \bB^{(1)} \times_2 \bB^{(2)} \times_3 \bB^{(3)}
=\llbracket \underline \bW; \bC \bW_{(1)}^{\dagger}, \bR \bW_{(2)}^{\dagger}, \bT  \bW_{(3)}^{\dagger}\rrbracket$.}
\label{Fig:CUR-tensor}
\end{figure}

The  CUR model, also called  skeleton Cross-Approximation, decomposes a data matrix $\bX \in \Real^{ I \times J}$ as \cite{Goreinov:1997,Goreinov:1997b} (see Fig. \ref{Fig:CUR}):
\be
\bX =  \bC \bU \bR + \bE,
\ee
where $\bC \in \Real^{I \times C}$ is a matrix
constructed from $C$ suitably selected columns of the data matrix $\bX$, $\bR \in
\Real^{R \times J}$ consists of $R$ rows of $\bX$, and the matrix $\bU \in
\Real^{C \times R}$ is chosen to minimize the norm of the error $\bE \in \Real^{I
\times J}$.
Since typically, $C \ll J$ and $R \ll I$,
%the objective is to  select  rows
%and columns of $\bX$ such that  the error cost function $||\bE||_F^2$ is minimized;
these columns and rows are chosen so as to exhibit high ``statistical
leverage'' and  provide the best low-rank fit to the data matrix, at the same time the error cost function $||\bE||_F^2$ is minimized.
For a given set of columns ($\bC$) and rows ($\bR$), the optimal choice for the core matrix is $\bU= \bC^{\dagger} \bX (\bR^{\dagger})^T$. This  requires access to all the entries of  $\bX$ and is not practical or feasible for large-scale data. A pragmatic choice for the core matrix would be $\bU =\bW^{\dagger}$, where the matrix  $\bW \in \Real^{R \times C}$  is defined from the intersections of the selected rows and columns. It should be noted that, if  rank$(\bX)\le C, R$, then the CUR approximation is exact.  For the general case, it has been proven that, when the intersection sub-matrix $\bW$ is of maximum volume (the volume of a sub-matrix $\bW$ is defined as $|\det(\bW)|$), this approximation is close to the optimal SVD solution \cite{Goreinov:1997b}.
%On the other hand, by choosing the set of rows and columns  with the ``statistical leverage'' criterion, the obtained CUR approximation is close to the optimal one (SVD) with high probability \cite{mahoney2008tensor}.

%\cite{Drineas-CUR}.
% (see Fig.  \ref{Fig-CUR}).
%
The concept of  CUR decomposition has been successfully generalized to tensors. In \cite{mahoney2008tensor} the matrix CUR decomposition was applied to one unfolded version of the tensor data, while in \cite{Caiafa-Cichocki-CUR} a reconstruction formula of a tensor having a low rank Tucker approximation was proposed, termed the Fiber Sampling Tucker Decomposition (FSTD), which is  a practical and fast technique. The FSTD takes into account the linear structure in all the modes of the tensor simultaneously.
Since real-life data  often have good  low multilinear rank approximations, the FSTD provides such a low-rank Tucker decomposition that is directly expressed in terms of a relatively small number of fibers  of the data tensor (see Fig. \ref{Fig:CUR-tensor}).

%By virtue of its construction from raw data elements,  FSTD preserves the original characteristics and features  of tensor data and can be used for efficient compressed representations of the original data tensor.

For a given 3rd-order tensor $\underline \bX \in \Real^{I_1\times I_2 \times I_3}$ for which an exact rank-$(R_1,R_2,R_3)$ Tucker representation exists, FSTD selects $P_n \geq R_n$ ($n=1,2,3$) indices in each mode, which determine an intersection sub-tensor $\underline \bW \in \Real ^{P_1\times P_2 \times P_3}$ so that the following exact Tucker representation can be obtained:
\begin{equation}\label{NTensorCUR}
    \underline \bX = \llbracket \underline \bU; \bC, \bR, \bT \rrbracket,
 \end{equation}
in which the core tensor is computed as $\underline \bU=\underline \bG=\llbracket \underline \bW;\mathbf{W}_{(1)}^{\dagger}, \mathbf{W}_{(2)}^{\dagger},  \mathbf{W}_{(3)}^{\dagger}\rrbracket$, and the factor matrices $\bC \in \Real^{I_1 \times P_2 P_3}, \bR \in \Real^{I_2 \times P_1 P_3}, \bT \in \Real^{I_3 \times P_1 P_2}$ contain the  fibers (columns, rows and tubes, respectively). This can also be written as a Tucker representation:
\begin{equation}\label{remark1}
 \underline \bX=\llbracket \underline \bW; \mathbf{C}\mathbf{W}_{(1)}^{\dagger},\mathbf{R}
 \mathbf{W}_{(2)}^{\dagger},\mathbf{T}\mathbf{W}_{(3)}^{\dagger} \rrbracket.
\end{equation}
Observe that for $N=2$ this model simplifies into the CUR matrix case, $\bX=\bC \bU \bR$,
% where $\mathbf{C}^{(1)}=\mathbf{C}$, $\mathbf{R}^{(2)}=\mathbf{R}^T$
and the core matrix is $\mathbf{U}=\llbracket \mathbf{W}; \mathbf{W}_{(1)}^{\dagger}, \mathbf{W}_{(2)}^{\dagger}\rrbracket=\mathbf{W}^{\dagger}\mathbf{W}\mathbf{W}^{\dagger}=\mathbf{W}^{\dagger}$.

In a more general case for an $N$th-order tensor, we can formulate the following Proposition \cite{Caiafa-Cichocki-CUR}.

{\bf Proposition 2}: If tensor $\underline \bX \in \Real^{I_1 \times I_2 \times \cdots \times I_N}$ has low multilinear rank $\{R_1,R_2,\ldots,R_N\}$, with $R_n \leq I_n, \;\; \forall n$, then it can be fully reconstructed via the CUR FSTD $\underline \bX = \llbracket \underline \bU; \bC^{(1)},\bC^{(2)},\ldots,\bC^{(N)}\rrbracket$, using only $N$  factor  matrices $\bC^{(n)} \in \Real^{I_n \times P_n}, \;\ (n=1,2,\ldots, N)$, built up from fibers of the data tensor, and a core tensor $\underline \bU=\underline \bG=\llbracket \underline \bW; \mathbf{W}_{(1)}^{\dagger}, \mathbf{W}_{(2)}^{\dagger}, \ldots, \mathbf{W}_{(N)}^{\dagger}\rrbracket
$, under the condition that subtensor $\underline \bW \in \Real^{P_1 \times P_2  \times \cdots \times P_N}$ with $P_n \geq R_n, \;\; \forall n$  has multilinear rank $\{R_1,R_2,\ldots,R_N\}$).
%For a given tensor $\underline \bX \in{\Real^{I_1\times I_2 \times \cdots \times I_N}}$ for which an exact rank-$(R_1,R_2,...,R_N)$ Tucker representation exists, FSTD selects $R$ indices in each mode, which determine an intersection sub-tensor $\underline \bW \in{\Real ^{R \times R \times \cdots \times R}}$ so that the following exact Tucker representation is obtained:
%\begin{equation}\label{NTensorCUR}
%    \underline \bX = \llbracket \underline \bU; \mathbf{C}^{(1)}, \mathbf{C}^{(2)},\ldots, \mathbf{C}^{(N)}\rrbracket,
% \end{equation}
%in which the core tensor is computed as $\underline \bU=\underline \bG=\llbracket \underline \bW;\mathbf{W}_{(1)}^{\dagger}, \mathbf{W}_{(2)}^{\dagger}, \ldots, \mathbf{W}_{(n)}^{\dagger}\rrbracket$, and the matrices $\mathbf{C}^{(n)}\in{\Real^{I_n \times I_{\bar n}}}$ contain the mode-$n$ fibers defined by restricting the indices in modes $m\neq n$ to belong to the selected subsets. This can also be written as a rank-$(R,R,...,R)$ Tucker representation:
%\begin{equation}\label{remark1}
% \underline \bX=\llbracket \underline \bW; \mathbf{C}^{(1)}\mathbf{W}_{(1)}^{\dagger},\mathbf{C}^{(2)}
% \mathbf{W}_{(2)}^{\dagger},...,\mathbf{C}^{(N)}\mathbf{W}_{(N)}^{\dagger} \rrbracket.
%\end{equation}
%

An efficient strategy for the selection of suitable
fibers, only requiring access to a partial (small) subset of entries of a data tensor through identifying the entries with maximum modulus within single fibers is given in \cite{Caiafa-Cichocki-CUR}. The indices are selected sequentially using a deflation approach making the FSTD algorithm  suitable for very large-scale but relatively low-order tensors (including tensors with missing fibers or entries).

\section{\bf Analysis of Coupled Multi-Block Tensor Data --  Linked Multiway Component Analysis (LMWCA)}

%\subsection{\bf Linked Multiway Component Analysis (LMWCA)}

Group analysis or multi-block data analysis  aims to identify links between  hidden components in  data making it possible
 to analyze the correlation, variability and consistency of  the components across  multi-block data sets. This equips us with enhanced flexibility: Some components do not necessarily need to be orthogonal or statistically independent, and can be instead sparse, smooth or non-negative  (e.g., for spectral components).
Additional constraints  can be used to reflect the
spatial distributions, spectral, or  temporal patterns \cite{Cichocki-SICE}.

Consider the analysis of multi-modal high-dimensional data collected  under the same or very similar conditions, for example,  a set of EEG and MEG or fMRI  signals recorded for different subjects over many trials and under the same experiment configuration and mental tasks. Such data  share some common latent (hidden) components but  can also have their own  independent features.
Therefore, it is quite important and necessary that they will be analyzed in a linked way instead of independently.

\begin{figure}[ht!]
 \centering
%(a) \\
% \includegraphics[width=8.6cm,height=8.5cm]{MLCAM.eps}
% \includegraphics[width=8.6cm,height=8.5cm]{Fig23.eps}
 \includegraphics[width=8.6cm,height=8.5cm]{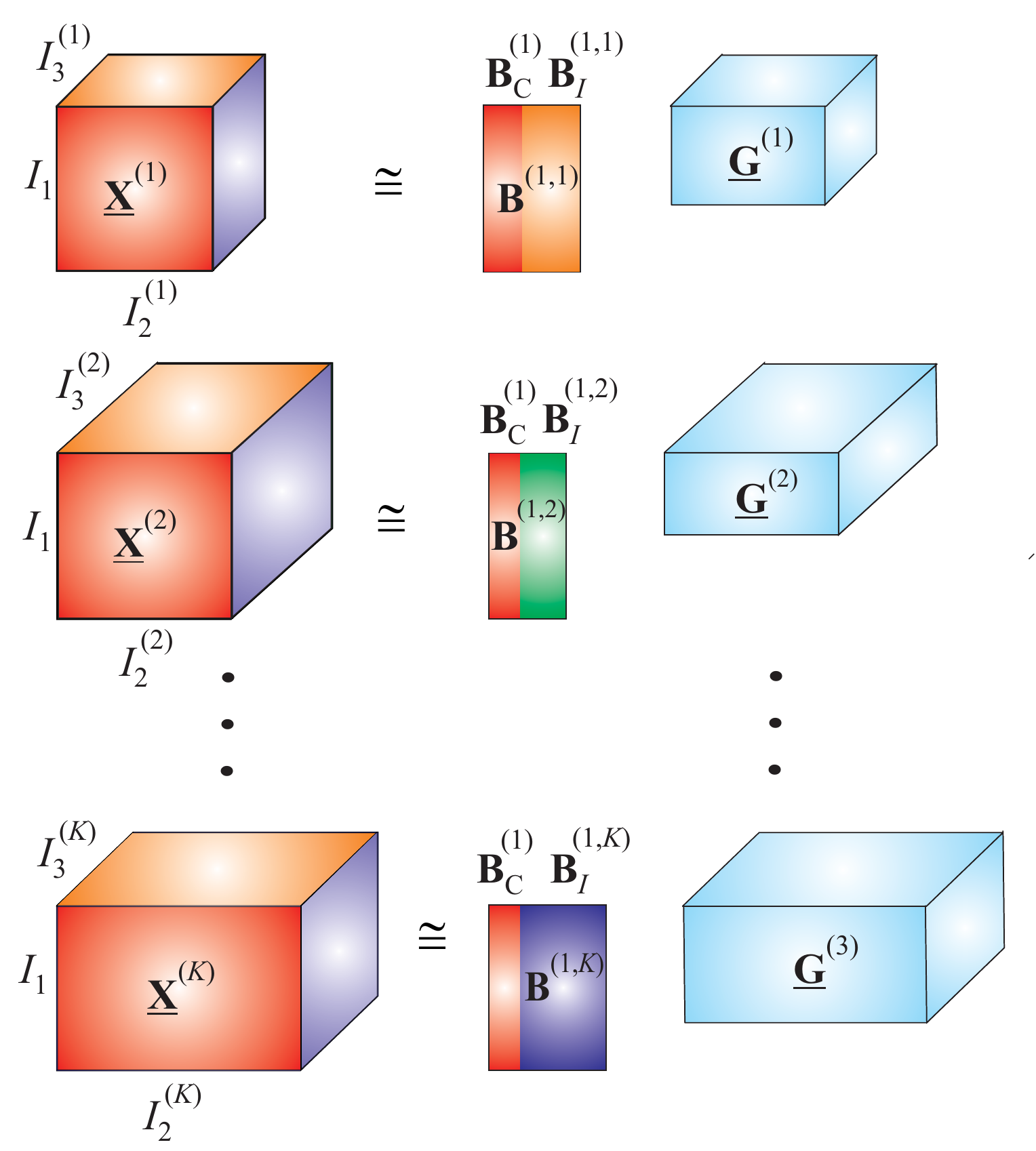}
%%\vspace{0.1cm}
%%\includegraphics[width=7.4 cm,height=8.8cm]{SPM-LMCA1T.eps}\\
%(b) \\
 %\includegraphics[width=7.99cm,height=9.99cm]{LMCA-equal.eps}
%\includegraphics[width=8.6cm,height=11.5cm]{Figure1.eps}
%
 \caption{Linked Multiway Component Analysis (LMWCA) for
  coupled multi-block 3rd-order tensors, with different dimensions in each mode except of the first mode. The objective is to find the common components $\bB^{(1)}_C \in \Real^{I_1 \times C_1}$, where $C_1 \leq R_1$ is the number of the common components in mode-1.}
 \label{Fig:LMCA}
\end{figure}

The linked multiway component analysis (LMWCA) for multi-block tensors data
is formulated as a set of approximate joint Tucker-$(1,N)$ decompositions of a set of data tensors $ \underline \bX^{(k)} \in  \Real^{I_1^{(k)} \times I_2^{(k)} \times \cdots \times I_N^{(k)}}$, with $I_1^{(k)}=I_1,\;\; \forall k$ $\;(k=1,2,\ldots,K)$ (see Fig.  \ref{Fig:LMCA}):
\be
   \underline \bX^{(k)} = \underline \bG^{(k)}  \times_{1} \bB^{(1,k)}, \quad (k=1,2,\ldots K)
 \label{LMBSS1}
\ee
where each factor (component) matrix $\bB^{(1,k)}=[\bB^{(1)}_C, \; \bB^{(1,k)}_I]
\in \Real^{I_n \times R_n}$ has  two sets of components: (1) Components $\bB^{(1)}_C \in \Real^{I_1 \times C} $ (with $0 \leq C \leq R$),
  which are common for  all available blocks and correspond to  identical or maximally correlated  components, and (2) components $\bB^{(1,k)}_I \in \Real^{I_1 \times (R_1-C_1)}$, which are  different independent processes, for example, latent variables  independent of excitations or stimuli/tasks. The objective is to estimate the common components $\bB_C^{(1)}$ and independent (distinctive) components $\bB^{(1,k)}_I $
  (see Fig. \ref{Fig:LMCA}) \cite{Cichocki-SICE}.

%The LWCA for multi-block tensors data
%can be formulated in  slightly different forms, if all tensors $\underline \bX^{(k)}$ has the same dimensions in all modes, as a set of approximate joint decompositions of a set of data tensors $ \underline \bX^{(k)} \in  \Real^{I_1 \times I_2 \times \cdots \times I_N}$, $\;(k=1,2,\ldots,K)$ (see Fig.  \ref{Fig:LMCA} (b)):
%%
%\be
%   \underline \bX^{(k)} = \underline \bG^{(k)}  \times_{1} \bB^{(1,k)}  \times_{2} \bB^{(2,k)}   \cdots  \times_{N} \bB^{(N,k)}, % + {\tE}^{(k)}, \nonumber
% \label{LMBSS}
%\ee
%where each factor (component) matrix $\bB^{(n,k)}=[\bB^{(n)}_C, \; \bB^{(n,k)}_I]
%\in \Real^{I_n \times R_n}$ has  two sets of components: (1) components $\bB^{(n)}_C \in \Real^{I_n \times C_n} $ (with $0 \leq C_n \leq R_n$)
%  which are common for  all available blocks and correspond to  identical or maximally correlated  components, and (2) components $\bB^{(n,k)}_I \in \Real^{I_n \times (R_n-C_n)}$, which are  different independent processes.  The objective is to estimate the common components $\bB_C^{(n)}$, independent components $\bB^{(n,k)}_I $,  and the most important interactions between them (in different modes) via the core tensors $\underline \bG^{(k)}$ for $k=1,2,\ldots,K$ and $n=1,2,\ldots,N$ \cite{Cichocki-SICE}.%.

%
 If $\bB^{(n,k)}= \bB^{(n)}_C \in \Real^{I_n \times R_n}$ for a specific mode $n$ (in our case $n=1$), under additional assumption that tensors are of the same dimension. Then  the problem simplifies into generalized Common Component Analysis  or tensor Population Value Decomposition (PVD) \cite{Phan2010TF} and can be solved by concatenating all data tensors along one mode, and perform constrained Tucker or CP  tensor decompositions (see \cite{Phan2010TF}).
%
 %In a more general case, when $C_n<R_n$,  we can  unfold  each data tensor $\underline \bX^{(k)}$ in common mode, and  perform a set of linked and constrained matrix factorizations: $\bX_{(1)}^{(k)}\cong \bB^{(1)}_C\matn[1,k]{A}_C + \matn[1,k]{B}_I\matn[1,k]{A}_I$ through solving constrained  optimization problems:
% \begin{equation}
%   \label{LWCA}
%   \begin{split}
%   \min \; & \sum_{k=1}^{K}\|\bX_{(1)}^{(k)}-\bB^{(1)}_C\matn[1,k]{A}_C-\matn[1,k]{B}_I\matn[1,k]{A}_I\|_F \\
%    & + f_1(\bB^{(1)}_C), \;\;
%    s.t. \;\; \bB_C^{(1)\;T}  \matn[1,k]{B}_I=\mathbf{0} \;\; \forall k,
%   \end{split}
% \end{equation}
% where $f_1$ are the penalty terms which impose additional constraints on common components $\bB_C^{(1)}$, in order to extract as many as possible unique and desired components. In a special case of orthogonality  constraints, the problem can be
% %on $\matn{A}_C$, (\ref{LWCA}) can be
% transformed to a generalized eigenvalue problem and solved by the power method \cite{Zhou-PAMI}. The key point is to assume that common factor sub-matrices $\bB_C^{(1)}$ are present in all multiple data blocks  and hence reflect structurally complex  (hidden) latent and intrinsic links between them. In practice, the number of common components $C_1$  in each mode is unknown and should be estimated  (see \cite{Zhou-PAMI} for detail).
 %
\begin{figure}[t!]
% \centering
(%a) \\
% \includegraphics[width=8.6cm]{TNLCTT.eps}
%%%\vspace{0.1cm}
%%\includegraphics[width=7.4 cm,height=8.8cm]{SPM-LMCA1T.eps}\\
%(b) \\
% \includegraphics[width=8.2cm]{TNLCTree.eps}
%%\includegraphics[width=8.6cm,height=11.5cm]{Figure1.eps}
a) \\
 \includegraphics[width=8.6cm]{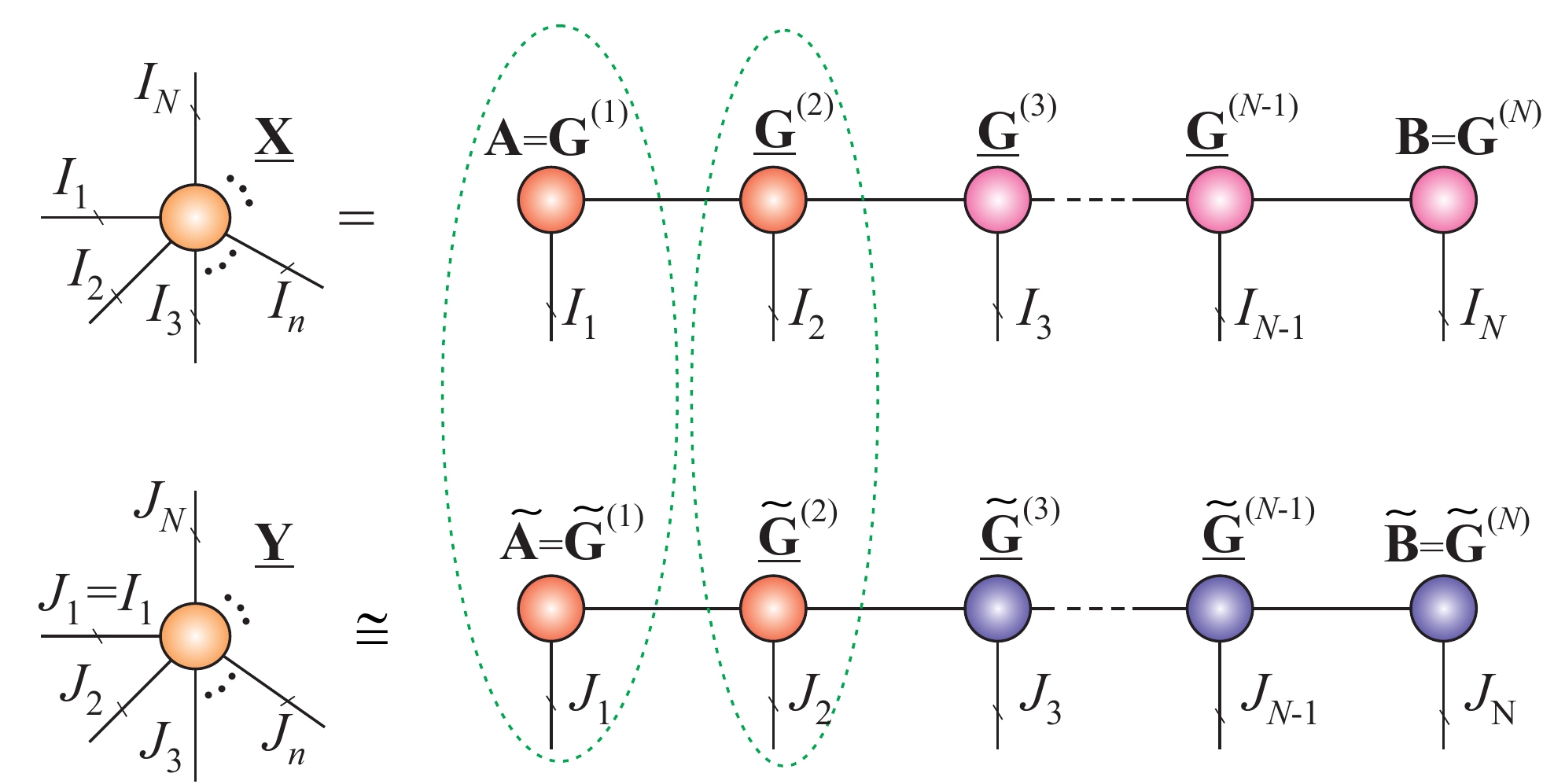}
%%\vspace{0.1cm}
%\includegraphics[width=7.4 cm,height=8.8cm]{SPM-LMCA1T.eps}\\
(b) \\
 \includegraphics[width=8.2cm]{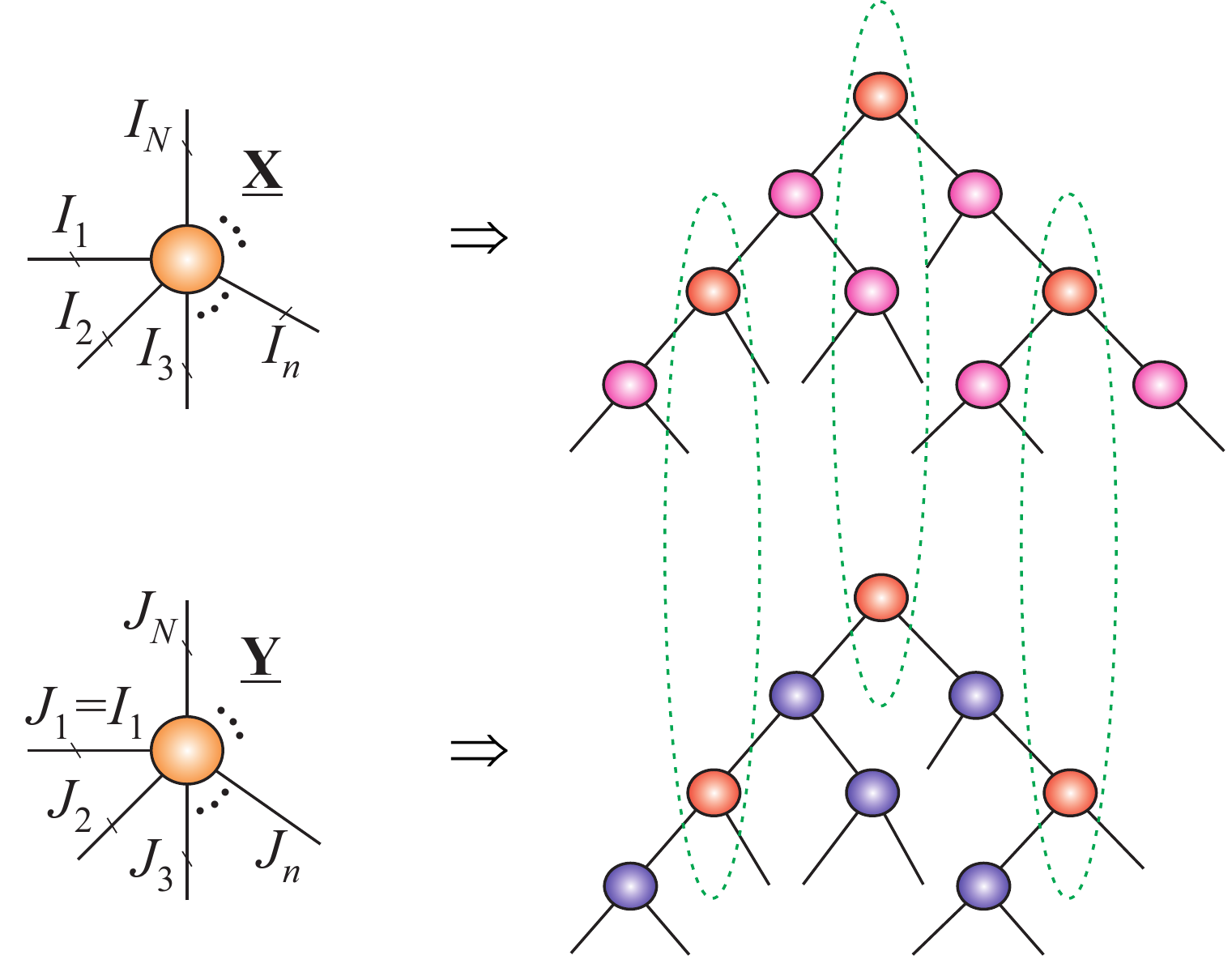}
 \caption{Conceptual models of generalized Linked  Multiway Component Analysis (LMWCA) applied to tensor networks: The objective is to find  core tensors which are maximally correlated for (a) Tenor Train and for (b) Tensor Tree States (Hierarchical Tucker).}
 \label{Fig:LMCA2}
\end{figure}

In a more general case, when $C_n<R_n$,  we can  unfold  each data tensor $\underline \bX^{(k)}$ in common mode, and  perform a set of linked and constrained matrix factorizations: $\bX_{(1)}^{(k)}\cong \bB^{(1)}_C\matn[1,k]{A}_C + \matn[1,k]{B}_I\matn[1,k]{A}_I$ through solving constrained  optimization problems:
 \begin{equation}
   \label{LWCA}
   \begin{split}
   \min \; & \sum_{k=1}^{K}\|\bX_{(1)}^{(k)}-\bB^{(1)}_C\matn[1,k]{A}_C-\matn[1,k]{B}_I\matn[1,k]{A}_I\|_F \\
    & + f_1(\bB^{(1)}_C), \;\;
    s.t. \;\; \bB_C^{(1)\;T}  \matn[1,k]{B}_I=\mathbf{0} \;\; \forall k,
   \end{split}
 \end{equation}
 where $f_1$ are the penalty terms which impose additional constraints on common components $\bB_C^{(1)}$, in order to extract as many as possible unique and desired components. In a special case, when we impose orthogonality  constraints, the problem can be
 %on $\matn{A}_C$, (\ref{LWCA}) can be
 transformed to a generalized eigenvalue problem and solved by the power method \cite{Zhou-PAMI}. The key point is to assume that common factor sub-matrices $\bB_C^{(1)}$ are present in all multiple data blocks  and hence reflect structurally complex  (hidden) latent and intrinsic links between them. In practice, the number of common components $C_1$  in each mode is unknown and should be estimated  (see \cite{Zhou-PAMI} for detail).

The linked multiway component analysis model
provides a  quite flexible and  general framework  and thus
supplements   currently available  techniques for group ICA
and  feature extraction for multi-block data.
The LWCA models are designed for blocks of $K$ tensors,
where dimensions naturally split into several  different modalities
(e.g., time, space and frequency). In this sense,
 a multi-block multiway CA  attempts to estimate both common  and independent or uncorrelated components, and is a natural extension of  group ICA, PVD, and  CCA/PLS methods
 (see \cite{Qibin-HOPLS,Cichocki-SICE,Zhou-PAMI,Yokota2012} and references therein).
 The concept of  LMWCA can be generalized to tensor networks as illustrated in Fig. \ref{Fig:LMCA2}.

\section{\bf Mathematical and Graphical Description of Tensor Trains (TT) Decompositions}

\begin{figure}[ht!]
\centering
%(a)\\
%\includegraphics[width=8.4cm{TTcolor1.eps}\\
%(b)\\
\includegraphics[width=4.6cm]{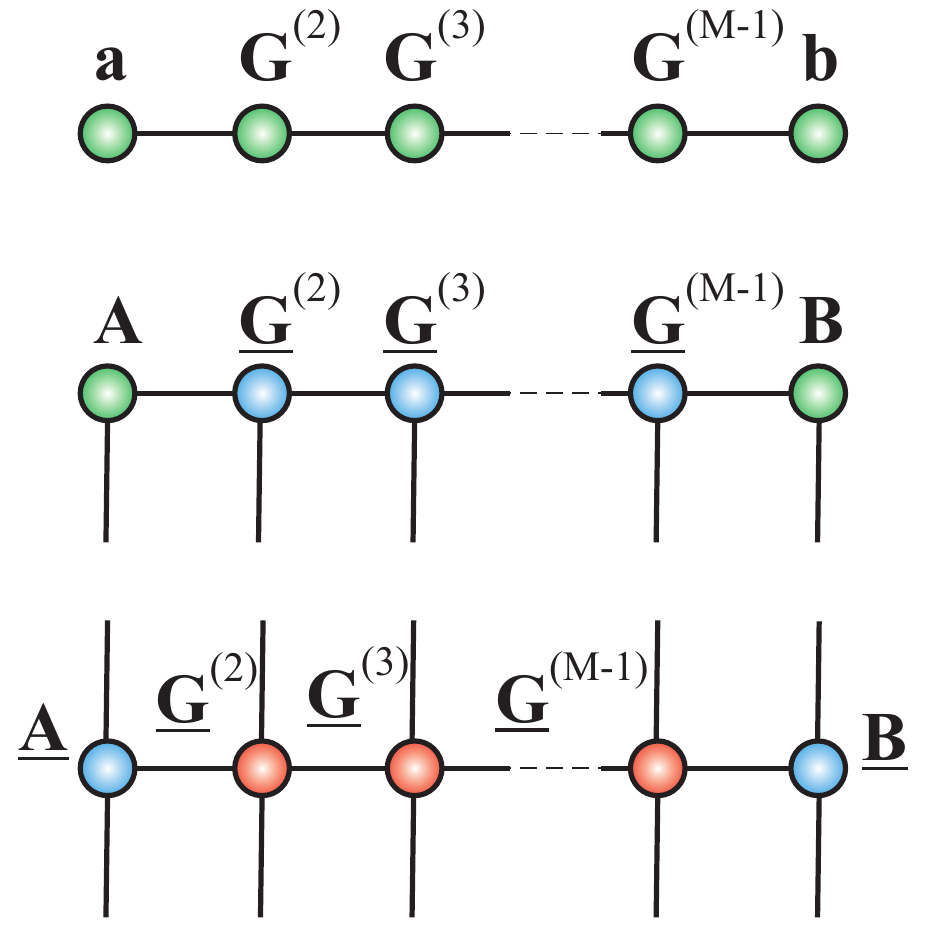}\\
\caption{Various forms of tensor train (TT) models: (Top) Scalar  function can be expressed as  $x=\ba^T \bG^{(2)}\bG^{(3)} \cdots \bG^{(M-1)} \bb$, (middle)  TT/MPS model of an $M$th-order data tensor (multidimensional vector) is expressed by 3rd-order tensors and two factor matrices as: $\underline \bX=\llbracket \bA, \underline \bG^{(2)}, \underline \bG^{(3)}, \ldots, \underline \bG^{(M-1)}, \bB \rrbracket$; (bottom) TT/MPO model of  $2M$th-order data tensor  (multidimensional matrix) can be expressed by the  chain of 3rd-order and 4th-order cores as: $\underline \bX=\llbracket \underline \bA, \underline\bG^{(2)}, \underline \bG^{(3)}, \ldots, \underline \bG^{(M-1)}, \underline \bB \rrbracket $.}
\label{Fig:TT3d}
\end{figure}

%\subsection{\bf Tensor Train (TT) Decompositions}
In this section we discuss in  more detail the Tensor Train (TT) decompositions which are
the simplest tensor networks.
Tensor train decomposition was introduced by Oseledets and Tyrtyshnikov \cite{OseledetsTT09,OseledetsTT11} and can take
various forms depending on the order of input data as illustrated in Fig. \ref{Fig:TT3d}.

 %\begin{figure*}[ht]
%\centering
%%\includegraphics[width=8.6cm,height=5.1cm]{TTOuterproda.eps}\\
%\includegraphics[width=14.6cm]{FigS25rc}\\
%\caption{Illustration of the tensor train decomposition  (TT/MPS) for a 4th-order tensor expressed via multilinear product of cores and the outer product of vectors (fibers) as sum of rank-1 tensors as: $\underline \bX \cong \underline \bG^{(1)} \times^1_3 \underline \bG^{(2)} \times^1_3 \underline \bG^{(3)} \times^1_3 \underline \bG^{(4)} = \sum_{r_1=1}^{R_1} \sum_{r_2=1}^{R_2} \sum_{r_{3}=1}^{R_{3}} (\bg_{\,1,r_1} \; \circ \; \bg^{(2)}_{\,r_1, r_2} \; \circ \; \bg^{(3)}_{\,r_2,r_3} \; \circ \; \bg^{(4)}_{\,r_{3},1})$ (for $R_1=3, R_2=4, R_3=5; R_0=R_4=1$).  All vectors (fibers) $\bg^{(n)}_{r_{n-1}r_n} \in \Real^{I_n}$ are considered as the column vectors.}
%\label{Fig:TTouter}
%\end{figure*}

 \begin{figure*}[ht]
%\centering
%\includegraphics[width=8.99cm,height=5.1cm]{TTslicesa.eps}
%\includegraphics[width=8.99cm,height=5.1cm]{MPSslices2.eps}
%\includegraphics[width=8.99cm,height=5.1cm]{FigS26.eps}\\
(a)\\
\begin{center}
\includegraphics[width=14.99cm]{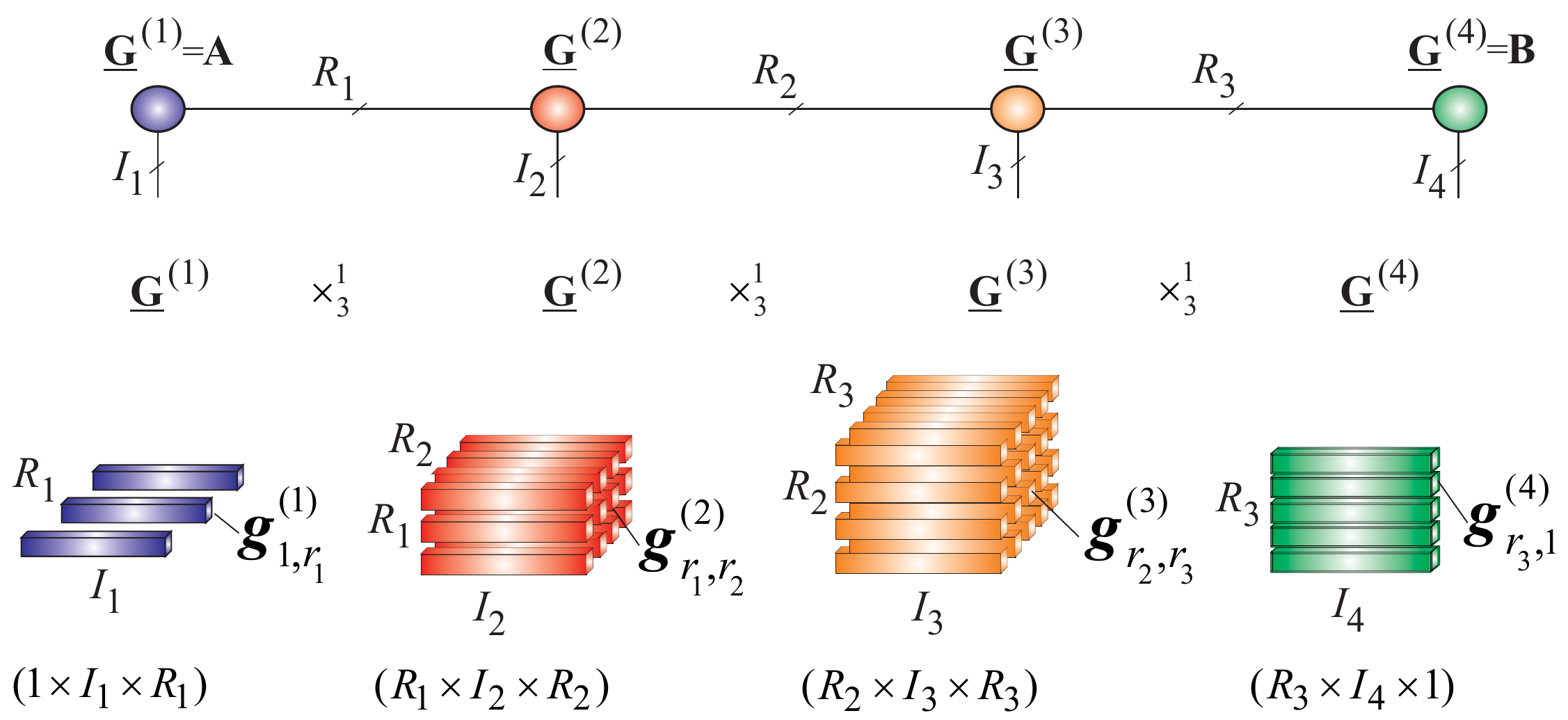}\\
\end{center}
(b)\\
\begin{center}
%\includegraphics[width=15.99cm]{Fig26b}\\
%\vspace{0.7cm}
\includegraphics[width=16.99cm]{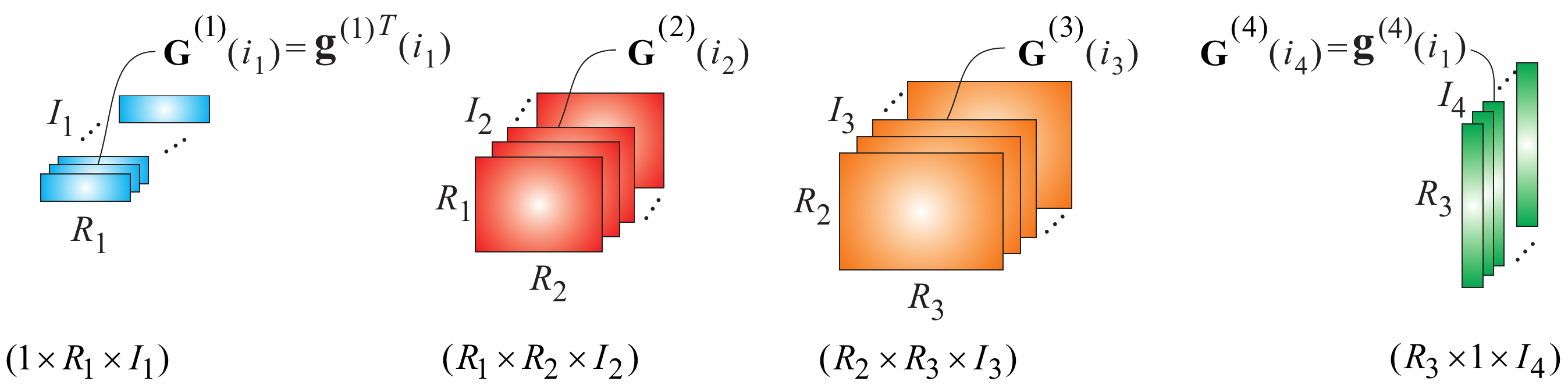}\\
\end{center}
\caption{Illustration of tensor train decomposition (TT/MPS) of a 4th-order  data tensor $\underline \bX \in \Real^{I_1 \times I_2 \times I_3 \times I_4}$. (a) Tensor form via multilinear product of cores and/or the outer product of vectors (fibers) as sum of rank-1 tensors as: $\underline \bX \cong \underline \bG^{(1)} \times^1_3 \underline \bG^{(2)} \times^1_3 \underline \bG^{(3)} \times^1_3 \underline \bG^{(4)} = \sum_{r_1=1}^{R_1} \sum_{r_2=1}^{R_2} \sum_{r_{3}=1}^{R_{3}} (\bg_{\,1,r_1} \; \circ \; \bg^{(2)}_{\,r_1, r_2} \; \circ \; \bg^{(3)}_{\,r_2,r_3} \; \circ \; \bg^{(4)}_{\,r_{3},1})$ (for $R_1=3, R_2=4, R_3=5; R_0=R_4=1$).  All vectors (fibers) $\bg^{(n)}_{r_{n-1}r_n} \in \Real^{I_n}$ are considered as the column vectors. (b)  Scalar form  via slice matrices as: $x_{i_1,i_2,i_3,i_4} \cong
%\bG^{(1)}_{1,R_1}(i_1) \; \bG^{(2)}_{R_1,R_2}(i_2) \; \bG^{(3)}_{R_2,R_3}(i_3) \; \bG^{(4)}_{R_{3},1}(i_{4}) =
\bG^{(1)}(i_1) \;  \bG^{(2)}(i_2) \; \bG^{(3)}(i_{3}) \;  \bG^{(4)}(i_4)=
  \sum_{r_1,r_2,r_3,r_4=1}^{R_1, R_2, \ldots, R_N} \; g^{(1)}_{1,i_1,r_1} \; g^{(2)}_{r_1,i_2,r_2} \; g^{(3)}_{r_2,i_3,r_3} \; g^{(4)}_{r_3,i_4,1}$.}
   %(for simplicity, we assumed that all core tensors
   %$\underline \bG_r \in \Real^{R_1 \times R_2 \times R_3}$ have the same dimensions).}
\label{Fig:TTslices}
\end{figure*}
%
 % \begin{figure*}[ht]
%\centering
%%\includegraphics[width=8.99cm,height=5.1cm]{TTslicesa.eps}
%%\includegraphics[width=8.99cm,height=5.1cm]{MPSslices2.eps}
%%\includegraphics[width=8.99cm,height=5.1cm]{FigS26.eps}\\
%\includegraphics[width=15.99cm]{Fig26b}\\
%\vspace{0.7cm}
%\includegraphics[width=15.99cm]{Fig26a}\\
%%
%\caption{Illustration of tensor train decomposition (TT/MPS) of a 4th-order  data tensor $\underline \bX \in \Real^{I_1 \times I_2 \times I_3 \times I_4}$ represented  in a scalar form  via slice matrices as: $x_{i_1,i_2,i_3,i_4} \cong
%%\bG^{(1)}_{1,R_1}(i_1) \; \bG^{(2)}_{R_1,R_2}(i_2) \; \bG^{(3)}_{R_2,R_3}(i_3) \; \bG^{(4)}_{R_{3},1}(i_{4}) =
%\bg^{(1)\;T}(i_1) \;  \bG^{(2)}(i_2) \; \bG^{(3)}(i_{3}) \;  \bg^{(4)}(i_4)=
%  \sum_{r_1,r_2,r_3,r_4=1}^{R_1, R_2, \ldots, R_N} \; g^{(1)}_{1,i_1,r_1} \; g^{(2)}_{r_1,i_2,r_2} \; g^{(3)}_{r_2,i_3,r_3} \; g^{(4)}_{r_3,i_4,1}$.}
%   %(for simplicity, we assumed that all core tensors
%   %$\underline \bG_r \in \Real^{R_1 \times R_2 \times R_3}$ have the same dimensions).}
%\label{Fig:TTslices}
%\end{figure*}
%%

\begin{figure*}[ht]
\centering
\includegraphics[width=16.8cm]{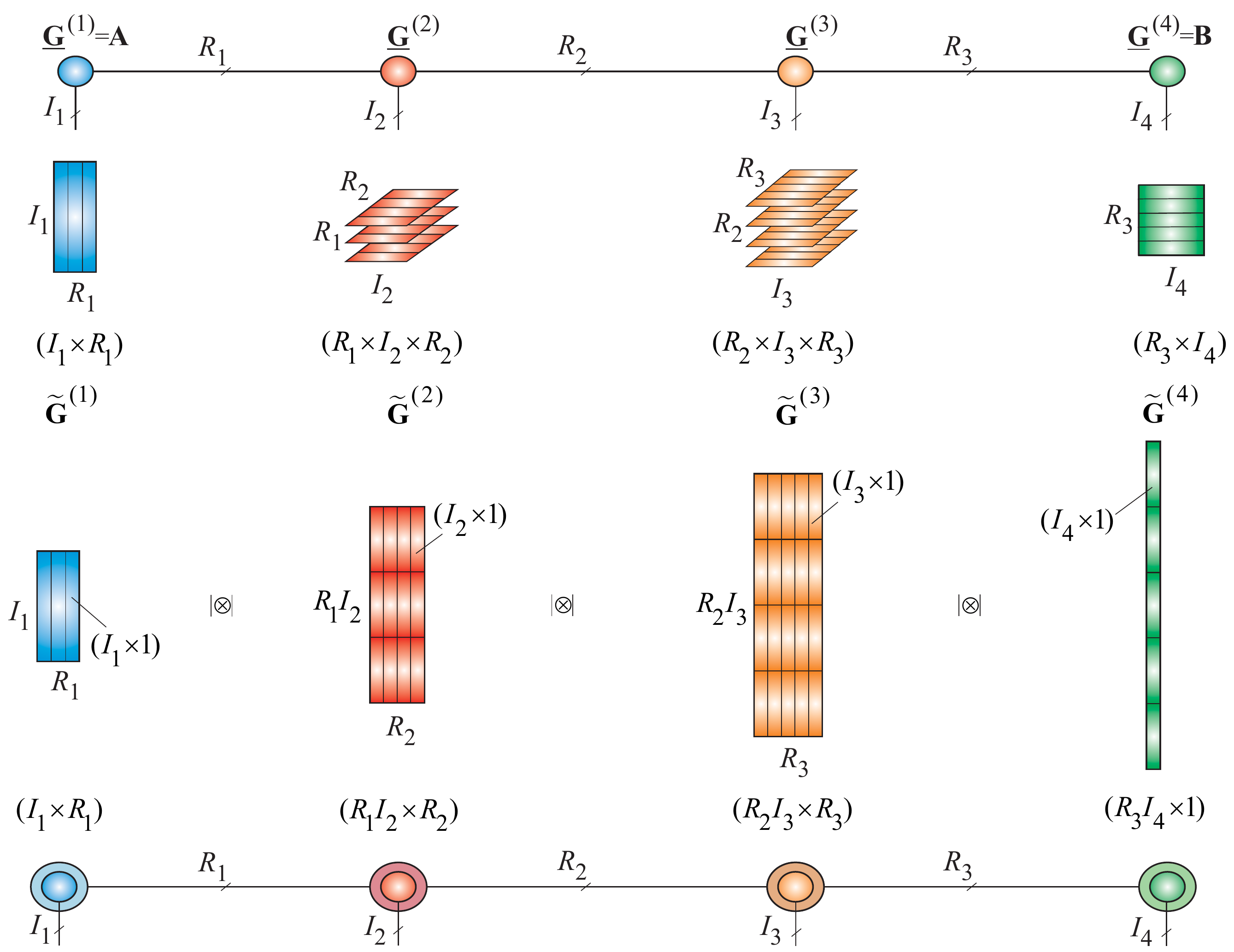}
%\includegraphics[width=16.8cm]{FigS27}
%(a)\\
%\includegraphics[width=8.6cm,height=4.1cm]{TTcolor1.eps}\\
%%(b)\\
%%\includegraphics[width=8.99cm,height=5.1cm]{TTslices.eps}\\
%(b)\\
%\includegraphics[width=8.99cm,height=8.1cm]{MPSSKV.eps}
%\includegraphics[width=15.0cm]{MPS5.eps}
%\includegraphics[width=8.99cm,height=8.1cm]{TTMPSkron.eps}
%\includegraphics[width=8.6Cm,height=6.2Cm]{TT51.eps}\\
%\vspace{0.4Cm}
%\includegraphics[width=8.6Cm,height=6.2Cm]{TT5m.eps}
\caption{Alternative representation of the tensor train  decomposition (TT/MPS) expressed via strong Kronecker products  of  block matrices in the form of a vector as: $\bx_{\overline{i_1, i_2, i_3, i_4}} \cong   \widetilde \bG^{(1)}  |\otimes| \; \widetilde \bG^{(2)} \; |\otimes| \; \widetilde \bG^{(3)} \; |\otimes| \; \widetilde \bG^{(4)} \in \Real^{I_1 I_2 I_3 I_4}$, where block matrices are defined as $\widetilde \bG^{(n)} \in \Real^{R_{n-1} I_{n} \times R_{n}}$, with block vectors $\bg^{(n)}_{r_{n-1},r_n} \in \Real^{I_n \times 1}$ for $n=1,2,3,4$  and $R_0=R_4=1$. For an illustrative purpose, we assumed that $N=4$, $R_1=3, R_2=4$ and $R_3=5$.}
   %(for simplicity, we assumed that all core tensors
   %$\underline \bG_r \in \Real^{R_1 \times R_2 \times R_3}$ have the same dimensions).}
\label{Fig:TT5}
\end{figure*}

   The  basic Tensor Train  \cite{OseledetsTT09,OseledetsTT11,Khoromskij-TT}, called also Matrix Product State (MPS), in quantum physics \cite{MPS2007,verstraete08-MPS,Orus2013,Schollwock13} decomposes the higher-order tensor into set of 3rd-order core tensors and factor matrices  as illustrated  in Figs. \ref{Fig:TTslices} -- \ref{Fig:TT5}.
    Note that the TT  model is equivalent to the MPS  only if the MPS  has the open boundary conditions (OBC) \cite{schollwock11-DMRG,Huckle2013}.

%
% \begin{figure}[t!]
%\centering
%\includegraphics[width=8.6cm,height=5.1cm]{TTOuterproda.eps}\\
%\caption{Illustration of the TT/MPS decomposition for the 4th-order tensor expressed via multolinear product of cores and the outer product of vectors (sum of rank-1 tensors): $\underline \bX \cong \underline \bG^{(1)} \times^1_2 \underline \bG^{(2)} \times^1_3 \underline \bG^{(3)} \times^1_3 \underline \bG^{(4)} =\sum_{r_1=1}^{R_1} \sum_{r_2=1}^{R_2} \sum_{r_{3}=1}^{R_{3}} (\bg_{r_1} \; \circ \; \bg^{(2)}_{r_1, r_2} \; \circ \; \bg^{(3)}_{r_2,r_3} \; \circ \; \bg^{(4)}_{r_{3}})$ (for $R_1=3, R_2=4, R_3=5$).  }
%\label{Fig:TTouter}
%\end{figure}
%
The tensor train (TT/MPS) for an $N$th-order data tensor $\underline \bX \in \Real^{I_1, \times I_2, \times \cdots \times I_N}$ can be described  in the  various equivalent mathematical forms as follows.

\begin{figure}[ht]
\includegraphics[width=8.6cm,height=3.0cm]{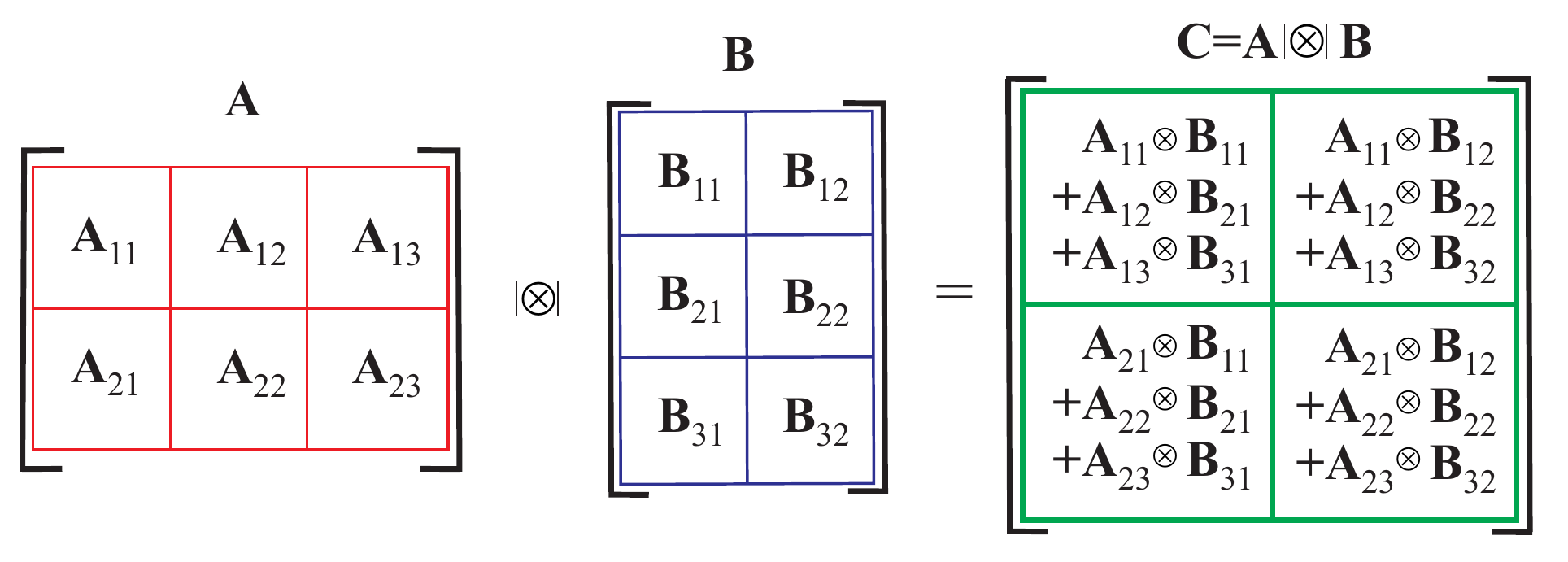}\\
\caption{Illustration of definition of the strong Kronecker product for two block matrices. The strong Kronecker product of two block matrices $\bA =[\bA_{r_1,r_2}] \in \Real^{R_1 I_1 \times R_2 J_1}$ and $\bB = [\bB_{r_2,r_3}] \in \Real^{R_2 I_2 \times R_3 J_2}$ is defined as the block matrix $\bC =\bA |\otimes| \bB \in \Real^{R_1 I_1 I_2 \times R_3 J_1 J_2}$, with blocks $\bC_{r_1,r_3} =\sum_{r_2=1}^{R_2} \bA_{r_1,r_2} \otimes \bB_{r_2,r_3} \in \Real^{ I_1 I_2 \times J_1 J_2}$, for $r_1=1,2;\;\; r_2=1,2,3$ and $r_3=1,2$.}
\label{Fig:strongKron}
\end{figure}

\begin{itemize}

\item In a compact tensor form using multilinear products:
\be
\underline \bX &\cong& \bA \times_2^1 \underline \bG^{(2)} \times_3^1 \underline \bG^{(3)} \times_3^1  \cdots \times_3^1 \underline \bG^{(N-1)} \times_3^1 \bB \nonumber \\
% &=&  \bA \times^1 \underline \bG^{(2)} \times^1 \underline \bG^{(3)} \times^1  \cdots \times^1 \underline \bG^{(N-1)} \times^1 \bB \nonumber \\
  &=& \llbracket \bA, \underline \bG^{(2)}, \underline \bG^{(3)}, \ldots, \underline \bG^{(N-1)}, \bB \rrbracket,
\ee
where 3rd-order cores are defined as $\underline \bG^{(n)} \in \Real^{R_{n-1} \times I_n \times R_n}$  for $n=2,3,\ldots,N-1$ (see Fig. \ref{Fig:TTslices} (a)).

\item By  unfolding of cores $\underline \bG^{(n)}$ and  suitable reshaping of matrices, we can obtain other  very useful  mathematical and graphical descriptions of the MPS, for example, as summation of rank-1 tensors  using outer (tensor) product (similar to CPD, Tucker and PARATREE formats):
\be
\underline \bX &\cong& \sum_{r_1,r_2,\ldots,r_{N-1}=1}^{R_1,R_2, \ldots,
 R_{N-1}} \bg^{(1)}_{\,1,r_1} \; \circ \; \bg^{(2)}_{r_1, r_2} \; \circ \; \bg^{(3)}_{r_2,r_3} \; \circ \cdots
\circ \; \bg^{(N)}_{\,r_{N-1},1} \notag \\
% &\cdots& \circ \; \bg^{(N-1)}_{r_{N-2}, r_{N-1}} \; \circ \; \bg^{(N)}_{r_{N-1}},
\ee
%\be
%\underline \bX &\cong& \sum_{r_1=1}^{R_1} \sum_{r_2=1}^{R_2} \cdots \sum_{r_{N-1}=1}^{R_{N-1}} \bg^{(1)}_{r_1} \; \circ \; \bg^{(2)}_{r_1, r_2} \; \circ \; \bg^{(3)}_{r_2,r_3} \; \circ \cdots
%\notag \\
% &\cdots& \circ \; \bg^{(N-1)}_{r_{N-2}, r_{N-1}} \; \circ \; \bg^{(N)}_{r_{N-1}},
%\ee
where
%$\ba_{r_1}$ are columns of the matrix $\bA=[\ba_1,\ba_2,\ldots,\ba_{R_1}] \in \Real^{I_1 \times R_1}$,
%analogously $\bb_{i_5}$ are the columns of the matrix
%$\bB =[\bb_1,\bb_2,\ldots,\bb_{I_5}] \in \Real^{R_4 \times I_5}$
$\bg^{(n)}_{r_{n-1},r_{n}} \in \Real^{I_n}$ are column vectors of  matrices $\bG^{(n)}_{(2)}=[\bg^{(n)}_{1,1}, \; \bg^{(n)}_{2,1},\ldots,\bg^{(n)}_{R_{n-1},1},\bg^{(n)}_{1,2},\ldots,\bg^{(n)}_{R_{n-1},R_n}] \in \Real^{I_n  \times R_{n-1} R_n}$ ($n=1,2,\ldots, N$), with $R_0=R_N=1$. Note that $\bg^{(1)}_{\,1,r_1}=\ba_{r_1}$ are columns of the matrix  $\bA=[\ba_1,\ba_2,\ldots,\ba_{R_1}] \in \Real^{I_1 \times R_1}$, while $\bg^{(N)}_{r_{N-1},1}=\bb_{r_{N-1}}$ are vector of the transposed factor  matrix $\bB^T = [\bb_1,\bb_2,\ldots,\bb_{R_{N-1}}] \in \Real^{I_{N} \times R_{N-1}}$ (see Fig. \ref{Fig:TTslices} (a)).

The minimal $(N-1)$ tuple $\{R_1,R_2,\ldots,R_{N-1}\}$  is called TT-rank (strictly speaking for the exact TT decomposition).

\item  Alternatively, we can use  the standard  scalar form:
\be
x_{i_1,i_2,\ldots,i_N} &\cong& \sum_{r_1,r_2,\ldots,r_{N-1}=1}^{R_1,R_2, \ldots,
 R_{N-1}} g^{(1)}_{\,1,i_1,r_1} \; g^{(2)}_{r_1,i_2,r_2}  \cdots g^{(n)}_{\,r_{N-1},i_N,1}, \notag \\
% &\cdots& g^{(N-1)}_{r_{N-2},i_{N-1},r_{N-1}} \; b_{r_{N-1},i_N},
\ee
or equivalently using slice representations (see Fig. \ref{Fig:TTslices}):
\be
x_{i_1,i_2,\ldots,i_N} &\cong& \bG^{(1)}(i_1) \; \bG^{(2)}(i_2)
 \cdots  \bG^{(N)}(i_N) \notag \\
&=& \bg^{(1)\;T}(i_1)  \; \bG^{(2)}(i_2) \cdots
  \bg^{(N)}(i_N), \notag \\
\ee
%\be
%x_{i_1,i_2,\ldots,i_N} &\cong& \bG^{(1)}(i_1) \; \bG^{(2)}(i_2)
% \cdots \bG^{(N-1)}(i_{N-1}) \; \bG^{(N)}(i_N) \notag \\
%&=& \bg^{(1)\;T}(i_1)  \bG^{(2)}(i_2) \cdots
%\bG^{(N-1)}(i_{N-1})  \bg^{(N)}(i_N), \notag \\
%\ee
%\be
%x_{i_1,i_2,\ldots,i_N} &\cong& \bG^{(1)}_{1,R_1}(i_1) \; \bG^{(2)}_{R_1,R_2}(i_2) \; \bG^{(3)}_{R_2,R_3}(i_3)\cdots \notag \\
% &\cdots& \bG^{(N-1)}_{R_{N-2},R_{N-1}}(i_{N-1}) \; \bG^{(N)}_{R_{N-1},1}(i_N)  \\
%&=& \bg^{(1)\;T}(i_1)  \bG^{(2)}(i_2) \cdots
%\bG^{(N-1)}(i_{N-1})  \bg^{(N)}(i_N),\notag
%\ee
where slice matrices $\bG^{(n)}(i_n)= \bG^{(n)}(:, i_n,:)$ $ = \bG^{(n)}_{R_{n-1},R_n}(i_n) \in \Real^{R_{n-1} \times R_n}$ (with $ \bG^{(1)}(i_1)=\bg^{(1)\;T}(i_1) \in \Real^{1 \times R_1}$ and  $ \bG^{(N)}(i_N)=\bg^{(N)}(i_N) \in \Real^{R_{N-1} \times 1}$) are  lateral slices of the  cores $\underline \bG^{(n)} \in \Real^{R_{n-1} \times I_n \times R_n}$ for $n=1,2,\ldots,N$ with $R_0=R_N=1$.

\item By representing  the cores $\underline \bG^{(n)} \in \Real^{R_{n-1} \times I_n \times R_n}$ by unfolding matrices $\widetilde \bG^{(n)} = (\bG^{(n)}_{(3)})^T \in \Real^{R_{n-1} I_n \times R_n}$ for $n=1,2,\ldots,N$ with $R_0=R_N=1$ and considering them as block matrices with  blocks $\bg^{(n)}_{r_{n-1},r_n} \in \Real^{I_n \times 1}$, we can express the TT/MPS in the matrix form via strong Kronecker products \cite{Seberry94,KazeevT13,kazeev2013LLA} (see Fig. \ref{Fig:TT5} (c) and Fig. \ref{Fig:strongKron}):
%

%If we reshape cores $\underline \bG^{(n)} \in \Real^{R_{n-1} \times R_n \times I_n}$ into block matrices $\widetilde \bG^{(n)} \in \Real^{R_{n-1} I_n \times R_n}$ with the blocks represented by column vectors of size $(I_n \times 1)$, then we can represent the TT=MPS in the vectorizing form via strong Kronecker products as:
\begin{equation}
\bx_{{\overline{i_1, i_2, \ldots, i_N}}} \cong  \widetilde \bG^{(1)} \; |\otimes| \;  \widetilde \bG^{(2)} \; |\otimes| \cdots  |\otimes| \; \widetilde \bG^{(N)}, \\
\end{equation}
where the vector $\bx_{\overline{i_1, i_2, \ldots, i_N}} =\bx_{(1:N)} \in \Real^{I_1 I_2 \cdots I_N}$ denotes vectorization of the tensor $\underline \bX$ in
lexicographical order of indices and $|\otimes|$ denotes strong Kronecker product.

\end{itemize}

The strong Kronecker product of two block matrices (e.g., unfolding cores):
\be
\bA = \begin{bmatrix} \bA_{1,1}&\cdots&  \bA_{1,R_2} \\
				\vdots & \ddots & \vdots \\
				\bA_{R_{1},1}&\cdots&  \bA_{R_1,R_2}
		\end{bmatrix} \in \Real^{R_1 I_{1} \times R_2 J_1}\notag
\ee
 and\\
\be
 \bB = \begin{bmatrix} \ \bB_{1,1}&\cdots& \bB_{1,R_{3}}\\
				\vdots & \ddots & \vdots \\
				 \bB_{R_{2},1}&\cdots& \bB_{R_{2},R_{3}}
		\end{bmatrix} \in \Real^{R_2 I_{2} \times R_{3} J_2 } \notag
 \ee
  is defined as a block matrix
\be
\bC = \bA \; |\otimes| \;  \bB \in \Real^{R_1 I_1 I_2 \times R_3 J_1 J_2},
\ee
with blocks $\bC_{r_{1},r_{3}}=\sum_{r_2=1}^{R_2}  \bA_{r_{1},r_2} \otimes   \bB_{r_2,r_{3}} \in \Real^{I_1 I_2 \times  J_1 J_2}$,    where $\bA_{r_{1},r_2} \in \Real^{I_{1} \times J_1}$ and  $\bB_{r_{2},r_{3}} \in \Real^{I_{2} \times J_2}$ are block matrices of $\bA$ and $ \bB$, respectively (see also Fig. \ref{Fig:strongKron} for graphical illustration).

The matrix  strong Kronecker product can be generalized to block tensors as follows:
Let
$\underline \bA = \begin{bmatrix}\underline \bA_{r_1,r_2} \end{bmatrix} \in \Real^{R_1 I_1 \times R_2 J_1 \times K_1}$ and
$\underline \bB = \begin{bmatrix}\underline \bB_{r_2,r_3} \end{bmatrix} \in \Real^{R_2 I_2 \times R_3 J_2 \times K_2}$ are
$R_1\times R_2$ and $R_2\times R_3$ block tensors, where
  blocks $\underline \bA_{r_1,r_2}
\in \Real^{I_1\times J_1 \times K_1}$ and
$\underline \bB_{r_2,r_3}
\in \Real^{I_2 \times J_2\times K_2}$
are 3rd order tensors,
then the strong Kronecker product of $\underline \bA$ and $\underline \bB$ is
defined by the $R_1\times R_3$   block tensor
\be
	\underline \bC
	= \begin{bmatrix} \underline \bC_{r_1,r_3}
	\end{bmatrix}
	=
	\underline \bA\ |\otimes|\ \underline \bB \in \Real^{R_1 I_1 I_2 \times R_3 J_1 J_2 \times K_1 K_2},
\ee
where
	$$
	\underline \bC_{r_1,r_3} = \sum_{r_2=1}^{R_2}
	\underline \bA_{r_1,r_2}\otimes \underline \bB_{r_2,r_3} \in \Real^{I_1 I_2 \times J_1 J_2 \times K_1 K_2},
	$$
for $r_1=1,2,\ldots,R_1$ and $r_3=1,2,\ldots,R_3$.
%\begin{equation}
%\bX_{(1)} \cong  \tilde \bG^{(1)} \; |\otimes| \; \tilde \bG^{(2)} \; |\otimes| \cdots |\otimes|\; \tilde \bG^{(N-1)} |\otimes| \; \tilde \bG^{(N)} \notag \\
%\end{equation}
%where $|\otimes|$ denotes strong Kronecker product. Note that the matrix $\tilde \bG^{(1)}=\bA$ is treated as a block vector with blocks $\ba_{r_1} \in \Real^{I_1 \times 1}$, $\tilde \bG^{(2)}$ is a block matrix with blocks $\tilde \bG_{r_1,r_2}^{(2)}$ of dimension $(1 \times I_2)$, $\tilde \bG^{(3)}$ is a block matrix with blocks $\tilde \bG_{r_2,r_3}^{(2)}$ of dimension $(1 \times I_3)$, etc., and finally $\tilde \bG^{(N)}$ is treated as a block column vector with the blocks of dimension $(1 \times I_N)$ (see Fig. \ref{Fig:TT5} (c)).
%

\begin{figure*}[ht]
\centering
%(a)\\
%\includegraphics[width=8.6cm,height=4.1cm]{TTcolor1.eps}\\
%(b)\\
%\includegraphics[width=8.8cm,height=5.9cm]{MPOSK.eps}
%\includegraphics[width=14.99cm]{TTMPOSKron.eps}
\includegraphics[width=16.99cm]{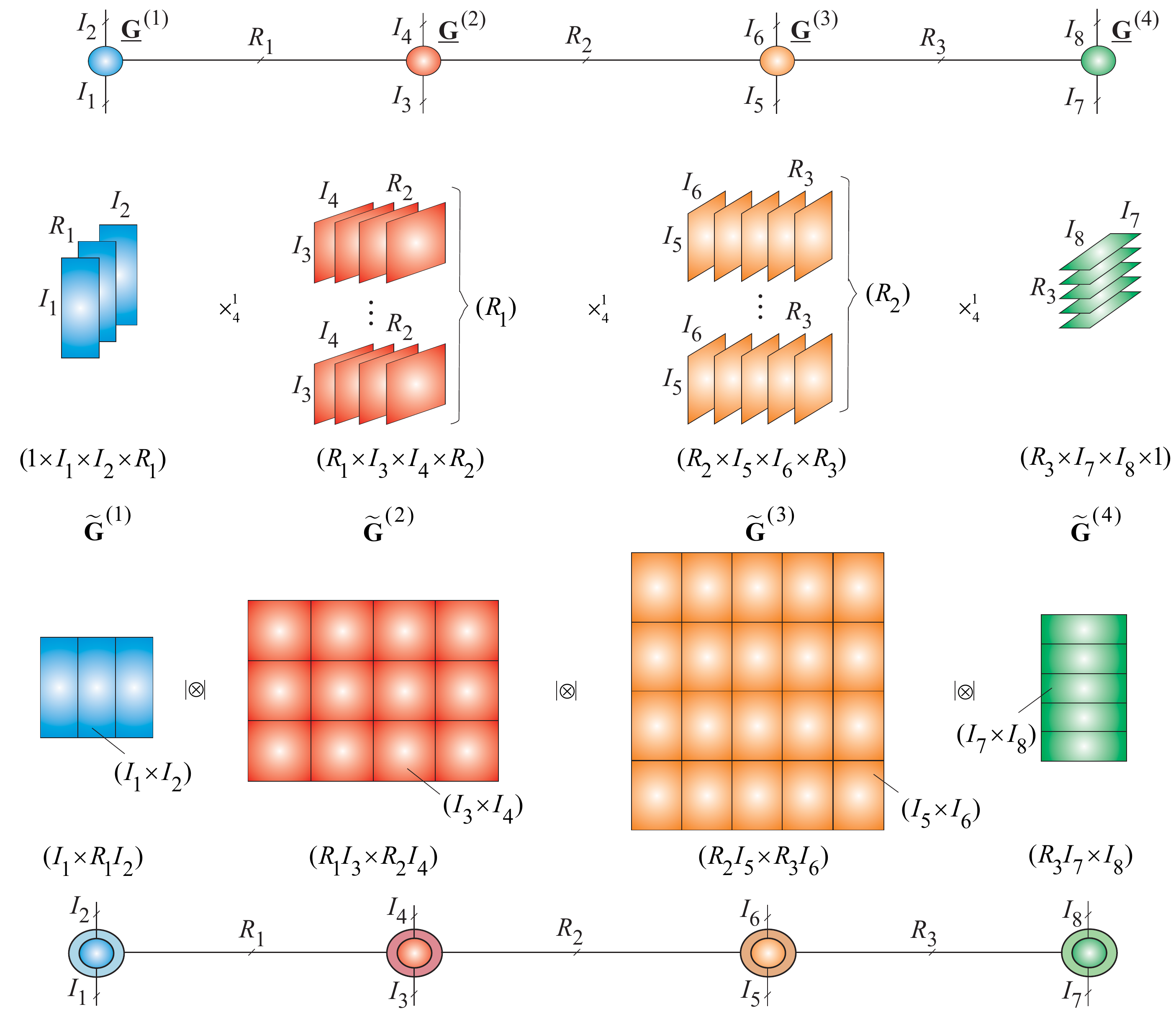}
\caption{The matrix tensor train decomposition (TT/MPO) for an 8th-order data tensor or equivalently multidimensional matrix
$\bX \in \Real^{\overline I_1 \times \overline I_2}$,
with $\overline I_1= I_1 I_3 I_5 I_7$ and $\overline I_2=I_2 I_4 I_6 I_8$, expressed by the  chain of  4th-order cores as: $\underline \bX  \cong \underline \bG^{(1)} \; \times_4^1 \; \underline \bG^{(2)} \; \times_4^1 \; \underline \bG^{(3)} \; \times_4^1 \; \underline \bG^{(4)} =
 \llbracket \underline \bG^{(1)}, \underline \bG^{(2)},  \underline \bG^{(3)}, \underline \bG^{(4)} \rrbracket$ or in a scalar  form  as $x_{i_1,i_2,\ldots,i_8} \cong \sum_{r_1=1}^{R_1} \sum_{r_2=1}^{R_2} \sum_{r_3=1}^{R_3} g^{(1)}_{\,1,i_1,r_1,i_2} \; g^{(2)}_{r_1,i_2,r_2,i_4} \; g^{(3)}_{r_2,i_5,r_3,i_6} \; g^{(4)}_{r_3,i_7,i_8,1}$. Alternatively, the TT/MPO decomposition can be expressed in a compact and elegant matrix form as strong Kronecker product of block matrices  $\widetilde \bG^{(n)} = \bG^{(n)}_{(\overline{r_{n-1}, \, i_{2n-1}}; \; \overline{r_n,  \,i_{2n}})} \in \Real^{R_{n-1} \,I_{2n-1} \times R_{n} \, I_{2n}}$ (with blocks $\bG^{(n)}_{r_{n-1},r_n} \in \Real^{I_{2n-1} \times I_{2n}}$) as: $\bX_{(\overline{i_1,i_3,i_5,i_7}\;;\;\overline{i_2,i_4,i_6,i_8})} \cong  \widetilde \bG^{(1)} \; |\otimes| \; \widetilde \bG^{(2)} \; |\otimes| \; \widetilde \bG^{(3)} \;|\otimes|\; \widetilde\bG^{(4)} \in \Real^{\overline I_1 \times \overline I_2}$. For simplicity, we assumed that $R_1=3$, $R_2=4$ and $R_3=5$, n=1,2,3,4.}
\label{Fig:MPOSK}
\end{figure*}

Another important TT model, called matrix TT  or MPO (Matrix Product Operator with Open Boundary Conditions),  consists of a chain (train) of 3rd-order and 4th-order cores, as illustrated in Fig. \ref{Fig:MPOSK}. Note that a 3rd-order tensor can be represented equivalently as a block (column or row) vector in which each element (block) is a matrix (lateral slice) of the tensor, while a 4th-order tensor can represented equivalently  as a block matrix. The TT/MPO model  for
$2N$th-order tensor $\underline \bX \in \Real^{I_1 \times I_2 \times \cdots I_{2N}}$ can be described mathematically in the following general forms (see also Table \ref{table:MPS-MPO}).

\begin{itemize}

\item A) In the tensor compact form using multilinear products
\be
\underline \bX & \cong& \underline \bG^{(1)} \; \times_4^1 \; \underline \bG^{(2)} \; \times_4^1 \; \cdots \times_4^1 \; \underline \bG^{(N)} \nonumber \\
% &=&  \underline \bG^{(1)} \; \times^1 \; \underline \bG^{(2)} \; \times^1 \cdots \times^1 \; \underline \bG^{(N)} \nonumber \\
  &=& \llbracket \underline \bG^{(1)}, \underline \bG^{(2)}, \dots, \underline \bG^{(N)} \rrbracket,
\ee
where the cores are defined as $\underline \bG^{(n)} \in \Real^{R_{n-1} \times I_{2n-1} \times I_{2n} \times R_{n}}$, with $R_0=R_N=1$, ($n=1,2,\ldots,N$).
%for $n=2,3,\ldots,N-1$ and two 3rd-order cores $\underline \bG^{(1)} \in \Real^{I_{1} \times I_{2} \times R_1}; \; \underline \bG^{(N)} \in \Real^{R_{N-1} \times I_{2N-1} \times I_{2N}}$.

\item B) Using  the standard (rather long and tedious)
 scalar form:
\be
x_{i_1,i_2,\ldots,i_{2N}} &\cong& \sum_{r_1=1}^{R_1} \sum_{r_2=1}^{R_2}\cdots \sum_{r_{N-1}=1}^{R_{N-1}} g^{(1)}_{1,i_1,i_2,r_1} \; g^{(2)}_{r_1,i_3,i_4,r_2} \cdots \notag \\
&\cdots& g^{(N-1)}_{r_{N-2},i_{2N-3},i_{2N-2},r_{N-1}} \; g^{(N)}_{r_{N-1},i_{2N-1},i_{2N},1}. \notag \\
\ee

\item C) By  matrix representations of cores, the TT/MPO decomposition can be expressed by strong Kronecker products (see Fig. \ref{Fig:MPOSK}):
\begin{equation}
\bX_{(\overline{i_1,i_3,\ldots,i_{2N-1}}\;;\;\overline{i_2,i_4,\ldots,i_{2N}
})} \cong  \widetilde \bG^{(1)} \; |\otimes| \; \widetilde \bG^{(2)} \; |\otimes| \cdots \;|\otimes|\; \widetilde\bG^{(N)},  \\
\end{equation}
%\begin{equation}
%\bX_{(\overline{1,3,\ldots,2N-1}\;;\;\overline{2,4,\ldots,2N})} \cong  \widetilde \bG^{(1)} \; |\otimes| \; \widetilde \bG^{(2)} \; |\otimes| \cdots \;|\otimes|\; \widetilde\bG^{(N)},  \\
%\end{equation}
where %$\bX_{(\overline{1,3,\ldots,2N-1}\;;\;\overline{2,4,\ldots,2N})}
$\bX_{(\overline{i_1,i_3,\ldots,i_{2N-1}}\;;\;\overline{i_2,i_4,\ldots,i_{2N}
})} \in \Real^{I_1 I_3 \cdots I_{2N-1} \times I_2 I_4 \cdots I_{2N}}$ is unfolding matrix of $\underline \bX$ in  lexicographical order of indices and $\widetilde \bG^{(n)} \in \Real^{R_{n-1} I_{2n-1} \times R_{n} I_{2n}}$ are block  matrices with blocks  $\bG^{(n)}_{r_{n-1},r_n} \in \Real^{I_{2n-1} \times I_{2n}}$ and the number of blocks $R_{n-1}\times R_{n}$.
 In the special case when ranks of the TT/MPO $R_n=1, \; \forall n$ the strong Kronecker products simplify to the standard Kronecker products.
 \end{itemize}

\begin{table*}[ht!]
\caption{Different forms of the Tensor Trains (TT): MPS and MPO (with OBC) representations of an $N$th-order tensor $\underline \bX \in \Real^{I_1 \times I_1 \times \cdots \times I_N}$ and a $2N$th-order tensor $\underline \bY \in \Real^{I_1 \times J_1 \times   I_2 \times J_2  \cdots \times I_N \times J_N}$, respectively. It is assumed that the TT rank is $\{R_1,R_2,\ldots,R_{N-1}\}$, with $R_0=R_N=1$ ($r_0=r_N=1$).}
% \centering
 {\small \shadingbox{
    \begin{tabular*}{1.01\textwidth}[t]{@{\extracolsep{\fill}}l@{\hspace{1em}}l} \hline  &  \\
{ \raisebox{0mm}[0mm][4mm]{\hspace{2.9cm}TT/MPS}}
&  \hspace{2.9cm}TT/MPO \\ \hline &   \\
 %\multicolumn{2}{c}{ Tensor representation, outer products}\\
% & \\
%%\begin{minipage}[t]{.5\textwidth}
%\begin{tabular*}{0.51\linewidth}[t]{@{\extracolsep{\fill}}cc@{}}
% $\tX = \sum\limits_{r=1}^{R} {\lambda_r \; \ba_{r} \circ \bb_{r} \circ \bc_{r}} $  & %\hspace{-6em}
%$\tX = \sum\limits_{r_1=1}^{R_1} \sum\limits_{r_2=1}^{R_2} \sum\limits_{r_3=1}^{R_3} {g_{r_1 \,r_2 \, r_3} \; \ba_{r_1} \circ \bb_{r_2} \circ \bc_{r_3}} $
%\end{tabular*}
%\end{minipage}
% \\  &   \\ \hline & \\
  \multicolumn{2}{c}{Scalar (standard) Representations}\\
  & \\
   $ x_{\;i_1,i_2,\ldots,i_N}  = \displaystyle{\sum_{r_1,r_2,\ldots,r_{N-1}=1}^{R_1,R_2, \ldots,
 R_{N-1}}} \;\;  g^{(1)}_{\;1,i_1, r_1} \; g^{(2)}_{\;r_1, i_2, r_2} \;
 g^{(3)}_{\;r_{2}, i_{3}, r_{3}}  \cdots
 g^{(N)}_{\;r_{N-1}, i_N,1}$  &  \hspace{0em}
 $y_{i_1,j_1, i_2, j_2, \ldots, i_N, j_N}  = \displaystyle{\sum_{r_1,r_2,\ldots,r_{N-1}=1}^{R_1,R_2, \ldots,
 R_{N-1}}} \;\; g^{(1)}_{\;1,i_1, j_1, r_1} \; g^{(2)}_{\;r_1, i_2, j_2, r_2} \cdots
  g^{(N)}_{\;r_{N-1}, i_{N}, j_{N},1}$
  \\ & \\
$g^{(n)}_{\;r_{n-1}, i_n, r_n}$ entries of a 3rd-order core $\underline \bG^{(n)} \in \Real^{ R_{n-1} \times I_n \times R_n}$ %$\;\;\;(r_0=r_N=1)$
&
\hspace{0em}
 $g^{(n)}_{\;r_{n-1}, i_n, j_n, r_n}$ entries of a 4th-order core $\underline \bG^{(n)} \in \Real^{ R_{n-1} \times I_n \times J_n \times R_n}$
  \\  &   \\ \hline & \\
  \multicolumn{2}{c}{Slice Representations}\\
  & \\
  $ x_{\;i_1,i_2,\ldots,i_N} =  \bG^{(1)}(i_1) \; \bG^{(2)}(i_2) \cdots \bG^{(N-1)}(i_{N-1}) \; \bG^{(N)}(i_N)$  &  \hspace{0em}
 $y_{i_1, j_1, i_2, j_2,\ldots, i_N, j_N}  =  \bG^{(1)}(i_1, j_1) \; \bG^{(2)}(i_2, j_2) \cdots \bG^{(N)}(i_N, j_N)$
  \\ & \\
$\bG^{(n)}(i_n) \in \Real^{R_{n-1} \times R_n}$  lateral slices of cores $\underline \bG^{(n)}$
&
\hspace{0em}
$\bG^{(n)}(i_n, j_n) \in \Real^{R_{n-1} \times R_n}$ slices of cores $\underline \bG^{(n)}$
% $g^{(n)}_{\;r_{n-1}, i_n, j_n, r_n}$ entries of a 4th-order core $\underline \bG^{(n)} \in \Real^{ R_{n-1} \times I_n \times J_n \times R_n}$
\\  &  \\ \hline  & \\
 \multicolumn{2}{c}{Tensor Representations: Multilinear Products (tensor contractions)}\\
  & \\
   $ \underline \bX = \underline \bG^{(1)} \times_3^1 \; \underline \bG^{(2)} \times_3^1 \; \cdots \times_3^1 \; \underline \bG^{(N-1)} \times_3^1 \; \underline \bG^{(N)} $  &  \hspace{2em}
 $ \underline \bY = \underline \bG^{(1)} \times_4^1 \; \underline \bG^{(2)} \times_4^1  \cdots \times_4^1 \; \underline \bG^{(N-1)} \times_4^1 \; \underline \bG^{(N)}$
 \\ & \\
 $\underline \bG^{(n)} \in \Real^{ R_{n-1} \times I_n \times R_n}$, $\; (n=1,2,\ldots,N)$ &  \hspace{2em}
 $\underline \bG^{(n)} \in \Real^{ R_{n-1} \times I_n \times J_n \times R_n}$
 \\  &   \\ \hline & \\
  \multicolumn{2}{c}{Tensor Representations: Outer Products}\\
  & \\
   $ \underline \bX  = \displaystyle{\sum_{r_1,r_2,\ldots,r_{N-1}=1}^{R_1,R_2, \ldots,
 R_{N-1}}} \;\; \bg^{(1)}_{\;1,r_1} \; \circ \; \bg^{(2)}_{\;r_1, r_2}  \circ \cdots
 \circ \; \bg^{(N-1)}_{\;r_{N-2}, r_{N-1}} \; \circ \; \bg^{(N)}_{\;r_{N-1},1} $  &  \hspace{1em}
 $  \underline \bY  = \displaystyle{\sum_{r_1,r_2,\ldots,r_{N-1}=1}^{R_1,R_2, \ldots,
 R_{N-1}}} \;\; \bG^{(1)}_{\;1,r_1} \; \circ \; \bG^{(2)}_{\;r_1, r_2}  \circ \cdots
 \circ \; \bG^{(N-1)}_{\;r_{N-2}, r_{N-1}} \; \circ \; \bG^{(N)}_{\;r_{N-1},1}$
   \\ & \\
$\bg^{(n)}_{\;r_{n-1}, r_n} \in \Real^{I_n}$ blocks of a matrix $\widetilde\bG^{(n)} =(\bG^{(n)}_{(3)})^T \in \Real^{ R_{n-1}  I_n \times  R_n}$  &  \hspace{1em}
 $\bG^{(n)}_{\;r_{n-1}, r_n} \in \Real^{I_n \times J_n} $ blocks of a matrix $\widetilde \bG^{(n)} \in \Real^{ R_{n-1} I_n  \times R_n J_n}$
 \\  &   \\ \hline & \\
  \multicolumn{2}{c}{Vector/Matrix Representations:
  Kronecker and Strong Kronecker Products} \\ & \\
    $ \bx_{(\overline{i_1, \ldots, i_N})}  = \displaystyle{\sum_{r_1,r_2,\ldots,r_{N-1}=1}^{R_1,R_2, \ldots,
 R_{N-1}}} \;\; \bg^{(1)}_{\;1,r_1} \; \otimes\; \bg^{(2)}_{\;r_1, r_2}  \otimes \cdots
 \otimes \; \bg^{(N)}_{\;r_{N-1},1} $  &  \hspace{-1em}
 $ \bY_{(\overline{i_1, \ldots, i_N}); \; \overline{j_1, \ldots, j_N})} = \displaystyle{\sum_{r_1,r_2,\ldots,r_{N-1}=1}^{R_1,R_2, \ldots,
 R_{N-1}}} \;\; \bG^{(1)}_{\;1,r_1} \; \otimes \; \bG^{(2)}_{\;r_1, r_2}  \otimes \cdots
\otimes \; \bG^{(N)}_{\;r_{N-1},1}$
   \\ & \\
   $ \bx_{(\overline{i_1, \ldots, i_N})} =  \widetilde \bG^{(1)} \; |\otimes| \;  \widetilde \bG^{(2)} \; |\otimes| \cdots |\otimes|\;  \widetilde \bG^{(N)} \in \Real^{I_1 I_2 \cdots I_N}$  &  \hspace{-2em}
 $ \bY_{(\overline{i_1, \ldots, i_N}); \; \overline{j_1, \ldots, j_N})} =  \widetilde \bG^{(1)} \; |\otimes| \; \widetilde \bG^{(2)} \; |\otimes| \cdots \;|\otimes|\; \widetilde\bG^{(N)} \in \Real^{I_1  \cdots I_N \; \times \; J_1  \cdots J_N}$
  \\ & \\
%  $ \bx_{(\overline{i_1 \cdots i_N})} =  [(\widetilde \bG^{(1)} \; |\otimes| \cdots |\otimes|\;  \widetilde \bG^{(N-1)}) \otimes \bI_{I_N}] \; \widetilde \bG^{(N)}$  &  \hspace{-1em}
%% $ \by_{\overline{i_1 \cdots i_N \; j_1 \cdots j_N}} =  [\bI_{I_N J_N} \; \otimes \;  (\widetilde \bG^{(1)} \; |\otimes| \cdots |\otimes|\;  \widetilde \bG^{(N-1)})] \; vec(\widetilde\bG^{(N)})$
%  \\ & \\
%  $ \bx_{(\overline{i_1 \cdots i_N})} =  [\widetilde \bG^{(1)} \otimes \bI_{I_2 I_3 \cdots I_N}](\widetilde \bG^{(2)} \; |\otimes| \cdots |\otimes|\;  \widetilde \bG^{(N)})$  &  \hspace{-2em}
% %$ \by_{\overline{i_1 \cdots i_N \; j_1 \cdots j_N}} =  [(\widetilde \bG^{(2)} \; |\otimes| \cdots |\otimes|\;  \widetilde \bG^{(N)}) \otimes \bI_{I_1 J_1}] \; vec(\widetilde\bG^{(1)})$
%  \\ & \\
 $\widetilde\bG^{(n)} \in \Real^{ R_{n-1} I_n \times  R_n}$ a block matrix with blocks $\bg^{(n)}_{r_{n-1}, r_n} \in \Real^{I_n}$; &  \hspace{-1em}
 $\widetilde \bG^{(n)} \in \Real^{ R_{n-1} I_n  \times R_n J_n}$ a block matrix with blocks $\bG^{(n)}_{r_{n-1}, r_n} \in \Real^{I_n \times J_n}$
  \\   &  \\ \hline
    \end{tabular*}
    }}
\label{table:MPS-MPO}
\end{table*}

%\begin{figure}[ht]
%\includegraphics[width=8.99cm,height=18.6cm]{FigS30}\\
%%\includegraphics[width=8.99cm]{MPSSVD5.eps}\\
%\caption{Illustration of the SVD algorithm for TT/MPS for a 5th-order tensor. Instead of the
%truncated SVD, we can employ any low-rank matrix factorizations, especially QR, CUR, SCA, ICA, NMF.}
%%
%\label{Fig:AlgSVDMPS}
%\end{figure}
%
%\begin{figure} %[ht]
%\includegraphics[width=8.6cm]{FigS31}
%%\includegraphics[width=8.99cm]{MPOSVD4.eps}\\
%\caption{Illustration of SVD algorithm for TT/MPO for a 6th-order tensor. Instead of the SVD we can use alternative Low-Rank Approximation (LRA) (constrained matrix factorizations, e.g., CUR or NMF. }
%%
%\label{Fig:AlgSVDMPS2}
%\end{figure}

%\subsection{\bf Hierarchical Tucker (HT) Decompositions}

\begin{figure}[ht]
\includegraphics[width=8.99cm,height=18.6cm]{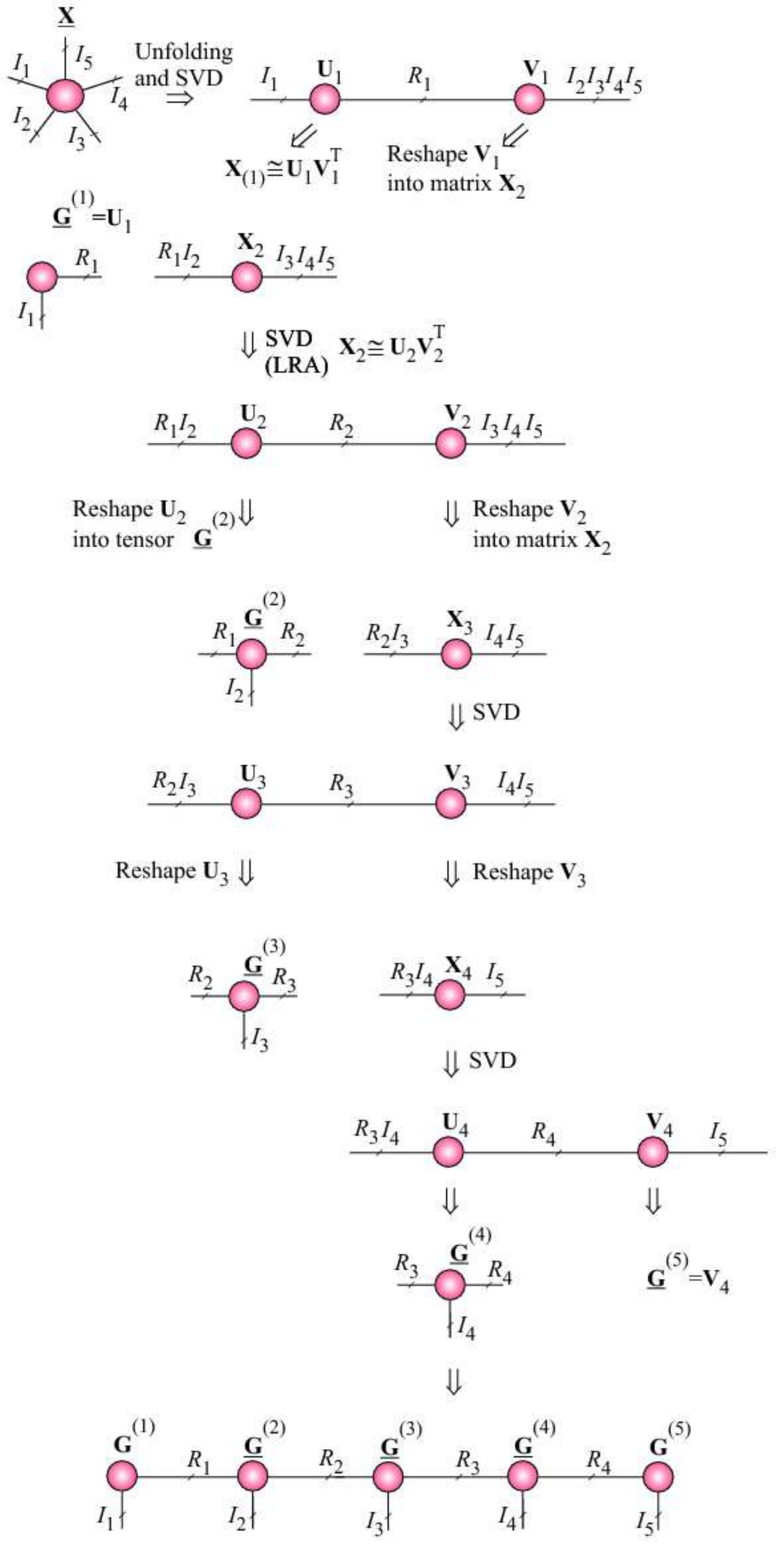}\\
\caption{Illustration of the SVD algorithm for TT/MPS for a 5th-order tensor. Instead of
truncated SVD, we can employ any low-rank matrix factorizations, especially QR, CUR, SCA, ICA, NMF.}
\label{Fig:AlgSVDMPS}
\end{figure}

\begin{figure} %[ht]
\includegraphics[width=8.6cm]{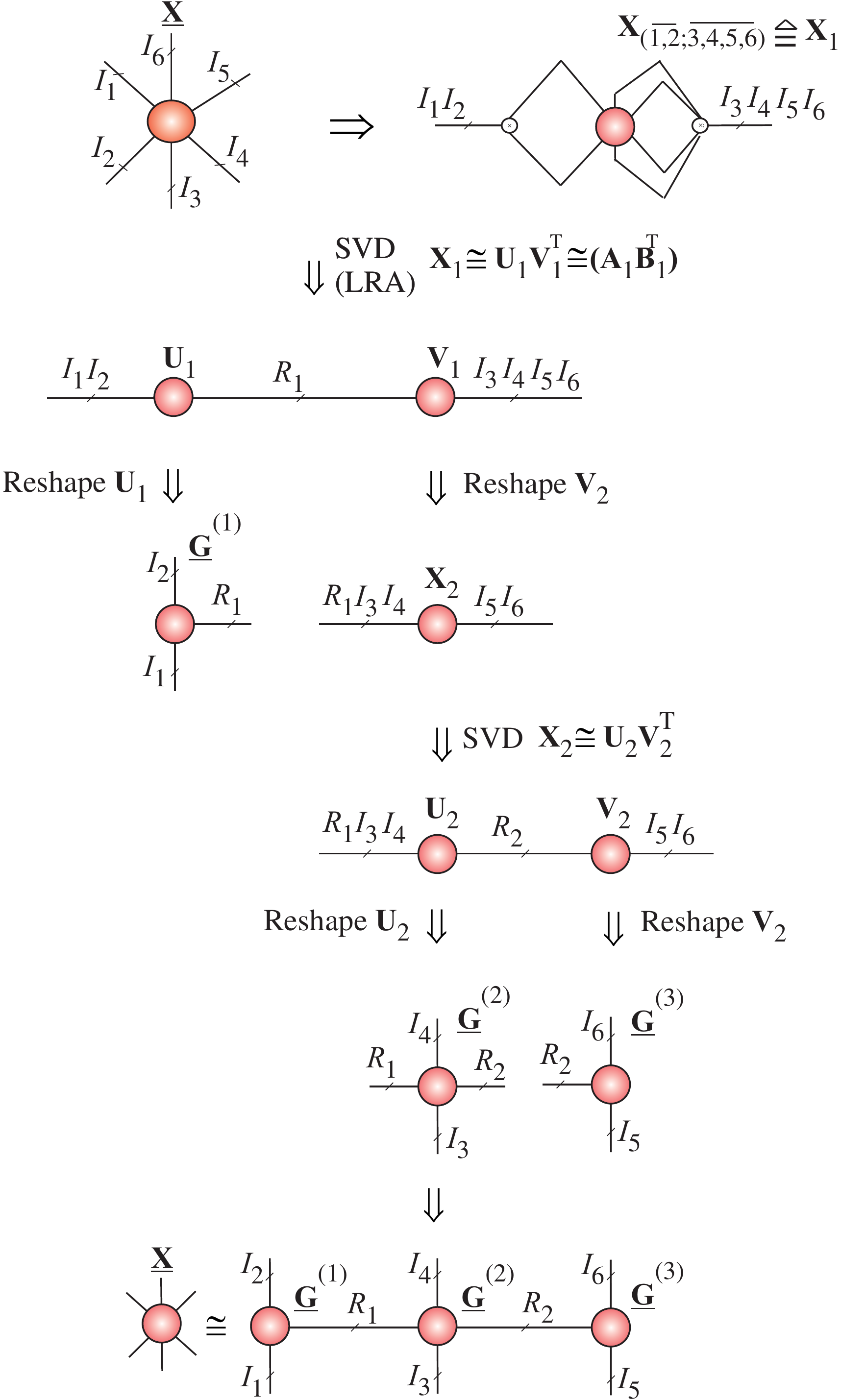}
\caption{SVD algorithm for TT/MPO for a 6th-order tensor. Instead of the SVD we can use alternative Low-Rank Approximations (constrained matrix factorizations, e.g., CUR or NMF). }
\label{Fig:AlgSVDMPS2}
\end{figure}

The Tensor Train (TT) format \cite{OseledetsTT11}, can be interpreted as
a special case of the HT \cite{Hackbush2012}, where all nodes of the underlying tensor
network are aligned
and where, moreover, the leaf matrices are assumed to be identities (and thus need not
be stored). An advantage of the TT format is its simpler practical implementation using SVD or alternative low-rank matrix approximations, as no binary tree need be involved \cite{Vidal03,OseledetsTT09} (see Figs. \ref{Fig:AlgSVDMPS} and \ref{Fig:AlgSVDMPS2}).

%For some very high-order data tensors  it has been observed
%that the ranks $R_n$ of 3rd-order tensors  increase rapidly with the order of the tensor,
%for any choice of tensor network that is a tree (including TT and  HT decompositions) \cite{kressner2012htucker}.
%

 Two different types of approaches to perform tensor approximation via TT exist \cite{Grasedyck-rev}.
The first class  of methods is based on combining standard iterative algorithms,
 with a low-rank decompositions, such as SVD/QR or CUR or Cross-Approximations.
Similar to the Tucker decomposition, the TT and HT decompositions are usually
based on low rank approximation of generalized unfolding matrices $\bX_{([n])}$, and  a good approximation in a  decomposition for a given TT/HT-rank can be obtained using the  truncated SVDs of the unfolding matrices \cite{Vidal03,OseledetsTT11}.

\begin{figure}[h]
\centering
\includegraphics[width=8.8cm]{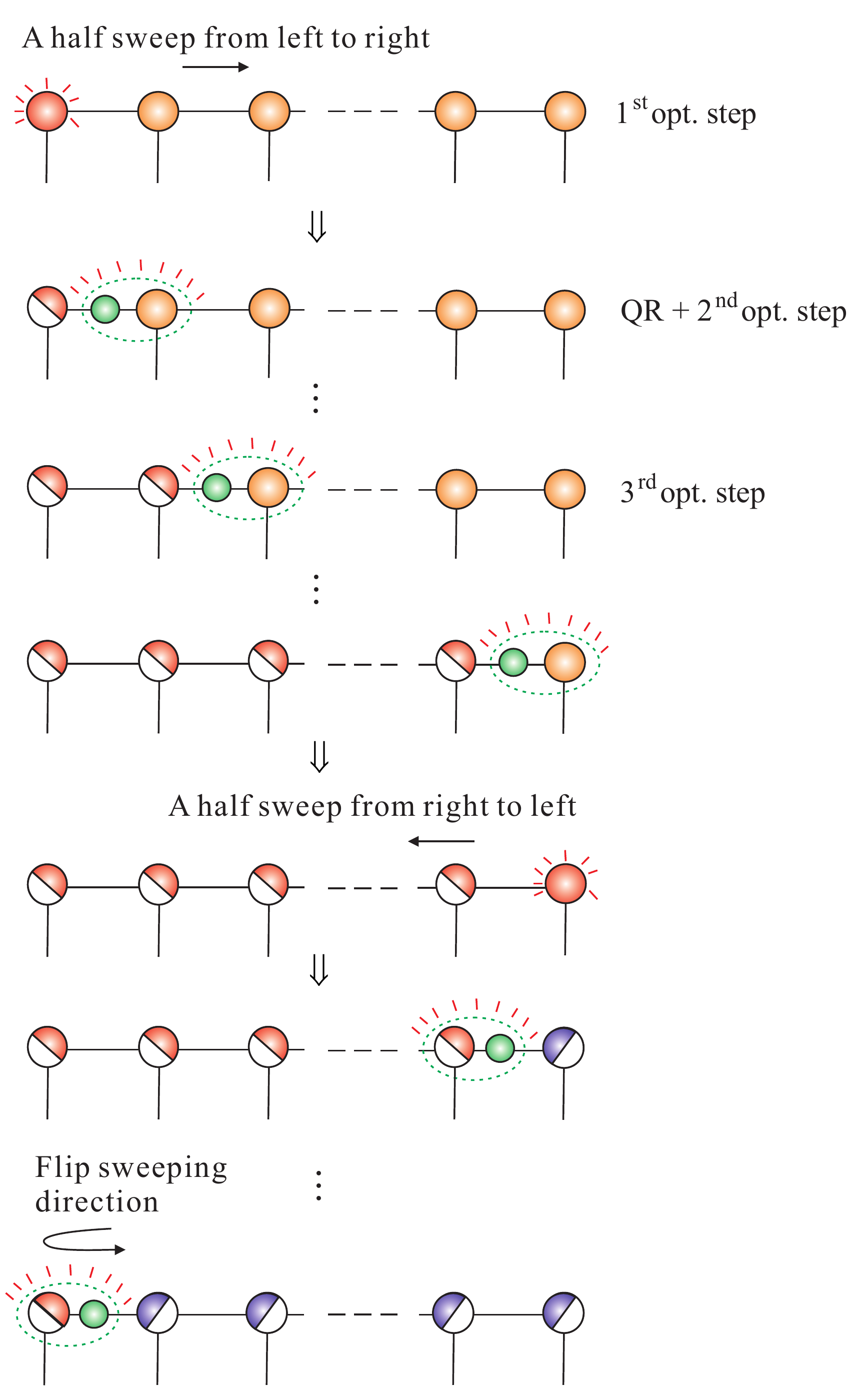}
\caption{Extension of the ALS algorithm for TT decomposition. The  idea is to optimize only one core tensor at a time (by a minimization of suitable cost function), while keeping the others fixed. Optimization of each core tensor is followed by an orthogonalization step via the QR or more expensive SVD decomposition. Factor matrices $\bR$ are absorbed (incorporated) into the following core.}
\label{Fig:TTALS}
\end{figure}

\begin{figure}[h]
\centering
\includegraphics[width=8.4cm]{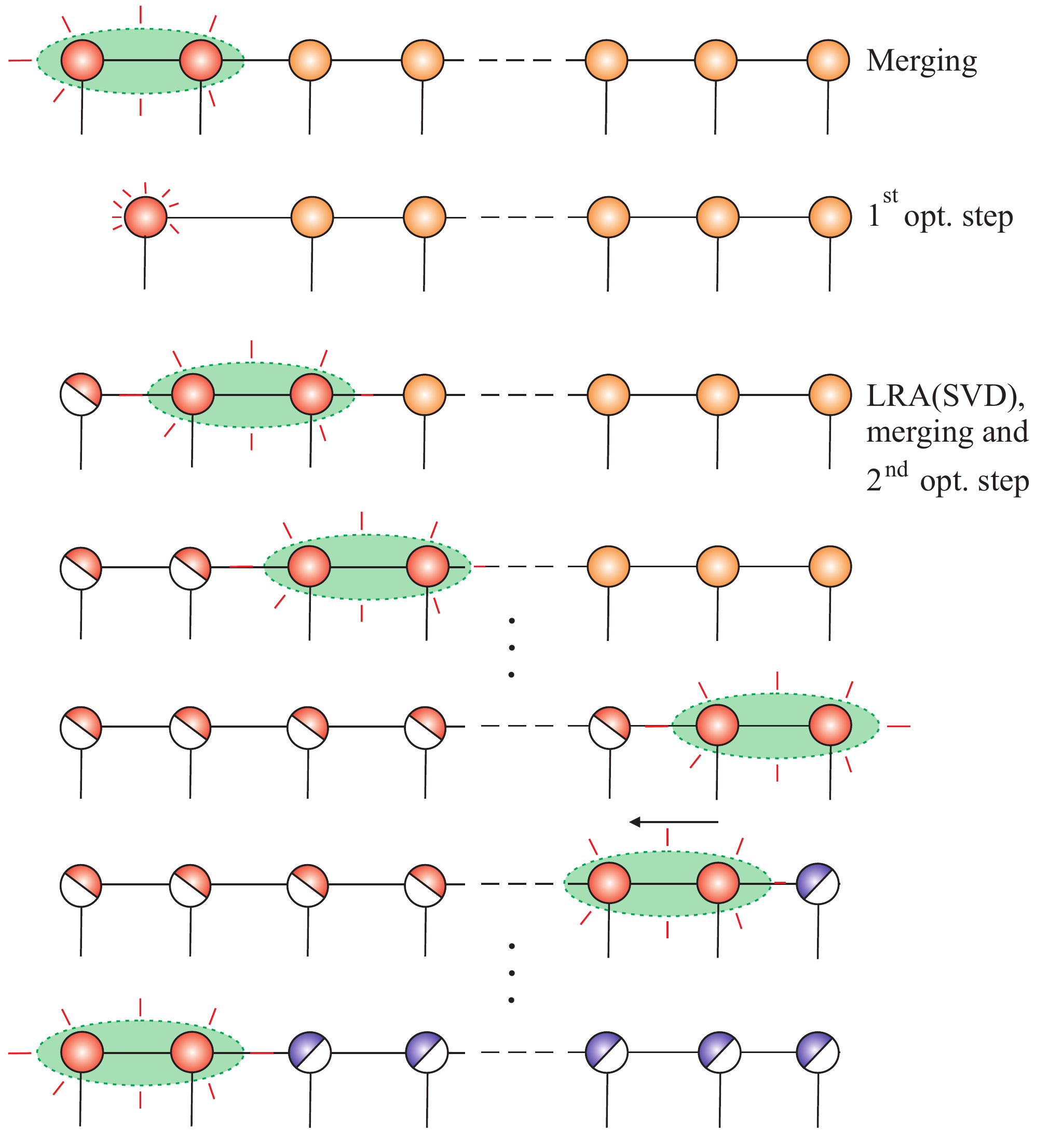}
\caption{Modified ALS (MALS) algorithm related to the Density Matrix Renormalization Group (DMRG)  for TT decomposition.  In each optimization step, two neighbor cores are merged. An optimization
performed over merged ``supercore''. After optimization in the next step we apply truncated SVD or other low-rank matrix factorizations (LRA) to separate the optimized supercore. For example, for nonnegative TT SVD steps can replaced by a non-negative matrix factorization (NMF) algorithm. Note that each optimization sub-problem is  more expensive than the standard ALS
and complexity increases but convergence speed may increase dramatically (see also \cite{Holtz-TT-2012,kressner2012htucker}).}
\label{Fig:MALS}
\end{figure}

In \cite{oseledets2010tt}  Oseledets and Tyrtyshnikov
 proposed for TT decomposition a new approximative  formula in
which a $N$th-order data tensor is interpolated using  special form of Cross-Approximation, which is  a modification of CUR algorithm.  The total number of entries and
the complexity of the interpolation algorithm depend linearly on the order of data tensor $N$, so the developed algorithm  does not suffer from the curse of dimensionality. The TT-Cross-Approximation is
analog to the SVD/HOSVD  like algorithms for TT/MPS, but uses  adaptive
cross-approximation instead of the computationally more expensive SVD.

The second class of algorithms based on optimization of suitable designed cost functions, often with additional penalty or regularization terms.
Optimization techniques include gradient descent, conjugate gradient, and Newton-like methods  (see \cite{Holtz-TT-2012,Grasedyck-rev} and references therein).
Gradient descent methods leads often to the Alternating Least Squares (ALS) type of algorithms,
which can be improved in various ways \cite{Holtz-TT-2012,Rohwedder2013} (see Fig. \ref{Fig:TTALS}).

A quite successful improvement in TNs (TT, HT) is called the DMRG method.  It
joins two neighboring factors (cores), optimize the resulting ``supernode'', and
splits the result into separate factors by a low-rank matrix factorization \cite{Holtz-TT-2012,kressner2012htucker,schollwock11-DMRG,Huckle2013}
(see Fig. \ref{Fig:MALS}).

{\bf Remark:} In most optimization problems it is very convenient  to present TT in a canonical form,
in which all cores are left or right orthogonal \cite{Holtz-TT-2012,QTT-Tucker} (see also Fig. \ref{Fig:TTALS} and Fig. \ref{Fig:MALS}).

The $N$-order core tensor is called right-orthogonal if
\be
\bG_{(1)} \bG^T_{(1)}= \bI.
\ee
Analogously the $N$-order core tensor is called left orthogonal if
\be
\bG_{(N)} \bG^T_{(N)}= \bI.
\ee
In contrast, for the  all-orthogonal core tensor we have $\bG_{(n)} \bG^T_{(n)}=\bI, \;\; \forall n$.\\

{\bf TT-Rounding } TT--rounding  (also called  truncation or recompression) \cite{oseledets2010tt} is post-processing procedure to
 reduce the TT ranks which in the first stage after applying low-rank matrix factorizations are usually not optimal with respect of desired approximation errors.
The optimal  computation of  TT-tensor is generally  impossible without TT-rounding.
The tensor  expressed already in TT format is approximated by another TT-tensor
with smaller TT-ranks but with prescribed accuracy of approximation  $\epsilon$.
The  most popular TT-rounding algorithm  is based on the QR/SVD
 algorithm, which requires $O(N I R^3)$  operations \cite{OseledetsTT09,oseledets2012tt}.
 In practice, we avoid  explicit construction of these matrices and the SVDs when truncating a tensor in TT decomposition to lower TT-rank. Such truncation algorithms for TT are described by Oseledets in \cite{OseledetsTT11}. In fact, the method exploits micro-iterations algorithm where the SVD is performed only on a relatively small core at each iteration. A similar approach has been developed by Grasedyck for the HT \cite{hTucker1}.
HT/TT algorithms that avoid the explicit computation of
these SVDs when truncating a tensor that is already in tensor network format are discussed in \cite{Grasedyck-rev,kressner2012htucker,espigtensorcalculus}.

TT Toolbox developed by Oseledets  (\url{http://spring.inm.ras.ru/osel/?page_id=24}) is focussed on TT structures, and deals with the curse of dimensionality \cite{oseledets2012tt}.
The Hierarchical Tucker (HT) toolbox  by Kressner and Tobler  (\url{http://www.sam.math.ethz.ch/NLAgroup/htucker_toolbox.html}) and Calculus library by Hackbusch, Waehnert and Espig, focuss mostly on HT and TT tensor networks \cite{oseledets2012tt,KressnerTobler14,espigtensorcalculus}.
See also  recently  developed  TDALAB (\url{http://bsp.brain.riken.jp/TDALAB} and TENSORBOX  \url{http://www.bsp.brain.riken.jp/~phan} that provide user-friendly interface and advanced algorithms for selected TD (Tucker, CPD) models  \cite{tdalab,tensorbox}. The \url{<http://www.esat.kuleuven.be/sista/tensorlab/>}{Tensorlab} toolbox builds upon the complex optimization framework and offers efficient numerical algorithms for computing the TDs
with various constraints (e.g. nonnegativity, orthogonality) and the possibility to combine and jointly factorize dense, sparse and incomplete tensors \cite{Sorber-tensorlab}.
The problems related to optimization  of
existing TN/TD algorithms are active area of research \cite{Grasedyck-rev,Holtz-TT-2012,Rohwedder2013,Lee-Cichocki,Wu-Cichocki}.

\section{\bf Hierarchical Outer Product Tensor Approximation (HOPTA) and Kronecker Tensor Decompositions}

%\section{\bf Extended Tensor Decomposition Models - Current Trends}

Recent advances in TDs/TNs include, TT/HT \cite{Khoromskij-TT,Holtz-TT-2012,Grasedyck-rev}, PARATREE \cite{VisaSP-09},
Block Term Decomposition (BTD) \cite{Lath-BCM12},
Hierarchical Outer Product Tensor Approximation (HOPTA) and Kronecker Tensor Decomposition (KTD) \cite{Phan2012-Kron,Ragnarsson-PHD,Phan_BTDLxR}.

\begin{figure}[ht]
\centering
 \includegraphics[width=6.8cm]{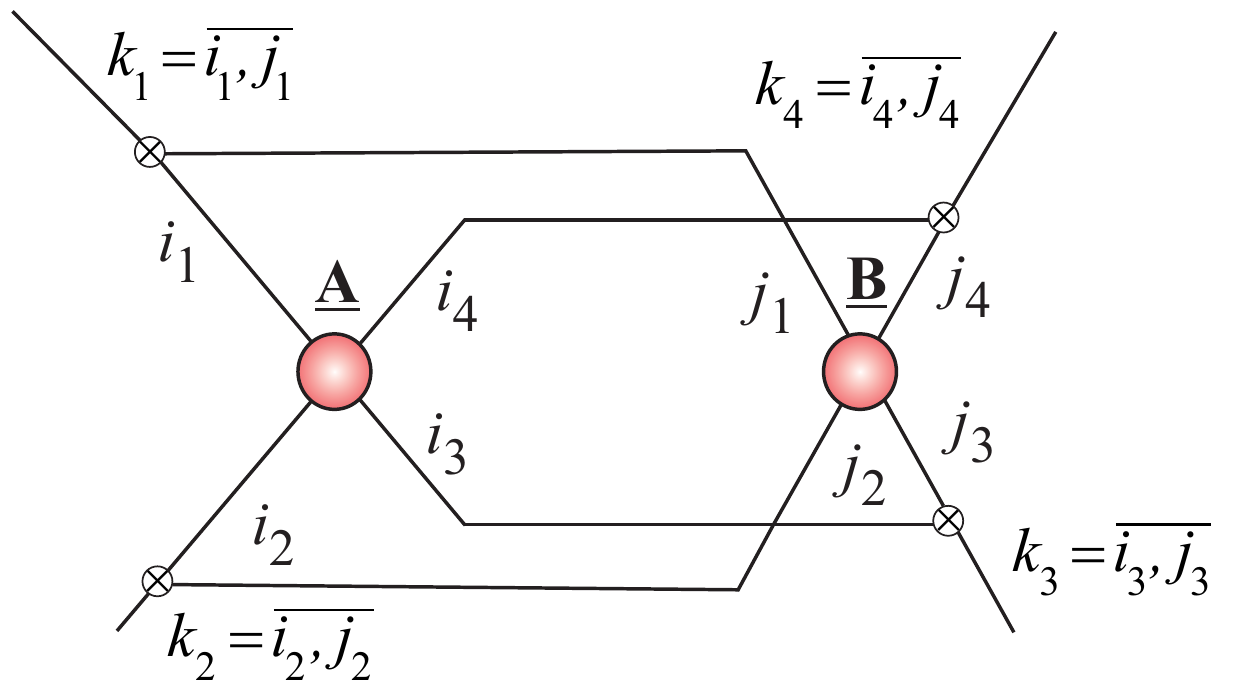}
\caption{Kronecker product of two 4th-order tensors yields a tensor $\underline \bC = \underline \bA \otimes \underline \bB \in \Real^{I_1 J_1 \times \cdots \times I_4 J_4}$,  with entries $c_{k_1,k_2,\ldots,k_4}  = a_{i_1, \ldots,i_4} \: b_{j_1, \ldots,j_4}$, where $k_n= \overline{i_n,j_n} =i_n \bar \otimes j_n = j_n+(i_n-1)J_n$ ($n=1,2,3,4$).}
\label{Fig:KP}
\end{figure}

\begin{figure}[ht!]
\centering
\includegraphics[width=8.6cm,height=9.1cm]{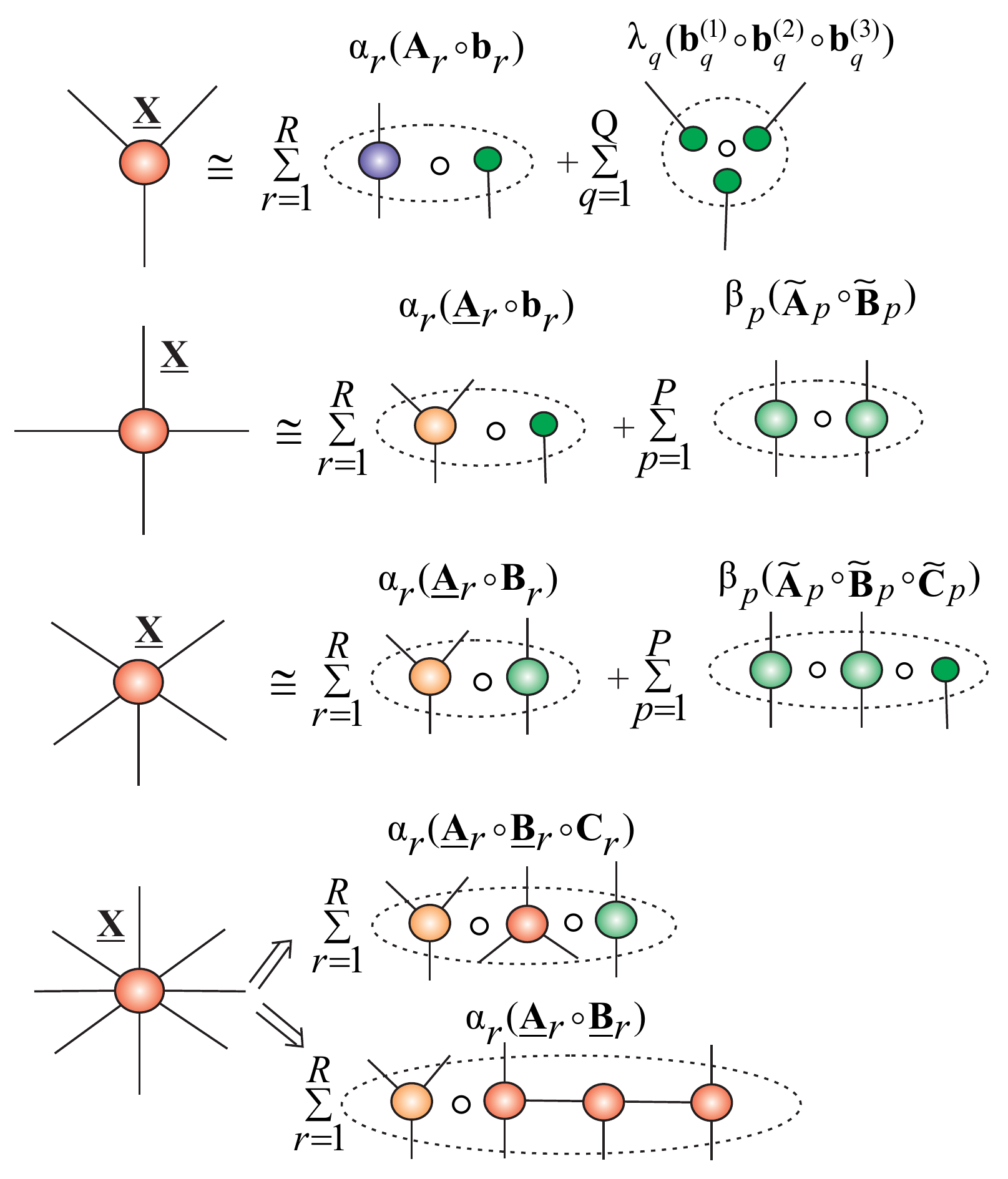}
\caption{Illustration of   Hierarchical Outer Product Tensor Approximation (HOPTA)
for higher-order data tensors of different orders. Each component tensor: $\underline
\bA_r$, $\underline \bB_r$  and/or $\underline \bC_r$ can be further decomposed using a
 suitable tensor network model.
%($\underline \bX =\sum_{r=1}^R \underline \bA \circ \underline \bB \circ \underline \bC$),
The model can be considered as an extension or generalization of the Block Term Decomposition (BTD) model to higher order tensors.}
\label{Fig:HOPTA}
\end{figure}

HOPTA and KTD models can be expressed mathematically in simple nested (hierarchical) forms, respectively
(see Fig. \ref{Fig:KP} and Fig. \ref{Fig:HOPTA}):
\be
\label{OPTD-model}
\underline \bX &\cong& \sum_{r=1}^R (\underline \bA_r \circ \underline \bB_r),\\
\underline {\tilde\bX} &=& \sum_{r=1}^R (\underline \bA_r \otimes \underline \bB_r),
\label{Phan-model2}
\ee
where each  factor tensor can be represented recursively as
$\underline \bA_r \cong  \sum_{r_1=1}^{R_1} (\underline \bA^{(1)}_{r_1} \circ \underline \bB^{(1)}_{r_1})$  or $\underline \bB_r \cong  \sum_{r_2=1}^{R_2} \underline \bA^{(2)}_{r_2} \circ \underline \bB^{(2)}_{r_2}$, etc.

The Kronecker product of two tensors: $\underline \bA \in \Real^{I_1 \times I_2 \times \cdots \times I_N}$ and $ \underline  \bB \in \Real^{J_1 \times J_2 \times \cdots \times J_N}$ yields $\underline  \bC = \underline \bA \otimes \underline \bB \in \Real^{I_1 J_2 \times \cdots \times I_N J_N}$, with entries $c_{i_1\otimes j_1,\ldots,i_N \otimes j_N} = a_{i_1, \ldots,i_N} \: b_{j_1, \ldots,j_N}$, where the operator $\bar \otimes$ for indices $i_n=1,2,\ldots, I_n$ and $j_n=1,2,\ldots,J_n$ is defined as follows $\overline{i_n,j_n} = j_n+(i_n-1)J_n$ (see Fig. \ref{Fig:KP}).

Note  that the $2N$th-order sub-tensors $\underline \bA_r \circ \underline \bB_r$ and $\underline \bA_r \otimes \underline \bB_r$ actually have the same elements,  arranged differently. For example,  if $\underline \bX = \underline \bA \circ \underline \bB$ and $\underline \bX' = \underline \bA \otimes \underline \bB$, where $\underline \bA  \in \Real^{J_1 \times J_2 \times \cdots \times J_N}$ and  $\underline \bB  \in \Real^{K_1 \times K_2 \times \cdots \times K_N}$, then $x_{j_1,j_2, \ldots,j_N,k_1,k_2, \ldots,k_N} = x'_{k_1+K_1(j_1-1), \ldots, k_N(K_N-1)}$.

%
%\begin{figure}[ht]
%\centering
%\includegraphics[width=8.6cm]{HOPTA2.eps}
%%\includegraphics[width=8.6cm]{Figure35.eps}
%%\includegraphics[width=8.6cm,height=9.1cm]{HOPTA.eps}
%%\includegraphics[width=8.8cm,height=9.1cm]{Fig36.eps}
%\caption{Illustration of   Hierarchical Outer Product Tensor Approximation (HOPTA)
%for higher-order data tensors with different orders.
%%($\underline \bX =\sum_{r=1}^R \underline \bA \circ \underline \bB \circ \underline \bC$).
%The model can be considered as extension or generalization of the BTD models for high-order tensors
% using tensor network diagrams.}
%\label{Fig:HOPTA}
%\end{figure}
%

\begin{figure}[ht]
\centering
\includegraphics[width=8.6cm,height=3.8cm]{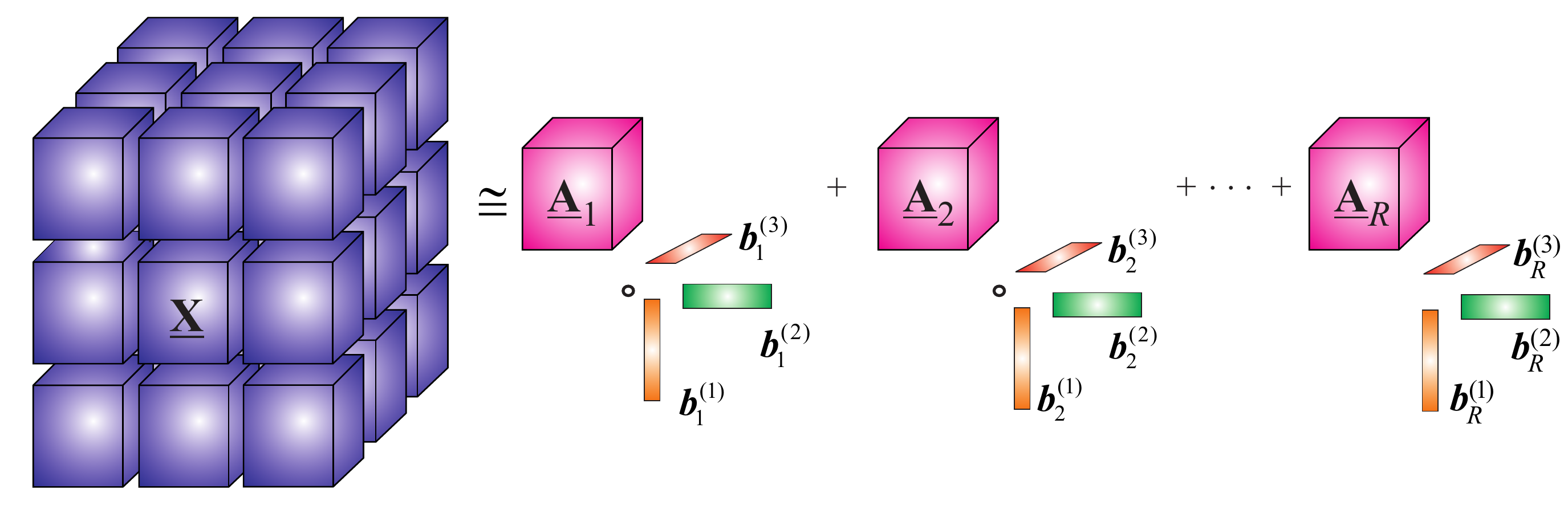}
\caption{Illustration of the decomposition of an 6th-order tensor using the BTD of  rank-$(L_r,L_r,1)$ as:
$\underline \bX =\sum_{r=1}^R \underline \bA_r \circ (\bb_r^{(1)} \circ \bb_r^{(2)} \circ \bb_r^{(3)} )$ \cite{Sorber2012b}, which can be considered as a special case of the HOPTA.}
\label{Fig:BTDS}
\end{figure}
%\begin{figure}[p]
%%\centering
%(a)\\
%\includegraphics[width=8.2cm,height=9.8cm]{TT-splitting-3.eps}\\
%(b)\\
%%\includegraphics[width=8.2cm,height=11.8cm]{Outerdec1.eps}
%\includegraphics[width=8.2cm,height=11.8cm]{Fig37.eps}
%\caption{Schematic representation of  a 9th-order tensor by (a) tensor train (TT=MPO or TT=MPS) and (b) HOPTA; in practice, we can use hierarchical decomposition which can be expressed, e.g., as $\underline \bX \cong (\sum_{r} \bA_{1r} \circ  \bA_{2r} \circ \bb_r) \circ (\sum_{p} \bB_{1p} \circ \bB_{2p})$.}
%\label{Fig:HOPTA9}
%\end{figure}

%Another promising tensor networks are Kronecker Tensor Decomposition
%(KTD) expressed as
%\be
%\underline \bX =\sum_{r=1}^R \underline \bA_r \otimes \underline \bB_r \otimes \underline \bC_r
%\ee
%and alternatively HOPTA
%\be
%\underline \bX =\sum_{r=1}^R \underline \bA_r \circ \underline \bB_r \circ \underline \bC_r
%\ee

 It is interesting to note that the KTD and HOPTA %\cite{Phan2012-Kron,Ragnarsson-PHD,Phan_BTDLxR}:
% \be
%\label{Phan-model}
%\underline \bX \cong\sum_{r=1}^R (\underline \bA_r \otimes \underline \bB_r),
%\underline \bX' &=& \sum_{r=1}^R (\underline \bA_r \circ \underline \bB_r),
%\label{Phan-model2}
%\ee
can be considered in special cases as a flexible form of Block Term Decomposition (BTD) introduced first by De Lathauwer \cite{Lath-BCM12,LathauwerTBSS,Lath-Nion-BCM3,Sorber2012b}.

%where $\underline \bX \in \Real^{I_1 \times I_2 \times \cdots \times I_N}$, $\underline \bB_r \in \Real^{K_{r1} \times K_{r2} \times \cdots \times K_{rN}}$
%%are of size ${K_{r1} \times K_{r2} \times \cdots \times K_{rN}}$,
%and $\underline \bA_r \in \Real^{J_{r1} \times J_{r2} \times \cdots \times J_{rN}}$, such that $I_n = J_{rn} \, K_{rn}$ \cite{Phan2012-Kron}.
%%Symbol $``\otimes''$ denotes
%The Kronecker tensor product of two tensors  $\underline \bA = [a_{\bj}]$ and $\underline \bB = [b_{\bk}]$ is defined as an $N$th-order tensor $\underline \bC = [c_{\bi}] \in \Real^{I_1 \times I_2 \times \cdots \times I_N}$, $\bi = [i_1, i_2, \ldots, i_N]$, $I_n = J_n K_n$ such that
%$c_{\bi} =  a_{\bj} \, b_{\bk}$, $\bj = [j_1, j_2, \ldots, j_N]$, $\bk = [k_1, k_2, \ldots, k_N]$, $i_n =  k_n + (j_n-1) K_n$.
%%

The definition of the tensor Kronecker product assumes that both core tensors $\underline \bA_r$
and $\underline \bB_r$  have the same order.
% thus  at first  KTD model  might  be seen as   too  restrictive. However,
It should be noted that vectors and matrices can be treated as tensors, e.g, matrix of dimension $I \times J$ can  be treated formally  as 3rd-order tensor of dimension $I \times J \times 1$.
In fact, from the KTD model, we can generate many existing and emerging TDs by
changing structures and orders of factor tensors: $\underline \bA_r$ and $\underline \bB_r$, for example:
\begin{itemize}
\item If $\underline \bA_r$ are rank-1 tensors of size $I_1 \times I_2 \times \cdots \times I_N$, and
$\underline \bB_r$  are scalars, $\forall r$, then (\ref{Phan-model2}) expresses the rank-$R$ CPD.
\item If $\underline \bA_r$ are rank-$L_r$ tensors in the Kruskal (CP) format, of size $I_1 \times I_2 \times \cdots \times I_R \times 1 \times \cdots \times 1$, and
$\underline \bB_r$ are rank-1 CP tensor of size $1 \times  \cdots \times 1 \times I_{R+1}\times \cdots \times I_N$, $\forall r$,  then (\ref{Phan-model2}) expresses the rank-($L_r \circ 1$) BTD \cite{Lath-BCM12}.
%
%%\item If $\underline \bA_r$ are rank-$L_r$ CP tensors and
%%$\underline \bB_r$ are rank-$M_r$ CP tensors, then $\tensor{\widehat X}_r$ are rank-$L_r M_r$ CPD tensors.
%%
%%\item If $\underline \bA_r$ and $\underline \bB_r$ are Tucker tensors, then $\tensor{\widehat X}_r$ are also  Tucker tensors.
%%
\item If $\underline \bA_r$ and $\underline \bB_r$ are expressed by KTDs, we have Nested Kronecker Tensor Decomposition (NKTD), where Tensor Train (TT) decomposition is a particular case  \cite{OseledetsTT11,Oseledets_TyrTT11,Khoromskij-TT}. In fact, the model (\ref{Phan-model2}) can be used for) the recursive TT-decompositions \cite{OseledetsTT11}.

\end{itemize}
In this way,  a large variety of  tensor decomposition models can be generated. However,  only some of them  yield unique decompositions and to date only a few have found concrete applications is scientific computing.

%Particulary promising are low-rank Block Term Decompositions  (BTDs)  (also called Block Component Decompositions -BCD) which have  already found application in blind source separation and wireless communications \cite{LathauwerTBSS,Lath-Nion-BCM3,Sorber2012b}.
 %%
% For example, a rank-$(L_r,L_r,1)$ BTD for 3rd-order data tensor $\underline \bX \in \Real^{I \times J \times K}$ decomposition is essentially unique if the matrices $[\bA_1,\bA_2,\ldots\bA_R]$ and
%$[\bB_1,\bB_2,\ldots\bB_R]$ are full column rank and the matrix $\bC = [\bc_1, \bc_2, \ldots, \bc_R]$ does not contain collinear columns, that admitting computation via generalized eigenvalue decomposition (GEVD).
%%
The advantage of HOPTA models over BTD  and KTD is that they are more flexible and can approximate very high order tensors with a relative small number of cores, and  they allow us to model more complex data structures. %\cite{VisaSP-09}.

%\begin{figure}[h!]
%\centering
%\includegraphics[width=8.6cm,height=3.8cm]{BTDS.eps}
%\caption{Illustration of decomposition of 6th-order tensor using BTD of  rank-$(L_r,L_r,1)$:
%$\underline \bX =\sum_{r=1}^R \underline \bA_r \circ (\bb_r^{(1)} \circ \bb_r^{(2)} \circ \bb_r^{(3)} )$ \cite{Sorber2012b}, which can be considered as a special case of the HOPTA.}
%\label{Fig:BTDS}
%\end{figure}

%\section{\bf Low-Rank Tensor Approximations via Tensor Networks}

\section{\bf Tensorization and Quantization -- Blessing of Dimensionality}
\label{sect:tensorization}

\subsection{\bf Curse of Dimensionality}

The term curse of dimensionality, in the context of tensors,  refers to the fact that the number of elements of an $N$th-order $(I \times I \times \cdots \times I)$ tensor, $I^N$, grows exponentially with the tensor order $N$.
 Tensors can easily become really big for very high order tensors since the size is exponentially growing  with  the number of dimensions ('ways', or 'modes').
For example, for the Tucker decomposition the number of entries of a original data tensor but also a core tensor  scales exponentially in the tensor order, for instance, the number of entries of an $N$th-order  $(R \times R \times \cdots \times R)$ core tensor is   $R^N$.

If all computations are performed on a CP tensor format and not on the raw data tensor itself, then instead of the original $I^N$  raw data entries, the number of parameters in a CP representation reduces to $N R I$, which scales linearly in $N$ and $I$  (see Table \ref{table_complexity}). This effectively bypasses the curse of dimensionality, however the CP approximation may involve numerical problems, since existing algorithms are not stable for high-order tensors. At the same time, existing algorithms for tensor networks, especially TT/HT ensure very good numerical properties (in contrast to CPD algorithms), making it possible to control an error of approximation i.e., to achieve a desired accuracy of approximation  \cite{Khoromskij-TT}.

 \subsection{\bf Quantized Tensor Networks}

  The curse of dimensionality can be  overcome through quantized tensor networks, which represents a tensor of possibly very
  high-order as a set of sparsely interconnected low-order and very low dimensions cores \cite{Khoromskij-TT,Oseledets10}.
  The concept of quantized tensor networks was first proposed by Khoromskij \cite{Khoromskij-SC} and Oseledets \cite{Oseledets10}.
  % (see also Table \ref{table_notation2}).
The very  low-dimensional  cores are interconnected via tensor contractions to provide an efficient, highly compressed low-rank representation of a data tensor.
\begin{figure}[t]
%\label{Fig:Tensorization}
%\begin{center}
(a)\\
 \includegraphics[width=8.6cm]{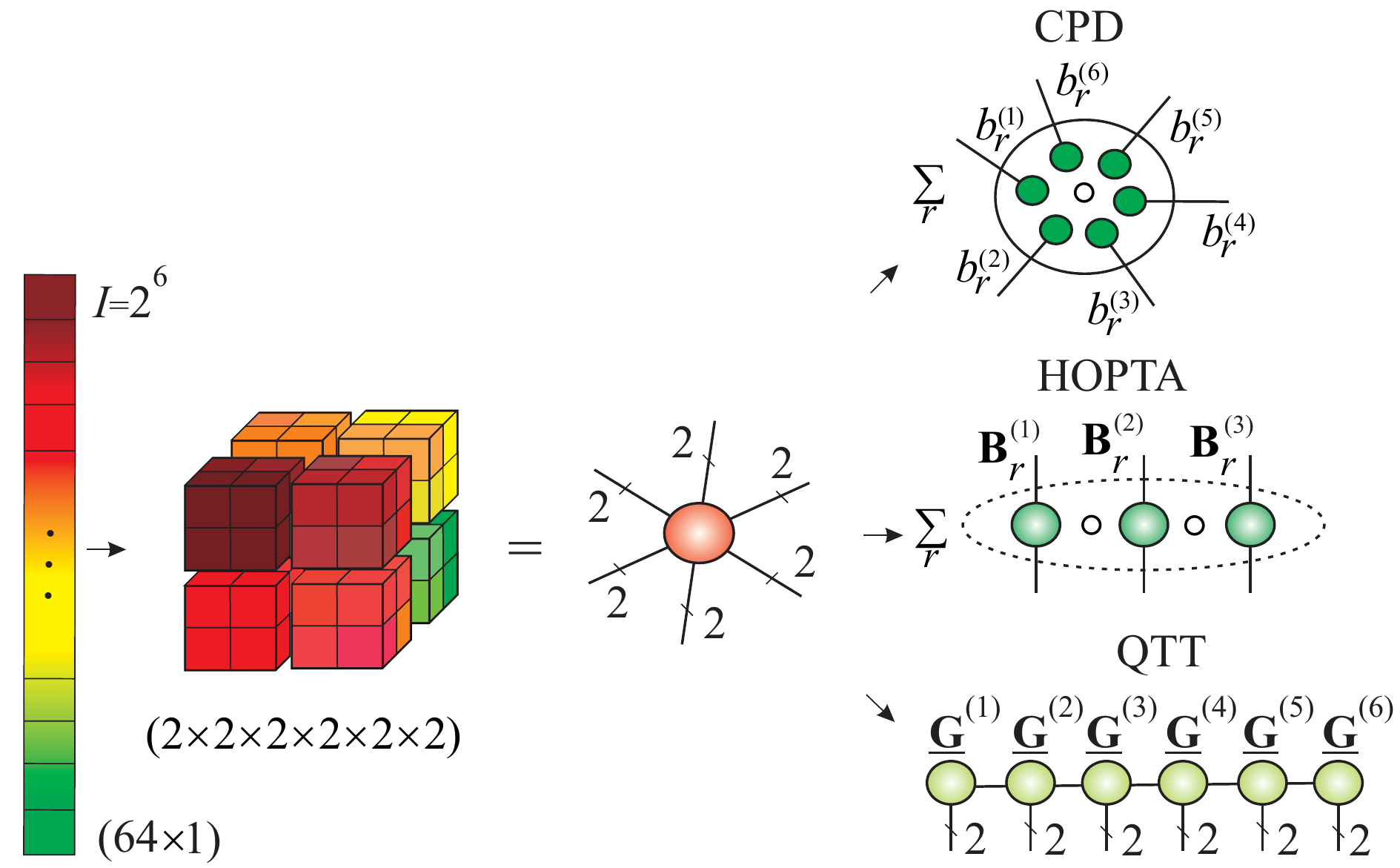}\\
 (b)\\
 \begin{center}
 \includegraphics[width=6.8cm]{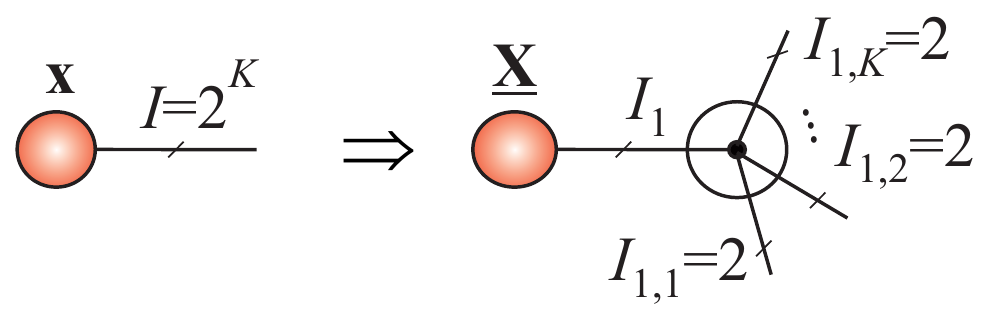}\\
 \end{center}
 (c)\\
 \begin{center}
   \includegraphics[width=7.8cm]{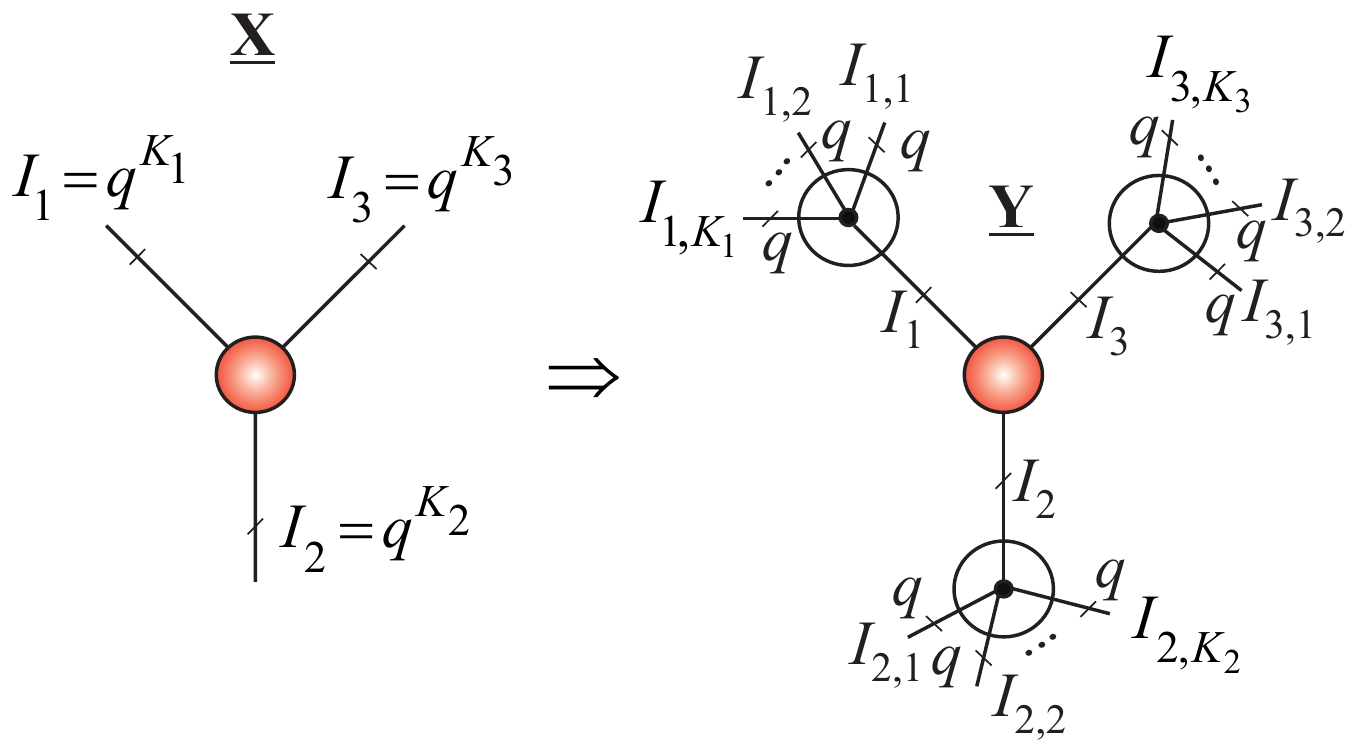}
 \end{center}
\caption{Tensorization. (a) Illustration of  the concept of tensorization of a (large-scale) vector ($I=2^K$) or matrix ($2^L \times 2^L$) into a higher-order tensor.
In order to achieve super-compression through a suitable quantized tensor decomposition (e.g., decomposition into rank-1 tensors $\underline \bX \cong \sum_{r=1}^R \bb_r^{(1)} \circ \bb_r^{(2)} \circ \cdots \circ \bb_r^{(6)}$ or rank-$q$  terms using  Hierarchical Outer Product Tensor Approximation (HOPTA) as: $\underline \bX \cong \sum_{\tilde r=1}^{\tilde R} \bA_{\tilde r} \circ \bB_{\tilde r} \circ \bC_{\tilde r}$ or Quantized Tensor Train (QTT). (b) Symbolic representation of tensorization of the vector $\bx \in \Real^{I}$ into $K$th-order
quantized tensor $\underline \bX \in \Real^{2 \times 2 \times \cdots \times 2}$. (c) Tensorization of a 3rd-order tensor $\underline \bX \in \Real^{I_1 \times I_2 \times I_3}$ into $(K_1+K_2+K_3)$th-order tensor
$\underline \bY \in \Real^{I_{1,1} \times \cdots \times I_{1,K_1} \times I_{2,1} \times \cdots \times I_{3,K_3}}$ with $ I_{n,k_n}=q$.}
\label{Fig:Tensorization}
\end{figure}

The procedure of creating a data tensor from lower-order original data is referred to as  tensorization.  In other words,  lower-order data tensors can be  reshaped (reformatted) into high-order tensors. The purpose of a such tensorization  is to achieve super compression \cite{Khoromskij-SC}.
 In general, very large-scale vectors or matrices can be easily  tensorized to higher-order tensors, then efficiently compressed by applying a suitable TT decomposition; this is the underlying principle for big data analysis  \cite{Oseledets10,Khoromskij-SC}.  For example,
the  quantization and tensorization  of a huge vector $\bx \in \Real^I$, $I = 2^K$ can be achieved through reshaping to give an
$(2 \times 2  \times \cdots \times 2)$  tensor $\underline \bX$ of order $K$,   as illustrated in Figure \ref{Fig:Tensorization} (a). Such a quantized tensor $\underline \bX$ often admits low-rank approximations, so that a good compression of a huge vector $\bx$ can be achieved by enforcing a maximum possible low-rank structure on the tensor network.

\begin{figure*}[ht]
(a)\\
 \begin{center}
 \includegraphics[width=8.6cm]{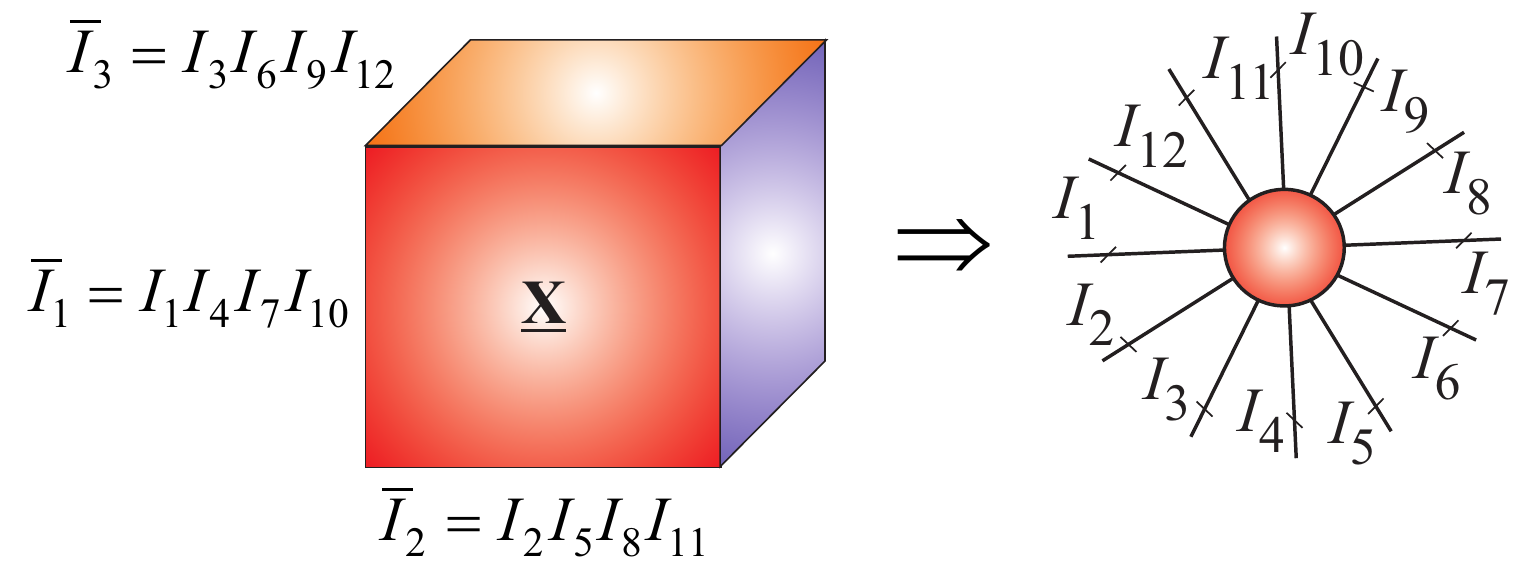} \\
  \end{center}
 (b)\\
 \begin{center}
 \includegraphics[width=15.6cm]{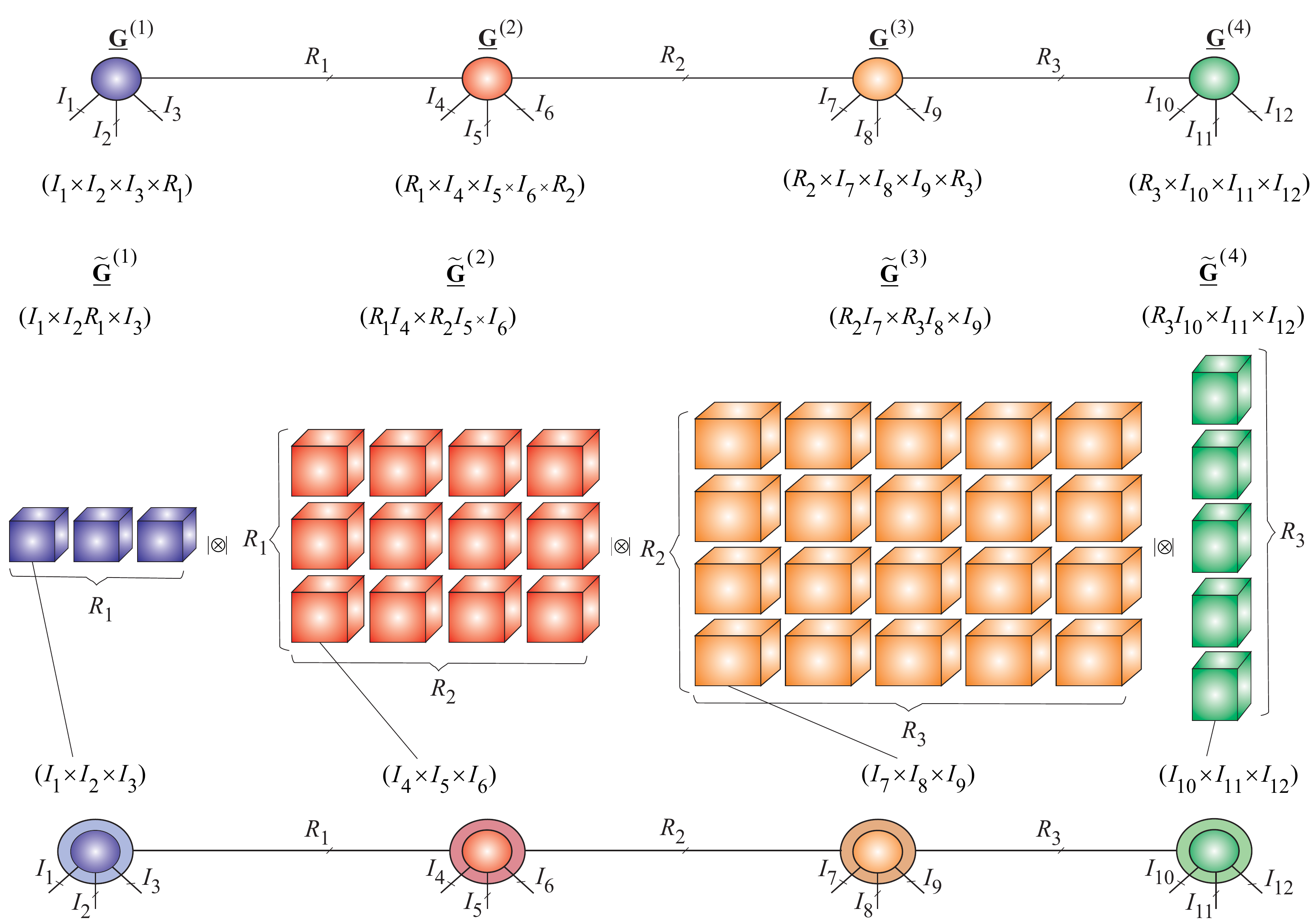}\\
 \end{center}
\caption{ Example of tensorization and decomposition of a large-scale 3rd-order tensor $\underline \bX \in \Real^{\overline I_1 \times \overline I_2 \times \overline I_3}$ into 12th-order tensor assuming that $\overline I_1 =I_1 I_4 I_7 I_{10}$, $\overline I_2 =I_2 I_5 I_8 I_{11}$ and $\overline I_3 =I_3 I_6 I_9 I_{12}$: (a) Tensorization and (b) representation of the higher-order tensor via generalized Tensor Train refereed to as the Tensor Product States (TPS). The  data tensor can be expressed by strong Kronecker product of block tensors as $\underline \bX \cong \underline {\widetilde \bG}^{(1)} \; |\otimes| \; \underline {\widetilde \bG}^{(2)} \;  |\otimes| \; \underline {\widetilde \bG}^{(3)} \; |\otimes| \; \underline {\widetilde \bG}^{(4)} \in
\Real^{\overline I_1 \times  \overline I_2  \times \overline I_3}$,
%\Real^{I_1 I_4 I_7 I_{10} \times I_2 I_5 I_8 I_{11} \times I_3I_6 I_9 I_{12}}, $
where each block of the core $\underline {\widetilde \bG}^{(n)} \in \Real^{R_{n-1} I_{3n-2} \times R_n I_{3n-1} \times I_{3n}}$ is a
3rd-order tensor of dimensions $(I_{3n-2} \times I_{3n-1} \times I_{3n})$, with $R_0=R_4=1$ for $n=1,2,3,4$.}
% The strong Kronecker product of two block cores  $\underline {\widetilde\bG}^{(n)} \in \Real^{R_{n-1} I_{3n-2} \times R_n I_{3n-1} \times I_{3n}}$ and
% $\underline {\widetilde\bG}^{(n+1)} \in \Real^{R_n I_{3n+1} \times R_{n+1} I_{3n+2} \times I_{3n+3}}$ is defined as the block tensor
%$\underline \bC = \underline {\widetilde\bG}^{(n)} |\otimes|  \underline {\widetilde\bG}^{(n+1)} =\sum_{r_n=1}^{R_n} \underline \bG^{(n)}_{r_{n-1},r_n} \otimes  \underline \bG^{(n+1)}_{r_n,r_{n+1}}$, where $\underline \bG^{(n)}_{r_{n-1},r_n} \in \Real^{I_{3n-2} \times I_{3n-1} \times I_{3n}}$ and  $\underline\bG^{(n+1)}_{r_{n},r_{n+1}} \in \Real^{I_{3n+1} \times I_{3n+2} \times I_{3n+3}}$ are block tensors of $\underline {\widetilde\bG}^{(n)}$ and $\underline {\widetilde\bG}^{(n+1)}$, respectively.}
\label{Fig:KTS}
\end{figure*}

Even more generally,  an $N$th-order tensor $\underline \bX \in \Real^{I_1 \times \cdots \times I_N}$, with  $I_n=q^{K_n}$,  can be quantized  in all modes simultaneously  to yield a
$(q \times q \times \cdots  q)$  quantized tensor $\underline \bY$ of  higher-order, with small $q$,  (see Fig. \ref{Fig:Tensorization} (c) and Fig. \ref{Fig:KTS}).

In the example shown in Fig. \ref{Fig:KTS} the Tensor Train of a huge 3rd-order tensor cab be represented by the strong Kronecker products of block tensors with relatively small 3rd-order blocks.

Recall that the strong Kronecker product of two block cores:
%\be
$\underline {\bG}^{(n)} = \begin{bmatrix}\underline \bG^{(n)}_{1,1}&\cdots& \underline \bG^{(n)}_{1,R_n}\\
				\vdots & \ddots & \vdots \\
				\underline \bG^{(n)}_{R_{n-1},1}&\cdots& \underline \bG^{(n)}_{R_{n-1},R_n}
		\end{bmatrix} \in \Real^{R_{n-1} I_{3n-2} \times R_n I_{3n-1} \times I_{3n}}\notag
$ %\ee
 and \\ \\
% \be
$ \underline {\bG}^{(n+1)} = \begin{bmatrix} \underline \bG^{(n+1)}_{1,1}&\cdots& \underline \bG^{(n+1)}_{1,R_{n+1}}\\
				\vdots & \ddots & \vdots \\
				\underline \bG^{(n+1)}_{R_{n},1}&\cdots& \underline \bG^{(n+1)}_{R_{n},R_{n+1}}
		\end{bmatrix} \in \Real^{R_n I_{3n+1} \times R_{n+1} I_{3n+2} \times I_{3n+3}} \notag
$ % \ee
  is defined as a block tensor
\be
\underline \bC &=& \underline {\bG}^{(n)} \; |\otimes| \; \underline {\bG}^{(n+1)} \notag \\
 && \in \Real^{R_{n-1}I_{3n-2} I_{3n+1} \times R_{n+1} I_{3n-1} I_{3n+2} \times I_{3n} I_{3n+3}},
\ee
with blocks $\underline \bG^{(n)}_{r_{n-1},r_{n+1}} =\sum_{r_n=1}^{R_n} \underline \bG^{(n)}_{r_{n-1},r_n} \otimes  \underline \bG^{(n+1)}_{r_n,r_{n+1}}$,    where $\underline \bG^{(n)}_{r_{n-1},r_n} \in \Real^{I_{3n-2} \times I_{3n-1} \times I_{3n}}$ and  $\underline\bG^{(n+1)}_{r_{n},r_{n+1}} \in \Real^{I_{3n+1} \times I_{3n+2} \times I_{3n+3}}$ are block tensors of $\underline {\bG}^{(n)}$ and $\underline {\bG}^{(n+1)}$, respectively.

 In practice, a fine ($q=2,3,4$ ) quantization is desirable to create as many virtual modes as possible, thus allowing us to implement efficient low-rank tensor approximations.
For example, the binary encoding ($q=2$) reshapes an $N$th-order tensor with $(2^{K_1} \times 2^{K_2} \times \cdots \times 2^{K_N})$ elements into a tensor of order $(K_1+K_2 +\cdots + K_N)$, with the same number of elements.
In other words, the idea of the quantized tensor is quantization of each $n$-th "physical" mode
(dimension) by replacing it with $K_n$ "virtual" modes, provided that the corresponding mode size $I_n$ are factorized as $I_n= I_{n,1} I_{n,2} \cdots I_{n, K_n}$. This corresponds to reshaping the $n$-th mode of size $I_n$ into $K_n$ modes of sizes $I_{n,1}, I_{n,2}, \ldots, I_{n,K_n}$.

The TT decomposition applied to quantized tensors is
referred to as the QTT; %the Quantics-TT or Quantized-TT (QTT);
it was first introduced as a compression scheme for large-scale matrices  \cite{Oseledets10}, and also independently  for  more general settings  \cite{Khoromskij-SC,QTT-Tucker,dolgovEIG2013,QTT-Laplace,kazeev2013LLA}.
The attractive property of QTT is that not only
its rank is typically small (below 10) but it is  almost  independent or at least uniformly bounded by data  size (even for  $I=2^{50}$), providing a logarithmic (sub-linear) reduction of storage requirements: ${\cal{O}}(I^N) \rightarrow {\cal{O}}(N \log_q(I))$ -- so-called super-compression  \cite{Khoromskij-SC}.

\minrowclearance 2ex
\begin{table}[t!]
\caption{Storage complexities of  tensor decomposition models for an $N$th-order  tensor $\underline \bX \in \Real^{I_1 \times I_2 \times \cdots \times I_N}$,
for which the original storage complexity is ${\cal{O}}(I^N)$, where $I=\max\{I_1,I_2,\ldots,I_N\}$, while $R$ is an upper bound on the ranks of tensor decompositions considered, that is  $R=\max\{R_1,R_2,\ldots, R_{N-1} \}$ or $R=\max\{R_1,R_2,\ldots, R_N \}$.}
%%i.e. for: CPD $(R_{CPD})$, Tucker $(R_{Tucker})$, TT $(R_{TT})$ and QTT $(R_{QTT})$ models, for
%which obey the following rank bounds:
%$R_{Tucker} \leq R_{CPD}, R_{TT} \leq R_{Tucker}, \;R_{TT} \leq R_{CPD}$ \cite{QTT-Tucker}.
 \centering
  {\small \shadingbox{
  %  \begin{tabular*}{0.5\linewidth}[t]{@{\extracolsep{\fill}}ll} \hline %\\
%\\[-2em]
\begin{tabular*}{0.99\linewidth}[t]{@{\extracolsep{\fill}}ll} \hline
1.  CPD   &  ${\cal{O}}(NIR)$ \\
2. Tucker   & ${\cal{O}}(NIR +R^N)$ \\
3. TT/MPS   & ${\cal{O}}(NIR^2)$ \\
4. TT/MPO   & ${\cal{O}}(NI^2R^2)$ \\
5.  Quantized TT/MPS (QTT)  &${\cal{O}}(N R^2 \log_q(I))$ \\
6.  QTT+Tucker  & ${\cal{O}}(N R^2 \log_q(I)+N R^3)$ \\
7. Hierarchical Tucker (HT) & ${\cal{O}}(N I R + N R^3)$  \\
\hline
    \end{tabular*}
   }}
%    \end{center}
\label{table_complexity}
\end{table}
\minrowclearance 0ex
Compared to the TT decomposition (without quantization), the QTT format often represents more deep structure in the data by introducing  some ``virtual'' dimensions.  The high compressibility of the QTT-approximation is a consequence of the noticeable separability properties in the quantized tensor for suitably structured data.

%It is interesting to emphasize  that the QTT-rank is usually relatively small, typically,  almost constant independent of a dimension of data. Moreover, the QTT provides a logarithmic (sub-linear) reduction of storage requirement -- so called super-compression: ${\cal{O}}(I^N) \rightarrow {\cal{O}}(N \log(I))$ and reduction of computational complexity \cite{Khoromskij-SC}.
%
The fact  that the TT/QTT ranks are often moderate or even low, e.g., constant or
growing linearly with respect to $N$ and constant or growing logarithmically with respect to $I$, is an important issue in the context of big data analytic and has been addressed so far mostly experimentally (see \cite{Khoromskij-TT,Khoromskij-SC,Grasedyck-rev} and references therein).
On the other hand, the high efficiency
of multilinear algebra in the TT/QTT algorithms based on the well-posedness of
the TT low-rank  approximation problems and the fact that such problems are solved quite efficiently  with the use of   SVD/QR, CUR and other cross-approximation techniques.

In general,  tensor networks  can  be considered as   distributed high-dimensional tensors built
up from many core tensors of low dimension through  specific tensor contractions. Indeed, tensor networks (TT, MPS, MPO, PEPS  and HT)  have already been successfully used
to solve  intractable problems in computational quantum chemistry  and in scientific computing \cite{Orus2013,DMRG2013,QTT-Laplace,kazeev2013LLA,KazeevT13,Kazeev-CME-2014}.

\begin{figure}[ht]
(a)\\
\begin{center}
\includegraphics[width=8.6cm]{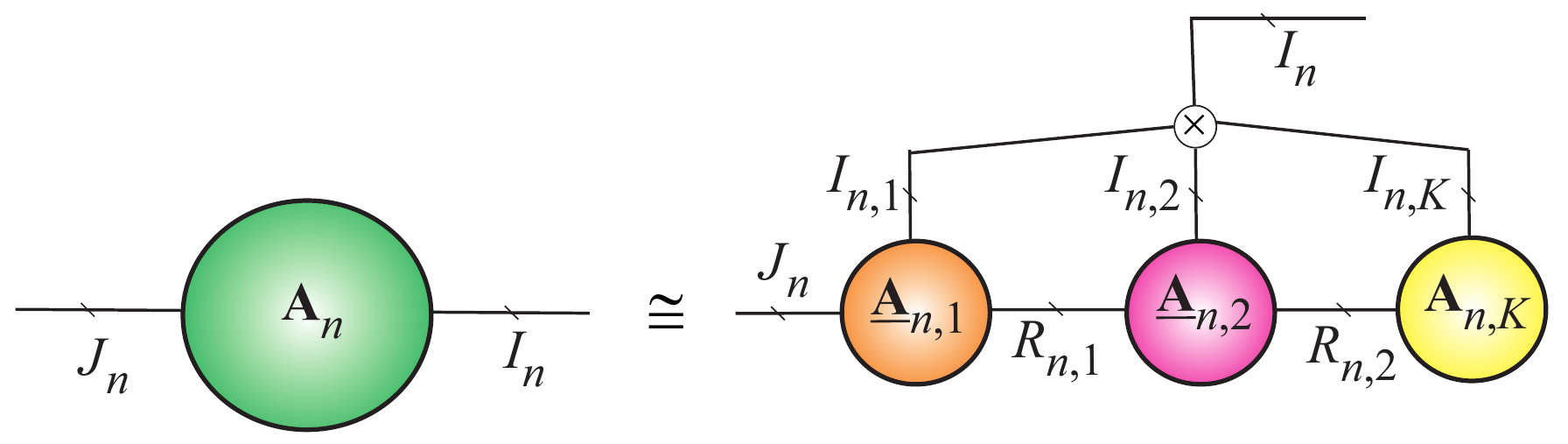}\\
\end{center}
(b)\\
\includegraphics[width=8.8cm]{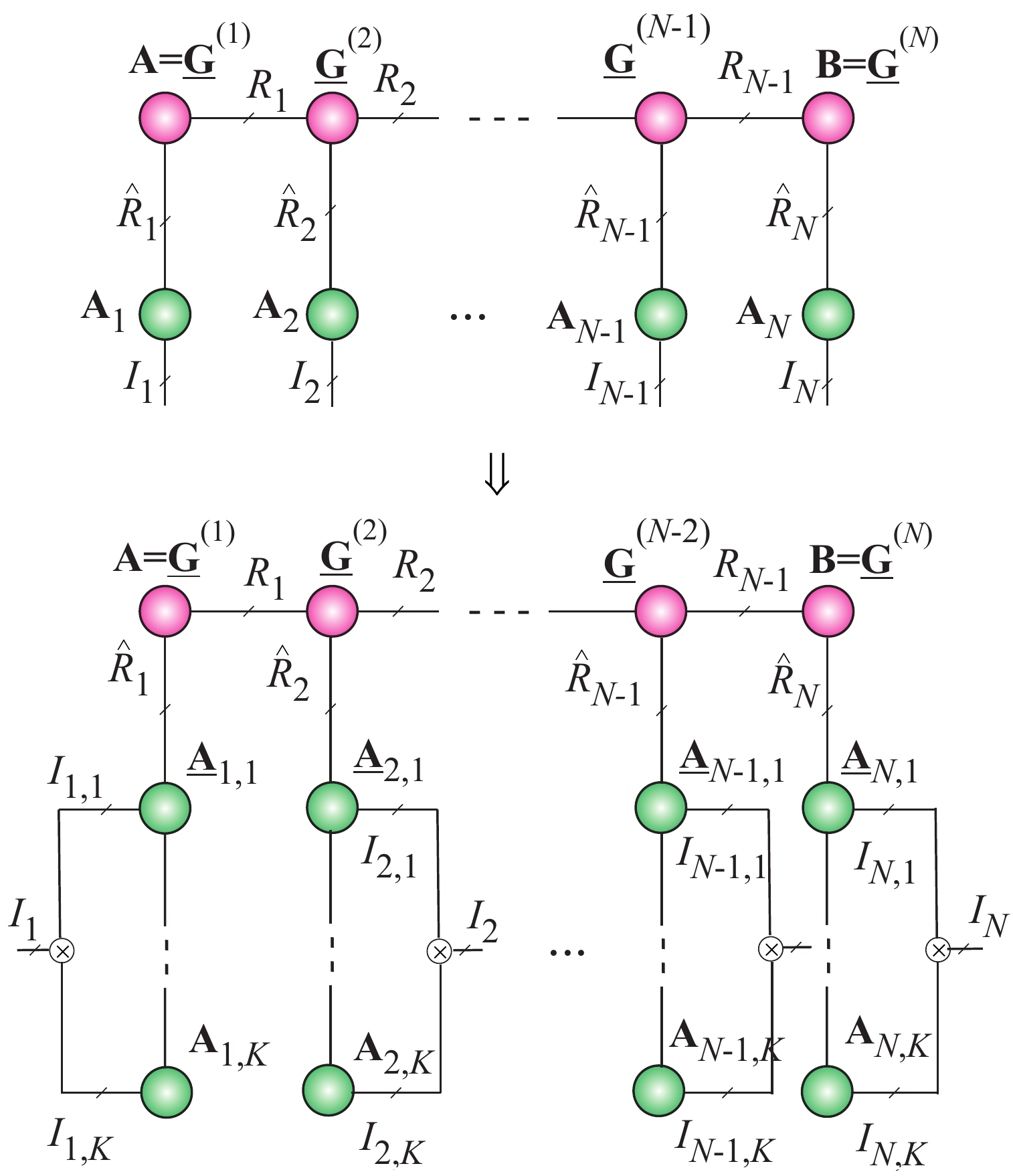}
%(a)\\
%\includegraphics[width=4.4cm,height=4.2cm]{matrixQTT.eps}
%(b)\\
%\includegraphics[width=8.6cm]{QTT-Tucker.eps}
\caption{QTT-Tucker format. (a) Distributed representation of a large matrix $\bA_n \in \Real^{I_n \times R_n}$ with large dimension of $I_n$ via QTT by tensorization of the matrix to high-order quantized tensor and next by performing QTT decomposition. (b) Distributed representation of a large-scale  $N$th-order Tucker model  via  a quantized TT model in which core tensor and all large-scale factor matrices $\bA_n$ ($n=1,2,\ldots,N$) are represented by tensor trains \cite{QTT-Tucker}.}
\label{Fig:QTT-Tucker}
\end{figure}

 However, in some cases, the ranks of the TT or  QTT formats
grow quite significantly  with the linearly increasing of  approximation accuracy. To overcome this
problem, new tensor models of  tensor approximation were developed, e.g., Dolgov and Khoromskij, proposed the QTT-Tucker format
\cite{QTT-Tucker} (see Fig. \ref{Fig:QTT-Tucker}), which
 exploits the TT approximation not  only for the Tucker core tensor, but also QTT  for the  factor matrices.
This model allows distributed computing, often  with  bounded ranks and to avoid
 the curse of dimensionality.
For very large scale tensors we can  apply a  more advanced approach in which factor matrices are tensorized to
higher-order tensors and then represented by TTs as illustrated in Fig. \ref{Fig:QTT-Tucker}.

Modern methods of  tensor  approximations combine many TNs and TDs formats including the CPD,  BTD,  Tucker, TT, HT decompositions and HOPTA (see Fig. \ref{Fig:HOPTA}) low-parametric tensor format.
The concept tensorization  and by representation  of a very high-order tensor in tensor network formats
(TT/QTT,  HT, QTT-Tucker) allows us to treat efficiently a very large-scale  structured  data that admit low rank tensor network approximations.
TT/QTT/HT tensor networks  have already found promising applications in very large-scale problems in scientific computing, such  as  eigenanalysis, super-fast Fourier transforms, and solving huge systems of large linear equations  (see \cite{dolgovEIG2013,KressnerEIG2014,QTT-Tucker,Huckle2013,Grasedyck-rev} and references therein).

In summary, the main concept or approach  is to apply
 a suitable tensorization and quantization of tensor data  and then perform approximative decomposition of this data into a
tensor network and finally perform all computations (tensors/matrix/vectors operations, optimizations) in tensor network formats.
%
%The employment  of the virtual tensorization (QTT, QCPD, Tucker-QTT, formats) allows us to treat efficiently even the cases of very large data  and  it is a  promising approach for big data analytic

\section{\bf Conclusions and Future Directions}

Tensor networks can be considered as a generalization and extension of TDs and are promising tools for the analysis of big data due to their extremely good compression abilities and distributed and parallel processing.
 TDs have already found  application in generalized multivariate regression, multi-way blind source separation,  sparse representation and coding, feature extraction, classification, clustering and data assimilation. Unique
advantages of tensor networks  include potential  ability of tera- or even peta-byte scaling and  distributed fault-tolerant computations.

Overall, the benefits of multiway  (tensor) analysis methods can be summarized as follows:
\begin{itemize}

\item ``Super'' compression of huge multidimensional, structured data which admits a low-rank approximation via TNs of  high-order tensors by extracting factor matrices and/or core tensors of low-rank and low-order and perform all mathematical manipulations in tensor formats (especially, TT and HT formats).

\item  A compact and very flexible  approximate representation of structurally rich data by accounting for their spatio-temporal and spectral dependencies.

 \item Opportunity to establish statistical links between cores, factors, components or hidden latent variables for blocks of data.

 \item Possibility to operate with noisy, incomplete, missing  data  by using powerful low-rank tensor/matrix approximation techniques.

\item  A framework to  incorporate  various diversities or constraints in different modes and thus naturally extend the standard (2-way) CA methods to large-scale multidimensional data.

\end{itemize}

Many  challenging problems related to low-rank tensor approximations remain  to be addressed.
\begin{itemize}

\item A whole new area emerges when several TNs which operate on different datasets are coupled or linked.

\item  As the complexity of big data increases, this requires more efficient iterative algorithms for their computation, extending beyond the ALS, MALS/DMRG, SVD/QR and CUR/Cross-Approximation class of algorithms.

\item Methodological approaches are needed to
determine the kind of constraints that should  be imposed on cores to extract desired hidden (latent) variables with meaningful physical interpretation.

\item We need  methods to reliably estimate the ranks of TNs,
 especially for structured data corrupted by noise and outliers.

\item The uniqueness of  various TN models under different constraints  needs to be investigated.

\item Special techniques  are needed for distributed computing  and  to save and process huge ultra large-scale tensors.

%\item Better understanding of the geometry of tensor
%in high dimensional spaces is necessary.

 \item Better visualization tools need to be developed to address large-scale tensor network representations.

\end{itemize}

In summary, TNs and TDs is a fascinating and perspective area of research with many   potential applications in multi-modal   analysis of massive big data sets.\\

{\bf Acknowledgments:}  The author wish to express his appreciation and gratitude to his colleagues, especially:  Drs. Namgill LEE,  Danilo MANDIC, Qibin ZHAO, Cesar CAIAFA, Anh Huy PHAN, Andr\'e ALMEIDA, Qiang WU, Guoxu ZHOU and Chao LI for reading the manuscript and giving him comments and suggestions.

\small
%\footnotesize
%\singlespacing

%\bibliographystyle{IEEEtran}
%\bibliography{SPM}

%% Generated by IEEEtran.bst, version: 1.13 (2008/09/30)

\end{document}